%% file: main.tex
\documentclass{article}

\usepackage[utf8]{inputenc}
\usepackage{hyperref}
\usepackage{graphicx}
\usepackage[numbers]{natbib}
\usepackage{amsmath}
\usepackage{amssymb}
\usepackage{subcaption}
\usepackage{multirow}
\usepackage{booktabs}
\usepackage{longtable}
\usepackage[a4paper, total={6.5in, 8.66in}]{geometry}
\usepackage[dvipsnames]{xcolor}
\usepackage{csquotes}
\usepackage{microtype}
\usepackage{rotating}
\usepackage{siunitx}
\usepackage{tikz}
\usetikzlibrary{arrows.meta, positioning, fit, calc, backgrounds}

\usepackage[linesnumbered,ruled,vlined]{algorithm2e}
\let\oldnl\nl
\newcommand{\nonl}{\renewcommand{\nl}{\let\nl\oldnl}}

\SetCommentSty{mycommfont}

\hypersetup{
    pdftitle={Reinforcement Learning in High-Energy Physics: A Survey from Accelerator Control to Theory Exploration},
    pdfauthor={Luca Anzalone},
    pdfkeywords={reinforcement learning, high-energy physics, particle physics,
                 accelerator control, jet physics, particle tracking, detector design,
                 beyond standard model, data quality monitoring},
    colorlinks=true,
    linkcolor=NavyBlue,
    citecolor=NavyBlue,
    urlcolor=NavyBlue,
}

\title{\huge Reinforcement Learning in High-Energy Physics: A Survey from Accelerator Control to Theory Exploration}

\author{
    \href{https://orcid.org/0000-0002-0399-8836}{%
        \includegraphics[scale=0.06]{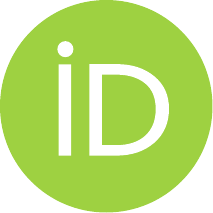}\hspace{1mm}Luca Anzalone}\\
    \small Independent researcher \\
    \texttt{luca.anzalone@proton.me}
}
\date{}

\begin{document}

\maketitle

% ------------------------------------------------------------------ Abstract
\begin{abstract}
Reinforcement learning~(RL) is a machine learning paradigm in which an agent
learns a decision-making policy through repeated interaction with an environment,
guided by a scalar reward signal rather than explicit supervision.
Supervised and unsupervised learning have become indispensable tools in
high-energy physics~(HEP), but they learn from fixed datasets, whereas many HEP
tasks are closed-loop decisions: the data an algorithm observes depend on the
choices it makes, the feedback is sparse or delayed, and the objective is often
measured on a real machine or through a stochastic simulation rather than
computed by a differentiable model.
This survey reviews the application of RL to such tasks across HEP: accelerator
control and feedback, data quality monitoring, jet grooming, clustering, and
representation, charged particle tracking, beyond-the-Standard-Model~(BSM) and
string model building, detector design, trigger control, analysis, and formal
quantum field theory computation.
It provides a self-contained introduction to the RL methods these applications
rely on, covering model-free, model-based and planning, multi-agent, offline,
inverse and imitation, and bandit approaches; it reviews each domain against its
traditional methods and summarizes the available head-to-head comparisons across
domains.
The applications span a gradient of maturity, from policies deployed on real
accelerators and evaluated on recorded collision data to exploratory studies in
formal theory.
RL is most advantageous for repeated tasks with a fast simulator, decisions
inside fast control loops, operation under faults and corrupted measurements,
physics-based rewards in place of labels, and sequential discrete constructions
with a reusable policy, while Bayesian optimization,
genetic algorithms, and differentiable methods remain competitive elsewhere.
We conclude with the open challenges that recur across domains, including reward
design, sim-to-real transfer, safety, the lack of shared benchmarks, and the
statistical validity of policy-driven decisions.
\end{abstract}

\tableofcontents
\clearpage

% 1. Introduction
\section{Introduction}
\input{sections/1_intro}

% 2. RL Background
\section{Reinforcement Learning}
\label{sec:rl}
\input{sections/2_rl}

\clearpage

% 3. Accelerator Control
\section{Accelerator Control and Beam Optimization}
\label{sec:accelerators}
\input{sections/3_accelerators}

\clearpage

% 4. Data Quality Monitoring
\section{Data Quality Monitoring}
\label{sec:dqm}
\input{sections/4_dqm}

% 5. Jet Physics
\section{Jet Physics}
\label{sec:jets}
\input{sections/5_jets}

% 6. Charged Particle Tracking
\section{Charged Particle Tracking}
\label{sec:tracking}
\input{sections/6_tracking}

% 7. BSM Model Building
\section{Beyond the Standard Model: Model Building and Theory Exploration}
\label{sec:bsm}
\input{sections/7_bsm}

% 8. Detector Design
\section{Detector and Instrument Design}
\label{sec:detectors}
\input{sections/8_detectors}

% 9. Analysis and Theoretical Computation
\section{Analysis, Event Processing, and Theoretical Computation}
\label{sec:analysis}
\input{sections/9_analysis}

% 10. Conclusions
\clearpage
\section{Conclusions and Outlook}
\label{sec:conclusions}
\input{sections/10_conclusions}

% ------------------------------------------------- Declaration of generative AI use
\section*{Declaration of Generative AI Use}
\addcontentsline{toc}{section}{Declaration of Generative AI Use}
During the preparation of this work, the author used Claude Opus 5.5 (Anthropic),
through the Claude Code assistant, to explore related and recent literature
starting from an initial set of articles selected and reviewed by the author, to
edit the language and structure of the manuscript, to draft and revise passages
of text and figure captions, and to design figures and generate their TikZ
source code.
All output of the tool was reviewed, edited, and verified by the author against
the original sources, and every bibliographic reference was checked by the
author.
The tool was not used to produce data or results.
The author takes full responsibility for the content of this publication.

% ------------------------------------------------------------------ References
\clearpage
\bibliographystyle{ieeetr}
\bibliography{main}

\end{document}

%% file: sections/1_intro.tex
% =============================================================================
%  1. Introduction
% =============================================================================

\subsection{Why Reinforcement Learning for High-Energy Physics?}
\label{sec:intro:motivation}

High-energy physics~(HEP) is the science of the fundamental constituents of
matter and their interactions, studied by colliding particles at high energies
and observing the resulting debris with large, specialized equipment, called
particle detectors.
At the Large Hadron Collider~(LHC) at CERN, proton bunches cross at a rate of
40~MHz, at a center-of-mass energy of 13.6~TeV in Run~3~(14~TeV by design), and
the collisions are recorded by detectors containing of order $10^8$ readout
channels, generating petabytes of data annually.
The fundamental challenge is to extract physics signals, such as new particles
or deviations from the Standard Model~(SM) predictions, from events produced by
known background processes that outnumber them by many orders of magnitude.
Extracting a signal requires a multi-stage pipeline.
Recorded data go through triggering, reconstruction, and event selection; the
backgrounds are then estimated, and the data are compared against simulated
samples produced with event generators such as
Pythia8~\cite{Bierlich2022Pythia8} and detector-response models such as
Geant4~\cite{Agostinelli2003GEANT4}.
A discovery is only claimed once the observed excess reaches a significance of five standard
deviations relative to the background-only hypothesis.

Machine learning has long been part of this pipeline, from boosted decision
trees for particle identification~\cite{Roe2005BDT} to deep learning for fast
detector simulation~\cite{ATLASCollaboration2022AtlFast3,
Buhmann2023CaloClouds}, anomaly detection~\cite{DAgnolo2019LearningNewPhysics,
Heimel2019QCDOrWhat,anzalone2024triggering}, and signal-background
classification~\cite{Baldi2016ParamNN,DBLP:journals/mlst/AnzaloneDB22}.
The vast majority of these applications, however, belong to two paradigms:
\textit{supervised learning}, which requires labeled examples, and
\textit{unsupervised} or \textit{self-supervised learning}, which exploits the
structure of unlabeled data.
In both, the model learns from a dataset whose distribution is fixed in advance
and does not depend on the outputs of the model itself.

Reinforcement learning~(RL)~\cite{Sutton2018RL} is a third paradigm,
complementary to the other two, in which an agent learns a \textit{policy},
i.e., a mapping from observations to actions, by interacting with an environment
and receiving scalar reward signals.
The agent is not told \textit{which} action is correct; it discovers \textit{how}
to act through trial and error, and the data it learns from depend on the
actions it takes.
This makes RL suited to HEP problems that can be formulated as
\textit{sequential decision-making under uncertainty}, in particular when the
reward is \textit{sparse}, i.e., most transitions carry zero or uninformative
reward, or \textit{delayed}, i.e., the consequences of an action appear only many
steps later, so that credit must be assigned across the whole sequence of
decisions.

Many HEP problems have this structure.
A particle accelerator must be tuned and stabilized as beam conditions change,
under hard safety constraints, with an objective such as beam transmission or
orbit stability that is measured on the real machine rather than computed by a
differentiable model.
A data quality monitoring system must decide, for each incoming histogram,
whether to classify it automatically or to query a human shifter, trading
accuracy against annotation effort.
A trigger must adapt its thresholds during a fill to keep its output rate within
the bandwidth budget as pileup and background composition change.
A jet grooming algorithm must decide, node by node along the clustering tree,
which soft emissions to remove.
A detector layout is built one layer at a time and scored only once complete,
through a stochastic full simulation.
A theorist searching for consistent beyond-the-Standard-Model~(BSM) models or
string compactifications must
navigate a discrete, combinatorially vast space of symmetry groups, charge
assignments, and compactification data, where consistency can only be checked
on a complete construction.
In each case, the problem maps onto a Markov decision process~(MDP): there are states,
actions, transition dynamics, a reward, and a horizon.

Sequential structure and a non-differentiable objective are not, by themselves,
sufficient reasons to prefer RL.
Bayesian optimization~\cite{Roussel2024BOReview}, evolution
strategies~\cite{Hansen2003CMAES}, and genetic
algorithms~\cite{Cole2021GAandRL, Abel2022GAvsRL} also optimize black-box
objectives, and gradients through an accurate differentiable model of the system
are more efficient than model-free exploration when such a model is available,
whether used directly~\cite{Dorigo2023MODEWhitePaper} or inside RL
training~\cite{Rajput2025DDRL}.
What distinguishes RL from these alternatives is that it returns a
\textit{policy}, a map from every state to an action, rather than a single
optimum, and that this policy acts in closed loop: each action changes the state
that later decisions observe.
The evidence reviewed in this survey identifies five conditions under which this
makes RL the more suitable choice:
\begin{itemize}
    \item \textbf{Repeated tasks with a fast simulator.} The training cost is
    paid once and amortized over many deployments of the same
    policy~\cite{Kaiser2024RLOvsBO}.
    \item \textbf{Decisions inside fast control loops.} A trained policy can act
    within loop periods too short for iterative optimizers, with microsecond
    inference on dedicated hardware~\cite{StJohn2021Booster,
    Scomparin2026KARA}.
    \item \textbf{Robustness to faults and corrupted measurements.} Policies
    trained over a distribution of operating conditions degrade more gracefully
    than classical controllers when the system itself
    degrades~\cite{Grech2022LHCTune}.
    \item \textbf{Physics as the training signal in place of labels.} The reward
    can encode first-principles knowledge where ground truth is unavailable,
    such as a scattering likelihood~\cite{Kortus2023Tracking} or a matrix
    element~\cite{Dillon2026TheoryRL}.
    \item \textbf{Sequential discrete construction with a reusable policy.} When
    the objective is defined only on a complete construction, a trained policy
    transfers across instances~\cite{Larfors2020LineBundle}.
\end{itemize}
In the last setting, learned policies have occasionally exposed structure that
experts had not articulated, as in a result on chiral families in string
compactifications that was subsequently proven~\cite{Halverson2019BranesBrains}.
Conversely, Bayesian optimization remains preferable in the opposite regime;
Section~\ref{sec:conclusions:closing} returns to these conditions with the
supporting evidence from the full review.

These properties have led to a growing body of work applying RL across the HEP
workflow, from accelerator operation and detector design to analysis and formal
theoretical computation.
We use the term HEP broadly: besides collider experiment and phenomenology, we
include accelerator physics, the operation of detectors in non-collider
experiments such as cryogenic dark matter searches~\cite{CRESST2023RL}, and formal
theory, from string phenomenology to the conformal bootstrap.

Existing reviews either catalog machine learning papers without a methodological
synthesis or focus on a single domain.
The HEP-ML Living Review~\cite{Feickert2021LivingReview} and its
successor~\cite{Krause2026LivingGuide} list RL papers within a broad catalog of
machine learning references for particle physics, but do not compare them.
Edelen and Huang~\cite{Edelen2024AnnualReview} review machine learning for
accelerator design and control, including notable RL applications, and
Ghribi~\cite{Ghribi2025AccelLivingReview} proposes a living review pipeline for
artificial intelligence and machine learning~(AI/ML) applications in accelerator
physics.
Ruehle~\cite{Ruehle2020StringDataScience} reviews data science methods for
string theory, RL included; Giarnetti and Meloni~\cite{Giarnetti2026FlavourRL}
focus on RL techniques for the flavor problem; and Martín-Guerrero and
Lamata~\cite{MartinGuerrero2021RLPhysics} review the interplay of RL and physics
in general, without a focus on HEP.
To the best of our knowledge, this is the first survey dedicated to RL across
HEP sub-fields: accelerator control, data quality monitoring, jet physics,
charged particle tracking, BSM and string model building, detector design,
trigger control, analysis, and theoretical computation.
Its contributions are: (i)~a self-contained RL background calibrated for HEP
researchers; (ii)~a per-domain contrast of RL with traditional methods, with the
available head-to-head evidence summarized across domains in
Table~\ref{tab:summary}; and (iii)~a discussion of open problems and future
research directions at the RL--HEP interface.\footnote{We use the term
\textit{reinforcement learning} broadly to include deep RL, tabular RL, bandit
algorithms, planning methods such as Monte Carlo Tree Search that use learned
policy and value functions, and the RL fine-tuning of pretrained generative
models, such as with group relative policy optimization~(GRPO)~\cite{Shao2024GRPO}.}

% -----------------------------------------------------------------------------
\subsection{Organization of the Paper}
\label{sec:intro:organization}

Section~\ref{sec:rl} covers the necessary reinforcement learning background.
It introduces the RL framework from classical principles to the main deep RL
algorithms, including model-based RL and planning with Monte Carlo Tree Search.
It then briefly covers the RL sub-fields most relevant to the surveyed
applications: multi-agent RL, offline RL, inverse RL and imitation learning,
bandit algorithms, and, more briefly, meta, multi-task, continual, hierarchical,
Bayesian, and multi-objective RL.
It closes with a discussion of practical challenges that are especially relevant
in the HEP setting: reward design, sim-to-real transfer, safety constraints,
sample efficiency, real-time inference and deployment, and physics-informed RL.

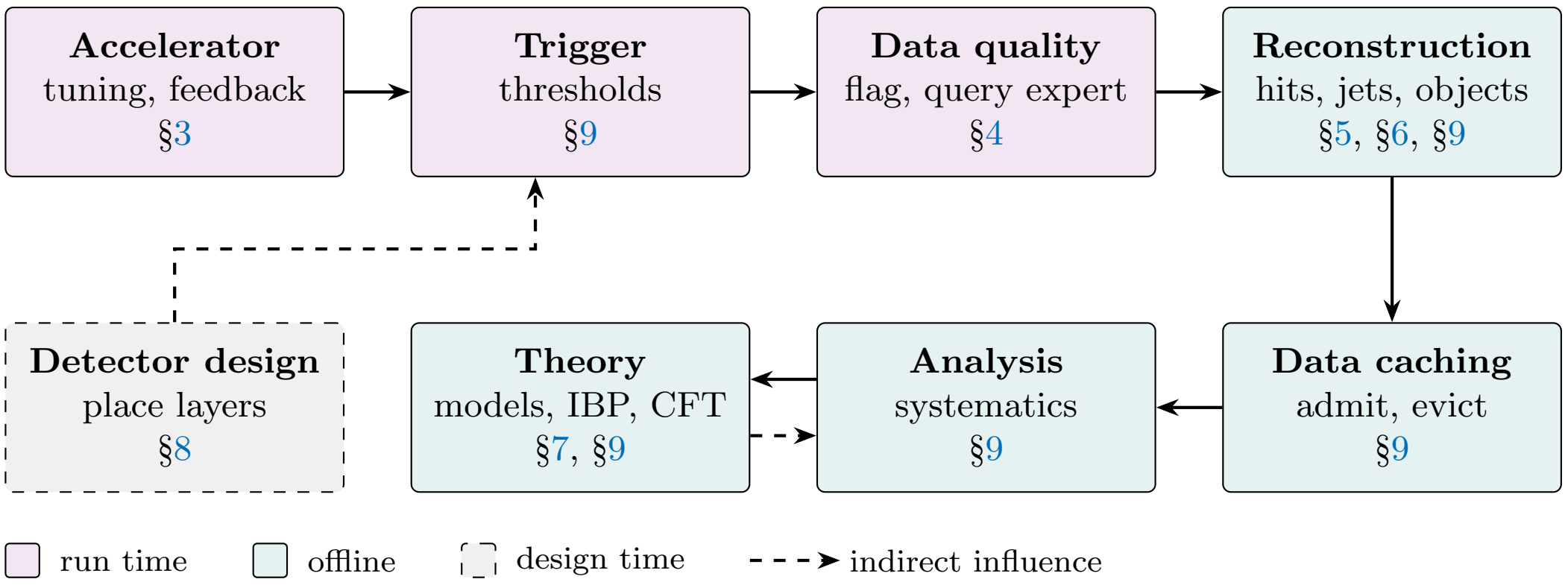
\begin{figure}[t]
    \centering
    \input{src/1/workflow.tex}
    \caption{RL decision points along the HEP workflow, each labeled with the
    decision taken by the agent and the section that reviews it.
    Run-time applications act on the accelerator and on the data stream,
    offline applications act on recorded data and on theory, and detector
    design precedes data taking.
    Dashed arrows denote indirect influence: the designed detector determines
    the data that reach the trigger, and theory predictions inform the
    analyses that constrain them.}
    \label{fig:workflow}
\end{figure}

Sections~\ref{sec:accelerators} to~\ref{sec:analysis} review RL applications
across seven domains of HEP.
Sections~\ref{sec:accelerators} to~\ref{sec:detectors} follow the same structure:
a background subsection introducing the physical problem and its traditional
algorithmic solutions, a detailed review of existing RL-based work, and a closing
subsection identifying open problems and promising future research directions.
Section~\ref{sec:analysis} collects heterogeneous applications and therefore
follows this structure within each of its subsections.
Figure~\ref{fig:workflow} places the decision points of the seven domains along
the HEP workflow, from data taking to theory.
The seven domains are:
\begin{itemize}
    \item \textbf{Section~\ref{sec:accelerators}}: Accelerator control and beam
          optimization, the most mature application area, covering beam
          steering and tuning, model-based and hybrid control, real-time
          feedback and stabilization, and the control of cryogenic and
          specialized equipment.
    \item \textbf{Section~\ref{sec:dqm}}: Data quality monitoring, where
          human-in-the-loop RL and bandits are used to reduce expert annotation
          effort and to detect detector malfunctions online.
    \item \textbf{Section~\ref{sec:jets}}: Jet physics, covering RL-based jet
          grooming, hierarchical clustering, and the ordering of particle
          constituents for downstream classifiers.
    \item \textbf{Section~\ref{sec:tracking}}: Charged particle tracking, where
          RL formulates track finding as a graph MDP with physics-informed
          rewards, extended to constrained multi-agent settings.
    \item \textbf{Section~\ref{sec:bsm}}: Beyond-the-Standard-Model model
          building, where RL explores the discrete space of consistent
          theories: Froggatt--Nielsen flavor and axion models, graph- and
          grammar-based model spaces, and string compactifications.
    \item \textbf{Section~\ref{sec:detectors}}: Detector and instrument design,
          where RL optimizes geometry and configuration with simulation-based
          rewards.
    \item \textbf{Section~\ref{sec:analysis}}: Analysis, event processing, and
          theoretical computation; a heterogeneous category covering trigger
          control, data caching for distributed analysis, background and
          systematics modeling, combinatorial event reconstruction, and formal
          quantum field theory~(QFT) computation (integral reduction, symbolic
          simplification, and the conformal bootstrap).
\end{itemize}

Section~\ref{sec:conclusions} summarizes the landscape, including the
cross-domain comparison of Table~\ref{tab:summary}, discusses the open challenges
that recur across domains, and outlines the conditions under which RL is
advantageous at the RL--HEP interface.

%% file: src/1/workflow.tex
% RL decision points along the HEP workflow (Section 1).
% Requires: \usepackage{tikz}
%           \usetikzlibrary{arrows.meta, positioning, fit, calc, backgrounds}
\begin{tikzpicture}[
    font=\small,
    >={Stealth[length=2mm]},
    block/.style={draw, rounded corners=2pt, align=center, minimum height=1.5cm,
                  minimum width=3.0cm, inner sep=3pt},
    runtime/.style={block, fill=violet!10},
    offline/.style={block, fill=teal!10},
    design/.style={block, fill=black!6, dashed},
    loop/.style={->, thick},
    flow/.style={->, thick, dashed},
    lbl/.style={font=\footnotesize, inner sep=2pt},
    swatch/.style={draw, rounded corners=1pt, minimum width=3mm, minimum height=3mm,
                   inner sep=0pt},
]

% --- Top row: run time, then reconstruction --------------------------------
\node[runtime] (acc)   at (1.6, 0)  {\textbf{Accelerator}\\tuning, feedback\\\S\ref{sec:accelerators}};
\node[runtime] (trig)  at (5.2, 0)  {\textbf{Trigger}\\thresholds\\\S\ref{sec:analysis}};
\node[runtime] (dqm)   at (8.8, 0)  {\textbf{Data quality}\\flag, query expert\\\S\ref{sec:dqm}};
\node[offline] (reco)  at (12.4, 0) {\textbf{Reconstruction}\\hits, jets, objects\\\S\ref{sec:jets}, \S\ref{sec:tracking}, \S\ref{sec:analysis}};

% --- Bottom row: design time and offline (data flow right to left) -------
\node[design]  (des)   at (1.6, -2.8)  {\textbf{Detector design}\\place layers\\\S\ref{sec:detectors}};
\node[offline] (theo)  at (5.2, -2.8)  {\textbf{Theory}\\models, IBP, CFT\\\S\ref{sec:bsm}, \S\ref{sec:analysis}};
\node[offline] (ana)   at (8.8, -2.8)  {\textbf{Analysis}\\systematics\\\S\ref{sec:analysis}};
\node[offline] (cache) at (12.4, -2.8) {\textbf{Data caching}\\admit, evict\\\S\ref{sec:analysis}};

% --- Data chain (solid) --------------------------------------------------
\draw[loop] (acc) -- (trig);
\draw[loop] (trig) -- (dqm);
\draw[loop] (dqm) -- (reco);
\draw[loop] (reco) -- (cache);
\draw[loop] (cache) -- (ana);
\draw[loop] ([yshift=2.5mm]ana.west) -- ([yshift=2.5mm]theo.east);

% --- Indirect influence (dashed) -----------------------------------------
\draw[flow] (des.north) -- ++(0, -0.65 + 2.8 - 1.5) -| ([xshift=-4mm]trig.south);
\draw[flow] ([yshift=-2.5mm]theo.east) -- ([yshift=-2.5mm]ana.west);

% --- Legend --------------------------------------------------------------
\path ($(des.south west) + (0, -0.6)$) coordinate (L);
\node[swatch, fill=violet!10, anchor=west] (s1) at (L) {};
\node[lbl, right=1mm of s1] (t1) {run time};
\node[swatch, fill=teal!10, right=5mm of t1] (s2) {};
\node[lbl, right=1mm of s2] (t2) {offline};
\node[swatch, fill=black!6, dashed, right=5mm of t2] (s3) {};
\node[lbl, right=1mm of s3] (t3) {design time};
\draw[flow] ([xshift=5mm]t3.east) -- ++(0.8, 0) node[right, lbl] {indirect influence};
\end{tikzpicture}

%% file: sections/2_rl.tex
% =============================================================================
%  2. Reinforcement Learning Background
% =============================================================================

Reinforcement learning~(RL) is a learning paradigm at the intersection of multiple fields like computer science, psychology, neuroscience, mathematics, economics, and engineering, whose aim is to solve \textit{decision-making} problems. A physical (e.g., a robot, a piece of equipment) or software RL agent discovers how to behave in an environment (i.e., a real or simulated world) by maximizing a cumulative reward signal obtained
through repeated interaction~\cite{Sutton2018RL}.
Unlike supervised learning, no labeled input-output pairs are provided; unlike
unsupervised learning, there is an explicit goal encoded in a reward function, either designed or provided.
RL has produced landmark results in board and video games~\cite{Silver2016Go,
Mnih2015DQN,Vinyals2019Starcraft}, robotics~\cite{Levine2016EndToEnd}, and
control~\cite{Degrave2022Tokamak}, and is increasingly applied in scientific
domains where the objective function is non-differentiable, or the data are
generated through interaction.

For a thorough treatment of classical and modern RL, we refer the reader to the
textbook by Sutton and Barto~\cite{Sutton2018RL} and the survey by
Li~\cite{Li2017DRLSurvey}.
This section provides the background necessary to follow the remaining sections of this survey.

% -----------------------------------------------------------------------------
\subsection{The RL Problem}
\label{sec:rl:problem}

The central objects in RL are the \textbf{agent} and the \textbf{environment}.
At each discrete time step $t$, the agent observes a \textit{state}
$s_t \in \mathcal{S}$ from the environment, selects an \textit{action}
$a_t \in \mathcal{A}$, and receives a scalar \textit{reward} $r_t \in \mathbb{R}$.
The environment transitions to a new state $s_{t+1}$ according to some (usually
unknown) \textit{dynamics}, and the cycle repeats until a termination criterion is met.
The goal of the agent is to learn a \textbf{policy} $\pi : \mathcal{S} \to
\Delta(\mathcal{A})$, a mapping from states to distributions over actions, that
maximizes the expected return $\mathbb{E}_\pi[G_0]$, where the \textbf{return}
from time $t$ is:
\begin{equation}
    G_t = \sum_{k=0}^{T-t-1} \gamma^k r_{t+k},
    \label{eq:return}
\end{equation}
where $\gamma \in [0, 1]$ is the \textbf{discount factor} controlling how much
future rewards are down-weighted relative to immediate ones~(with $\gamma < 1$
required when $T = \infty$), and $T$ is the horizon (or episode length, which
may be infinite).
This interaction defines a \textbf{trajectory} or \textbf{episode}:
$\tau = (s_0, a_0, r_0, s_1, a_1, r_1, \ldots, s_T)$.
When episodes terminate after a finite number of steps, the task is called
\textit{episodic}; otherwise it is \textit{continuing}.

The agent is not told which (optimal) actions to take; it must discover how to act
through experience.
This makes RL fundamentally different from supervised and unsupervised learning.
Four structural properties distinguish sequential decision-making problems as
natural targets for RL:
\begin{itemize}
    \item \textbf{Implicit supervision.} The optimal actions are generally not
          known in advance. The agent can only learn them by trial and error,
          guided by a reward signal that provides indirect, and often imperfect,
          feedback. Hence, even if expert demonstrations are available, a policy resulting from cloning such actions would still lack robustness in unseen situations.

    \item \textbf{Delayed feedback.} The reward at time $t$ evaluates the
          immediate goodness of a single action, but that action may have
          consequences that unfold over many subsequent steps. A greedy policy
          that maximizes the instantaneous reward is therefore suboptimal;
          the return $G_t$ captures the long-term effects that $r_t$ alone
          cannot.

    \item \textbf{Correlated data.} Standard machine learning assumes
          independent and identically distributed (i.i.d.) data.
          Transitions in an RL trajectory are temporally correlated by
          construction: the state $s_{t+1}$ depends on $s_t$ and $a_t$.
          The actions predicted by the agent shape the future data distribution once executed in the environment,
          introducing non-stationarity that naive supervised learning
          cannot handle, since it violates the i.i.d. hypothesis.

    \item \textbf{The reward hypothesis.} RL is applicable whenever the task
          can be expressed as the maximization of a cumulative reward signal.
          If no such signal can be designed, RL cannot be directly applied;
          conversely, if the hypothesis holds, RL provides a principled
          framework for solving the problem.
\end{itemize}

An RL agent may include one or more of the following internal components.
A \textbf{policy} $\pi(a \mid s)$ is the behavior function of the agent: stochastic
policies output a probability distribution over actions, while deterministic
policies output a single action $\mu(s)$.
Depending on the action space, a policy can be parameterized as a Bernoulli
distribution (two binary actions), a categorical distribution (finite discrete
actions), or a Gaussian or squashed Gaussian (continuous and possibly
bounded actions\footnote{Alternatively, the Beta distribution~\cite{chou2017improving} can be used for bounded continuous actions.}).
A \textbf{value function} estimates the expected return from a given state or
state-action pair, and is used to guide policy improvement.
A \textbf{model} is the representation of the environment dynamics held by the agent,
$p(s_{t+1}, r_t \mid s_t, a_t) \approx P$.
A \textbf{replay buffer}~\cite{Lin1992Replay} stores past transitions for reuse during training,
breaking temporal correlations and improving sample efficiency.

% -----------------------------------------------------------------------------
\subsection{Markov Decision Processes}
\label{sec:rl:mdp}

The standard mathematical framework for RL is the \textbf{Markov Decision
Process}~(MDP)~\cite{Bellman1957DP}, defined by the tuple
$\mathcal{M} = (\mathcal{S}, \mathcal{A}, P, R, \gamma)$, where:
\begin{itemize}
    \item $\mathcal{S}$ is the \textbf{state space}: the set of all possible
          states the agent may observe.
          States can be discrete or continuous, scalar or multi-dimensional;
          in general, even if the state space is enumerable, the agent does
          not know all possible states in advance and discovers them
          incrementally through interaction.

    \item $\mathcal{A}$ is the \textbf{action space}: the set of actions the
          agent is allowed to take, which can be binary, discrete, or
          continuous.

    \item $P(s' \mid s, a)$ is the \textbf{transition probability}\footnote{Also called the environment dynamics, or transition model.}: the
          distribution over next states given the current state and action.
          When $P$ is neither known nor provided, the agent operates in the
          \textit{model-free} setting, the most common in practice.
          When $P$ is given or learned, \textit{model-based} agents can
          use it to plan future actions without interacting with the real
          environment.

    \item $R : \mathcal{S} \times \mathcal{A} \times \mathcal{S} \to \mathbb{R}$
          is the \textbf{reward function}, providing the agent with an immediate
          scalar reward $r_t = R(s_t, a_t, s_{t+1})$ at each step.
          This is the most general formulation; it is often simplified to
          $R(s_t, a_t)$ when the next state is not needed. In general,
           rewards can be sparse (zero almost everywhere, non-zero only at
          episode termination), dense (available at every step), noisy,
          and even stochastic.

    \item $\gamma \in [0, 1]$ is the \textbf{discount factor}, which
          down-weights rewards by a factor of $\gamma^t$ as time progresses.
          Discounting is mandatory for continuing environments to prevent
          the infinite sum of rewards from diverging.
          It also defines an \textit{effective horizon}: rewards received
          after $1/(1-\gamma)$ timesteps are down-weighted by a factor of
          about $e^{-1}$, and their weight decays exponentially thereafter.
          Intuitively, $\gamma \to 0$ leads to \textit{myopic} evaluation
          that prioritizes immediate rewards, while $\gamma \to 1$ leads to
          \textit{far-sighted} evaluation that values long-term outcomes.
\end{itemize}

\paragraph{The Markov property.}
The MDP framework relies on the \textbf{Markov property}:
\begin{equation}
    P(s_{t+1} \mid s_t, a_t) = P(s_{t+1} \mid s_0, a_0, \ldots, s_t, a_t),
    \label{eq:markov}
\end{equation}
stating that the future state $s_{t+1}$ is conditionally independent of the
history $(s_0, a_0, \ldots, s_{t-1}, a_{t-1})$ given the present state $s_t$
and action $a_t$.
This property is critical for algorithmic efficiency: the agent only needs
the current state to predict the optimal action, not the full history.
In practice, when raw observations are not Markovian (e.g., a single video
frame in an Atari game, or a single beam position reading at an accelerator),
the Markov property can be approximately restored by incorporating sufficient
history into the state representation, typically by stacking a fixed number
of past observations.

\paragraph{Full and partial observability.}
MDPs assume \textit{full observability}: the agent directly observes the
internal state of the environment, so $o_t = s_t$.
Many real-world problems violate this assumption; if the internal state of the
environment is only partially visible to the agent, it receives an
\textit{observation} $o_t \in \mathcal{O}$ rather than the full state: an observation can be seen as a partially hidden, or corrupted state.
This setting is formalized as a \textbf{Partially Observable MDP}
(POMDP)~\cite{Kaelbling1998POMDP}, in which the agent must infer the
underlying state from a history of observations.
In some HEP applications, partial observability arises naturally: an accelerator
operator sees beam diagnostic readouts rather than the full magnet current
and beam phase-space state; a tracking algorithm observes detector hits
rather than the true particle trajectories.
Common mitigation includes recurrent neural networks (e.g., long short-term
memory~(LSTM) networks~\cite{Hochreiter1997LSTM}) as policy
networks, belief-state representations, and observation stacking.

\paragraph{Value functions.}
Two key quantities measure the quality of a policy $\pi$.
The \textbf{state-value function}:
\begin{equation}
    V^\pi(s) = \mathbb{E}_\pi\!\left[G_t \mid s_t = s\right]
    \label{eq:vfunction}
\end{equation}
computes the expected return from state $s$ under $\pi$.
It quantifies how good it is to \textit{be} in a given state.
The \textbf{action-value function} (Q-function):
\begin{equation}
    Q^\pi(s, a) = \mathbb{E}_\pi\!\left[G_t \mid s_t = s, a_t = a\right]
    \label{eq:qfunction}
\end{equation}
measures the expected return after taking action $a$ in state $s$ and then
following $\pi$.
It quantifies how good it is to take a specific \textit{action} in a given
state.
The \textbf{advantage function}:
\begin{equation}
    A^\pi(s, a) = Q^\pi(s, a) - V^\pi(s)
    \label{eq:advantage}
\end{equation}
quantifies how much better action $a$ is compared to the average action
taken by $\pi$ in state $s$.
A positive advantage indicates that $a$ is better than average (or better than the average action); a negative
advantage indicates it is worse, instead.
Note that $V^\pi(s) = \mathbb{E}_{a \sim \pi(\cdot|s)}[Q^\pi(s, a)]$, so
the state-value function is the expectation of the Q-function under the policy.

\paragraph{The RL objective.}
In deep RL, both the policy and the value functions are parameterized by neural
networks with parameters $\theta$.
The RL objective is to find the optimal model parameters:
\begin{equation}
    \theta^* = \arg\max_{\theta \in \Theta}\;
    \mathbb{E}_{\tau \sim p_\theta}\!\left[\sum_{t=0}^{T-1}
    \gamma^{t} r(s_t, a_t)\right],
    \label{eq:rl_objective}
\end{equation}
where $p_\theta(\tau)$ is the probability of a trajectory induced by the
policy $\pi_\theta$:
\begin{equation}
    p_\theta(\tau) = p(s_0) \prod_{t=0}^{T-1}
    \pi_\theta(a_t \mid s_t)\, P(s_{t+1} \mid s_t, a_t).
    \label{eq:trajectory_prob}
\end{equation}
This objective cannot be differentiated directly by backpropagation, because
the expectation is taken over trajectories generated by unknown and generally
non-differentiable environment dynamics; gradients must instead be estimated
from sampled interactions.
The two main families of algorithms that address this are value-based methods
and policy gradient methods, described in
Sections~\ref{sec:rl:valuebased} and~\ref{sec:rl:policygradient}.

% -----------------------------------------------------------------------------
% -----------------------------------------------------------------------------
\subsection{Value-Based Methods}
\label{sec:rl:valuebased}

Value-based methods learn the Q-function and derive a policy implicitly as
$\pi(s) = \arg\max_a Q_\theta(s, a)$.
The prototypical example is Q-learning~\cite{Watkins1992QLearning}, which
updates the Q-function toward the one-step Bellman optimality target
$r + \gamma \max_{a'} Q(s', a')$ and, in the tabular case, converges to the
optimal Q-function.
They are most natural for \textit{discrete} action spaces, though variants for
continuous actions have been proposed by parameterizing the advantage function
as a closed-form quadratic over the action, as in normalized advantage
functions~(NAF)~\cite{GuLSL16NAF}, enabling
$\arg\max_a Q(s,a)$ to be computed analytically.
Learning is generally \textit{off-policy}: transitions used to update the Q-function are
not required to be recent or generated by the current policy, making reuse of
old experience via a replay buffer possible and sample-efficient.

\paragraph{DQN.}
The \textbf{Deep Q-Network}~(DQN)~\cite{Mnih2015DQN} was the first deep RL
algorithm to achieve human-level performance across a broad set of tasks,
demonstrated on Atari 2600 games of the Arcade Learning Environment~(ALE).
The key insight is that a neural network $Q_\theta(s, a)$ can approximate the
optimal Q-function across a large, continuous state space.
Given a Q-network, the policy is derived as the greedy policy: $\pi(s) =
\arg\max_a Q_\theta(s, a)$, which always exploits the current value estimates.
To ensure sufficient exploration during training, the agent acts according to an
\textit{$\varepsilon$-greedy} policy: with probability $\varepsilon$ a random
action is selected; otherwise the greedy action is taken.
The value of $\varepsilon$ is usually decayed over training from 1 to a small value such
as 0.05, shifting the agent from exploration to exploitation as learning
progresses.

Training proceeds by minimizing the temporal-difference~(TD)~\cite{Sutton1988TD} loss:
\begin{equation}
    \mathcal{L}(\theta) = \mathbb{E}_{(s,a,r,s') \sim \mathcal{D}}\!
    \left[\bigl(y - Q_\theta(s,a)\bigr)^2\right],
    \quad y = r + \gamma \max_{a'} Q_{\theta^-}(s', a'),
    \label{eq:dqn_loss}
\end{equation}
where transitions $(s, a, r, s')$ are sampled from a \textbf{replay buffer}
$\mathcal{D}$, and $\theta^-$ are the parameters of a periodically updated
\textbf{target network}.
Two instabilities arise without these components: (i)~consecutive transitions
are temporally correlated, violating the i.i.d.\ assumption of gradient descent;
and (ii)~the regression target $y$ depends on $Q_\theta$ itself, creating
\textit{moving targets} that destabilize learning.
The replay buffer breaks temporal correlations by sampling transitions uniformly
at random; the target network fixes the regression targets for a designed number of
gradient steps before being updated by copying $\theta^- \leftarrow \theta$.
These two innovations are the core stabilizing contributions of DQN, enabling it to solve problems beyond toy tasks.

\paragraph{DQN extensions.}
Several targeted modifications substantially improve the performance of DQN and have
become standard components of modern value-based agents:
\begin{itemize}
    \item \textbf{Double DQN}~\cite{VanHasselt2016DDQN} addresses the systematic
          overestimation of Q-values that arises from using $\max_{a'}
          Q_{\theta^-}(s', a')$ for both action selection and value estimation.
          It decouples the two steps: the online network $Q_\theta$ selects the
          greedy action, and the target network $Q_{\theta^-}$ evaluates it.
          The target becomes $y = r + \gamma Q_{\theta^-}(s',
          \arg\max_{a'} Q_\theta(s', a'))$, which significantly reduces
          the overestimation bias of the action-value and improves policy quality.

    \item \textbf{Dueling DQN}~\cite{Wang2016Dueling} introduces a network
          architecture that explicitly decomposes the Q-function as:
          $Q(s,a) = V(s) + A(s,a) - \frac{1}{|\mathcal{A}|}\sum_{a'}A(s,a')$,
          where $V(s)$ and $A(s,a)$ are computed by separate network streams
          sharing a common feature backbone.
          This factorization is particularly beneficial in states where the
          choice of action has little effect on the outcome; the value stream
          learns accurately without requiring every action to be sampled.

    \item \textbf{Prioritized Experience Replay}~(PER)~\cite{Schaul2015PER}
          replaces uniform buffer sampling with sampling proportional to the
          TD error magnitude $|\delta_t| = |y - Q_\theta(s,a)|$, where the
          temporal-difference error $\delta_t = r_t + \gamma \max_{a'}
          Q_{\theta^-}(s_{t+1}, a') - Q_\theta(s_t, a_t)$~\cite{Sutton1988TD}
          measures the discrepancy between the current Q-estimate and the
          Bellman target.
          Transitions with high TD error are sampled more frequently, since they
          represent states where the current Q-estimate is most inaccurate and
          thus provide the most informative gradient signal.
          Importance-sampling weights correct for the resulting bias in the
          gradient estimator.

    \item \textbf{Rainbow}~\cite{Hessel2018Rainbow} systematically combines
          six orthogonal improvements to DQN: Double DQN, dueling networks,
          PER, multi-step returns~\cite{Sutton2018RL} (replacing the one-step TD target with an
          $n$-step return for better credit assignment), distributional RL \cite{bellemare2017distributional}
          (learning a distribution over returns rather than their mean), and
          noisy networks~\cite{Fortunato2018NoisyNets} (replacing $\varepsilon$-greedy exploration with
          learned stochastic weights).
          The combined agent substantially outperforms each individual
          modification, establishing it as a standard reference for value-based
          deep RL.

    \item \textbf{IQN}~\cite{Dabney2018IQN}~(Implicit Quantile Network) extends
          distributional RL \cite{bellemare2017distributional} by learning the full return distribution through
          implicit quantile regression: the network takes a quantile level
          $\tau \in [0,1]$ as additional input and outputs the corresponding
          quantile of the return distribution.
          This enables \textit{risk-sensitive} behavior by adjusting $\tau$:
          a conservative agent can target low quantiles, while a risk-seeking
          agent can target high ones.
          This property is particularly relevant in HEP settings with hard
          operational constraints, where avoiding worst-case outcomes matters
          as much as maximizing expected return.
\end{itemize}

% -----------------------------------------------------------------------------
\subsection{Policy Gradient Methods}
\label{sec:rl:policygradient}

Policy gradient~(PG) methods optimize a parameterized policy $\pi_\theta$
directly by gradient ascent on the expected return $J(\theta) =
\mathbb{E}_{\tau \sim p_\theta}[G_0]$.
Unlike value-based methods, PG agents learn the policy \textit{explicitly} and
can handle both discrete and continuous action spaces without modification,
simply by embedding a suitable probability distribution in the policy network.
The trade-off is that the PG methods introduced in this section are
\textit{on-policy}: the gradient is estimated from experience collected by the
current policy.
REINFORCE discards each batch after a single update; proximal policy
optimization~(PPO, introduced below) reuses the same batch for several gradient
epochs, because the probability ratio between
the updated policy and the policy that collected the data corrects for the
change, but it still discards the data once the policy has moved away from that
policy.
Both are therefore less sample-efficient than off-policy methods, which reuse
old experience from a replay buffer; off-policy actor-critic methods for
continuous control are discussed in Section~\ref{sec:rl:continuous}.

\paragraph{REINFORCE.}
The \textbf{REINFORCE} algorithm~\cite{Williams1992REINFORCE} derives the
policy gradient by applying the likelihood ratio trick to $\nabla_\theta
J(\theta)$, yielding:
\begin{equation}
    \nabla_\theta J(\theta) = \mathbb{E}_{\tau \sim p_\theta}\!\left[
    \sum_{t=0}^{T-1} \nabla_\theta \log \pi_\theta(a_t \mid s_t) \cdot G_t
    \right].
    \label{eq:reinforce}
\end{equation}
This gradient formulation has an intuitive interpretation: it increases the log-probability
of actions that led to high returns and decreases it for actions associated with
low returns, formalizing trial-and-error learning.
In practice, the estimator is computed over $N$ sampled trajectories, making it a
Monte Carlo estimate with high variance.
The variance is reduced by subtracting a \textbf{baseline} $b(s_t)$ from the
return: any function of the state alone leaves the gradient unbiased, and the
value function $V(s_t)$ is the standard choice.
Replacing $G_t$ with $\hat{A}_t = G_t - V(s_t)$, a Monte Carlo estimate of the
advantage $A(s_t, a_t)$, yields the REINFORCE-with-baseline estimator.

\paragraph{Actor-Critic methods.}
\textbf{Actor-Critic}~(AC) architectures~\cite{Konda2000ActorCritic} make
the baseline idea practical by learning the value function alongside the policy.
The \textit{actor} is the policy network $\pi_\theta$; the \textit{critic} is a
separate value network $V_\phi \approx V^\pi$, learned by regression toward
sampled returns.
The advantage is then estimated as:
\begin{equation}
    \hat{A}(s_t, a_t) = r_t + \gamma V_\phi(s_{t+1}) - V_\phi(s_t),
    \label{eq:td_advantage}
\end{equation}
which is the one-step TD residual and can be seen as an estimate of $Q(s_t,a_t)
- V(s_t)$.
More powerful estimators include $n$-step returns and
\textbf{Generalized Advantage Estimation}~(GAE)~\cite{schulman2015high}, which
interpolates between the high-variance Monte Carlo return and the low-variance
(but biased) one-step TD estimate through a parameter $\lambda \in [0,1]$:
\begin{equation}
    \hat{A}_t^{\mathrm{GAE}} = \sum_{k=0}^{\infty} (\gamma\lambda)^k
    \delta_{t+k}, \quad \delta_t = r_t + \gamma V_\phi(s_{t+1}) - V_\phi(s_t).
    \label{eq:gae}
\end{equation}

\textbf{Advantage Actor-Critic}~(A2C)~\cite{Mnih2016A3C} collects experience
from multiple parallel environment workers synchronously, averaging their
gradients, and then performing a single parameter update.
Parallelism serves two purposes: it decorrelates experience across workers
(each of them exploring a different trajectory, in principle), and it increases the batch size of each
gradient step, reducing variance without requiring a replay buffer.

\paragraph{Proximal Policy Optimization.}
\textbf{PPO}~\cite{Schulman2017PPO} is the most widely used on-policy RL
algorithm in practice, combining strong empirical performance with implementation
simplicity.
The core problem it addresses is the \textit{step-size dilemma} of on-policy
gradient ascent: a large update step can move the policy so far that subsequent
data collection degrades catastrophically, while too small a step wastes the
collected experience.
PPO controls the update size by clipping the probability ratio $\rho_t(\theta) =
\pi_\theta(a_t \mid s_t) / \pi_{\theta_\mathrm{old}}(a_t \mid s_t)$:
\begin{equation}
    \mathcal{L}^{\mathrm{CLIP}}(\theta) = \mathbb{E}_t\!\left[
    \min\!\bigl(\rho_t A_t,\;
    \mathrm{clip}(\rho_t, 1-\varepsilon, 1+\varepsilon) A_t\bigr)
    \right],
    \label{eq:ppo}
\end{equation}
where $\varepsilon \approx 0.2$.
The clipped surrogate objective removes the incentive for the policy ratio to move outside
the interval $[1-\varepsilon, 1+\varepsilon]$, preventing destructively large
policy updates without the computational complexity of the second-order
trust-region constraint used in trust region policy
optimization~(TRPO)~\cite{Schulman2015TRPO}.
Multiple gradient steps can be taken on the same collected batch, improving data
efficiency relative to REINFORCE.

% -----------------------------------------------------------------------------
\subsection{Learning in Continuous Action Spaces}
\label{sec:rl:continuous}

Many HEP control tasks involve \textit{continuous} action spaces, such as
setting a quadrupole magnet current or a radio-frequency phase to a real-valued
setpoint.
For these settings, $\arg\max_a Q(s,a)$ is intractable, ruling out standard
DQN-family methods.
Two approaches address this: deterministic actor-critic methods that learn a
closed-form greedy policy, and maximum-entropy methods that maintain a
stochastic policy over continuous actions.

\paragraph{DDPG.}
\textbf{Deep Deterministic Policy Gradient}~(DDPG)~\cite{Lillicrap2016DDPG}
extends Q-learning to continuous actions by learning a deterministic policy
$\mu_\theta : \mathcal{S} \to \mathcal{A}$ jointly with a Q-function
$Q_\phi(s, a)$.
Because the policy is deterministic and differentiable, the intractable
$\arg\max$ over actions is replaced by gradient ascent on $Q_\phi$ through the
actor, using the \textbf{deterministic policy gradient}
theorem~\cite{Silver2014DPG}:
\begin{equation}
    \nabla_\theta J \approx \mathbb{E}_s\!\left[
    \nabla_a Q_\phi(s,a)\big|_{a=\mu_\theta(s)}
    \nabla_\theta \mu_\theta(s)
    \right].
    \label{eq:dpg}
\end{equation}
DDPG is off-policy and uses a replay buffer like DQN.
Exploration is similarly achieved by adding noise to the deterministic action at training
time.

\paragraph{TD3.}
\textbf{Twin Delayed DDPG}~(TD3)~\cite{Fujimoto2018TD3} addresses three failure
modes of DDPG through targeted modifications.
First, \textbf{clipped double Q-learning} maintains two independent critic
networks $Q_{\phi_1}, Q_{\phi_2}$ and uses $\min(Q_{\phi_1}, Q_{\phi_2})$ for
the TD target, mitigating the overestimation that arises from the interaction
between function approximation error and the deterministic policy gradient.
Second, \textbf{delayed policy updates} update the actor less frequently than
the critic~(every 2 critic steps by default), allowing the value estimates to
stabilize before they are used to improve the policy.
Third, \textbf{target policy smoothing} adds small clipped noise to the target
policy action before computing the TD target, acting as a regularizer that
prevents the critic from overfitting to narrow peaks in the Q-function.
TD3 substantially improves stability and final performance over DDPG.

\paragraph{SAC.}
\textbf{Soft Actor-Critic}~(SAC)~\cite{Haarnoja2018SAC} reframes the RL
objective by augmenting the return with an entropy bonus:
\begin{equation}
    J(\pi) = \mathbb{E}_\tau\!\left[\sum_t r_t + \alpha
    \mathcal{H}\!\left(\pi(\cdot \mid s_t)\right)\right],
    \label{eq:sac_obj}
\end{equation}
where $\mathcal{H}(\pi(\cdot \mid s)) = -\mathbb{E}_{a \sim \pi}[\log
\pi(a \mid s)]$ is the Shannon entropy of the policy and $\alpha > 0$ is a
temperature parameter.
Maximizing entropy alongside return produces policies that are as random as
possible while still achieving high reward: this encourages thorough exploration,
prevents premature convergence to suboptimal deterministic behaviors, and
produces more robust policies.
The temperature $\alpha$ can be automatically tuned during training by treating it
as a Lagrange multiplier with a target entropy constraint $\mathcal{H}(\pi
(\cdot \mid s)) \geq \mathcal{H}_\text{target}$, removing the need for manual
tuning.
SAC maintains two critic networks~(like TD3) and a stochastic actor implemented
as a squashed Gaussian: the action is sampled from a Gaussian distribution and
passed through a $\tanh$ to bound it within $[-1, 1]$, with the log-probability
corrected accordingly to account for the applied transform.
SAC is off-policy and learns from a replay buffer, combining the sample
efficiency of off-policy learning with the stability and exploration quality of
entropy regularization.
It is one of the most widely used algorithms for continuous control.

% -----------------------------------------------------------------------------
\subsection{Model-based RL and Planning}
\label{sec:rl:modelbased}

In \textbf{model-based RL}~(MBRL), the agent uses either a given or learned
explicit model of the environment dynamics $\hat{P}(s' \mid s, a)$ to find a
policy, without being forced to interact with the real environment for every
gradient step~\cite{DBLP:journals/ftml/MoerlandBPJ23}.
MBRL is particularly relevant in HEP settings where real interactions are
expensive or slow: accelerator tuning shifts have limited machine time, Geant4
simulations take hours per detector configuration, and live detector equipment
carries physical risk.
A key advantage of having an environment model is that it provides
\textit{reversible} access to the MDP dynamics: the agent can reason forward or
backward from any state without committing to an irreversible real transition;
a capability model-free RL lacks entirely.

\paragraph{Dynamics model learning.}
A dynamics model is typically learned by supervised regression on a dataset of
observed transitions $\{(s_t, a_t, r_t, s_{t+1})\}$.
Three types of dynamics functions can be learned, each with different uses:
\textit{forward dynamics} $(s_t, a_t) \to s_{t+1}$ predicts the next state and
enables look-ahead planning;
\textit{backward dynamics} $s_{t+1} \to (s_t, a_t)$ predicts which state-action
pair preceded a given state, allowing backward planning and faster propagation of
value information through the state space;
\textit{inverse dynamics} $(s_t, s_{t+1}) \to a_t$ predicts which action caused a
transition, useful for representation learning without a reward signal.
In practice, the most common choice is to learn a forward model
$p_\theta(s', r \mid s, a)$ jointly predicting next state and reward.
The reliability of learned models decreases as prediction steps accumulate:
multi-step rollouts propagate approximation errors that grow rapidly, causing
the imagined (predicted) states to diverge from the true dynamics.
Uncertainty estimation is therefore a central concern in MBRL.

\paragraph{Dyna-style methods.}
The \textbf{Dyna} architecture~\cite{Sutton1991Dyna} is the canonical framework
for combining model learning and model-free RL.
After each real environment step, the learned dynamics model generates $K$
synthetic transitions from previously visited states; these are added to the
replay buffer alongside real transitions and used to train the value function or
policy.
The planning budget, i.e., the ratio of simulated to real steps per update,
controls the trade-off between model exploitation and real data collection.
Too large a budget allows the agent to overfit to model errors and fail in the
real environment; too small a budget wastes the model capacity and converges to
model-free sample complexity.
Modern variants such as model-based policy
optimization~(\textbf{MBPO})~\cite{Janner2019MBPO} learn a neural network
ensemble as the dynamics model and use short model rollouts to limit error
accumulation, while initializing rollouts from real
states in the replay buffer to ensure coverage.

\paragraph{Bayesian ensembling and uncertainty.}
Representing the model as an ensemble of $M$ independently trained neural
networks provides estimates of \textit{epistemic uncertainty}: regions of state
space where the models disagree indicate areas of limited data coverage.
This uncertainty can be used to terminate planning before errors accumulate
(stopping when model disagreement exceeds a threshold), and to add a pessimistic
penalty to imagined rewards that discourages over-optimistic value estimates in
poorly explored regions.
Probabilistic ensembles with trajectory sampling~(\textbf{PETS})~\cite{Chua2018PETS}
implement this idea by propagating model uncertainty through model-predictive
planning.
This approach is particularly valuable in accelerator control
(Section~\ref{sec:accelerators:modelbased}), where distributional shift between
the simulation used for training and the real machine is unavoidable.

\paragraph{Planning budget and strategy.}
Three design decisions characterize any MBRL planner.
First, the \textit{starting state} for planning: the most common choice is to
plan from the current state, though priority-based methods start from states with
the highest estimated value error.
Second, the \textit{planning budget}: approaches such as Dyna and AlphaGo plan
for a fixed number of model steps after each real step; others, such as
PILCO~\cite{DeisenrothR11PILCO}, collect a full real episode before re-planning the
entire trajectory.
Third, the \textit{planning strategy}: discrete planning enumerates state-action
sequences in a tree or table; differentiable planning exploits the fact that if
both the dynamics model and policy are differentiable, the gradient of the
return objective can be propagated directly through imagined trajectories.
\textbf{Dreamer}~\cite{HafnerLB020Dreamer} exemplifies the latter: it encodes
observations into compact latent representations with a Variational Auto-Encoder (VAE) \cite{kingma2013auto}, learns a recurrent latent dynamics model, and trains the policy entirely by back-propagating through
imagined latent trajectories via stochastic backpropagation.

\paragraph{Monte Carlo Tree Search.}
\textbf{Monte Carlo Tree Search}~(MCTS)~\cite{Coulom2006MCTS} is a discrete
planning algorithm that incrementally builds a search tree over action sequences,
using Monte Carlo rollouts to estimate the value of leaf nodes.
Each iteration consists of four steps: \textit{selection}~(traverse the tree using a Upper Confidence Bound (UCB)-style \cite{Kocsis2006UCT} criterion:
$\text{UCB}(n) = \hat{Q}(n) + c\sqrt{\ln N(n_p) / N(n)}$, where $\hat{Q}(n)$
is the mean return estimate at node $n$, $N(n)$ its visit count, $N(n_p)$ the
parent visit count, and $c$ a constant balancing exploitation and exploration),
\textit{expansion} (add a new node for an unexplored action), \textit{simulation}
(roll out a trajectory from the new node using a default policy to obtain a
return estimate), and \textit{backpropagation} (update value estimates along the
path from root to leaf).
When a learned neural policy and value function replace the random default policy
and leaf evaluation (as in AlphaZero~\cite{Silver2018AlphaZero}), MCTS achieves
strong performance in combinatorial domains at a fraction of the search budget
required by exhaustive methods.
In HEP, MCTS has been applied to hierarchical jet
clustering~(Section~\ref{sec:jets:clustering}), where the tree search over
constituent merging sequences reconstructs the most probable parton shower
history.

% -----------------------------------------------------------------------------
\subsection{Advanced Topics}
\label{sec:rl:advanced}

\subsubsection{Multi-Agent Reinforcement Learning}
\label{sec:rl:marl}

In \textbf{multi-agent RL}~(MARL), $N$ agents each observe a local state
$o^i_t$ and take an action $a^i_t$, producing a joint action
$\mathbf{a}_t = (a^1_t, \ldots, a^N_t)$ that determines the transition and the
reward signals.
Agents may be cooperative~(sharing a global reward), competitive~(adversarial
rewards), or mixed.
The core difficulty of MARL is the exponential growth of the joint action space:
$|\mathcal{A}^1 \times \cdots \times \mathcal{A}^N| = \prod_i |\mathcal{A}^i|$,
which makes centralized Q-learning over joint actions infeasible even for
moderate $N$.
A further challenge is \textit{non-stationarity}: from the perspective of any
individual agent, the environment is non-stationary because the other agents are
also learning and changing their policies simultaneously, breaking the
stationarity assumption underlying single-agent convergence guarantees.

The dominant solution paradigm is \textbf{Centralized Training with
Decentralized Execution}~(CTDE): during training, agents share information
(observations, actions, or critic inputs) through a centralized component, but
at deployment, each agent acts using only its local observation.
\textbf{QMIX}~\cite{Rashid2018QMIX} factorizes the joint Q-function as a
monotonic mixing of per-agent Q-values, enabling centralized training while
guaranteeing that decentralized greedy action selection is globally optimal under
the factorization.
Multi-agent DDPG~(\textbf{MADDPG})~\cite{Lowe2017MADDPG} extends DDPG to
multi-agent settings by giving the centralized critic of each agent access to
the observations and actions of all agents during training, while the actors
remain decentralized.
In HEP, MARL has been applied to collaborative charged particle tracking, where
several agents reconstruct different tracks on the same hit graph under a
hit-uniqueness constraint~(Section~\ref{sec:tracking:marl}); its scaling to
high-pileup environments has not yet been evaluated.

\subsubsection{Offline Reinforcement Learning}
\label{sec:rl:offline}

\textbf{Offline RL}~(also called batch RL or offline batch learning) learns a
policy from a fixed dataset $\mathcal{D} = \{(s_t, a_t, r_t,
s_{t+1})^{(i)}\}_{i=1}^N$ of pre-collected transitions, without any further
interaction with the environment~\cite{Levine2020OfflineRL}.
The dataset is generated by some behavioral policy $\pi_\beta$, which may be a
prior heuristic, a human operator, or a previously deployed RL policy.
The goal is to learn a policy $\pi$ that attains higher return than $\pi_\beta$
when deployed, effectively extracting the best possible behavior from the
historical data.

Two distinct forms of distributional shift undermine naive offline Q-learning.
At \textit{test time}, the trained policy visits states that may differ from
those in $\mathcal{D}$, incurring reward loss that grows quadratically with the
horizon under covariate shift.
At \textit{training time}, the Bellman backup $y = r + \gamma \max_{a'}
Q_\theta(s', a')$ queries the Q-function at actions $a'$ sampled from the current
policy, which may be out-of-distribution~(OOD) with respect to the data; the
Q-function has not been trained on these actions and can return arbitrarily
optimistic values, causing \textit{extrapolation error} that cascades through
the policy improvement step.

Conservative methods address this by penalizing or ignoring OOD actions.
\textbf{Conservative Q-learning} (CQL)~\cite{Kumar2020CQL} adds a regularization term to the Q-learning
loss that minimizes Q-values for OOD actions while maximizing them for actions in
the dataset, producing Q-estimates that are guaranteed to be lower bounds on the
true values.
Implicit Q-learning~(\textbf{IQL})~\cite{Kostrikov2021IQL} avoids querying OOD actions entirely by
replacing the Bellman max with an expectile
regression\footnote{An \textit{expectile} is a generalization of the mean
obtained by minimizing an asymmetrically weighted sum of squared errors:
$e_\tau = \arg\min_e\,\mathbb{E}[|\tau - \mathbf{1}[X < e]|\,(X-e)^2]$, where
$\tau \in (0,1)$ controls the asymmetry. At $\tau = 0.5$ the expectile coincides
with the mean; as $\tau \to 1$ it approximates the maximum of the distribution.}
that approximates the maximum over in-distribution actions, enabling stable offline training with no
explicit constraint on the policy.
Offline RL is relevant in HEP wherever large historical datasets exist but
online interaction is costly: logged accelerator operational data and archived
detector calibration histories are natural candidate datasets.

\subsubsection{Inverse Reinforcement Learning and Imitation Learning}
\label{sec:rl:irl}

\textbf{Imitation learning}~(IL) covers methods that learn a policy from expert
demonstrations $\mathcal{D}_E = \{(o_i, a_i)\}_{i=1}^N$ without a reward
function.
The simplest form is \textbf{behavioral cloning}~(BC): supervised regression of
actions on observations, treating the problem as a standard classification or
regression task.
BC is sample-efficient but fragile: small deviations from the training
distribution compound over time because the agent experiences states never seen
during training, causing the predicted trajectory to diverge quadratically in the
horizon~(covariate shift).
Mitigation strategies include data augmentation, collecting recovery
demonstrations, and dataset aggregation~(DAgger)~\cite{Ross2011DAgger}, which queries the expert to
label states visited by the current learned policy in an active learning loop.

\textbf{Inverse RL}~(IRL)~\cite{Ng2000IRL} instead infers the reward function
$r_\psi(s, a)$ that the expert is implicitly optimizing, after which standard RL
can be applied to the recovered reward.
This is valuable when the reward is difficult to specify manually: a common
situation in accelerator control, where the notion of a ``good'' beam is encoded
in the intuition gathered by expert operators over decades.
The core challenge is that many reward functions are consistent with the same
demonstrations; the maximum entropy formulation~\cite{Ziebart2008MaxEnt}
resolves this ambiguity by selecting the reward that induces the highest-entropy
policy matching the feature expectations of the expert.
\textbf{Generative Adversarial Imitation Learning}~(GAIL)~\cite{ho2016generative}
re-frames IRL as a two-player game: a discriminator $D(s,a)$ learns to
distinguish the trajectories of the current policy from expert trajectories,
while the policy is trained to fool the discriminator using $\log D(s,a)$ as a
reward signal.
The policy is optimal when the discriminator cannot separate the two, i.e., when
the state-action occupancy of the policy matches that of the expert.

\subsubsection{Bandit Algorithms}
\label{sec:rl:bandits}

\textbf{Multi-armed bandits} are a special case of RL in which each episode
consists of a single decision step: the agent selects an action~(``arm'') and
receives a reward, with no state transitions or temporal credit assignment.
The fundamental challenge is the \textit{exploration--exploitation} tradeoff:
exploiting the arm with the highest known expected reward may miss a better arm
that has been sampled too rarely.

The \textbf{Upper Confidence Bound}~(UCB1) algorithm~\cite{Auer2002UCB}
addresses this by adding an optimism bonus to the estimated value of each arm:
\begin{equation}
    a_t = \arg\max_{a}\!\left[\hat{\mu}_a + c\sqrt{\frac{\ln t}{n_a}}\right],
    \label{eq:ucb}
\end{equation}
where $\hat{\mu}_a$ is the empirical mean reward of arm $a$, $n_a$ is its pull
count, $t$ is the total number of rounds, and $c$ controls the exploration
strength.
UCB achieves near-optimal regret~(the gap between cumulative reward and the
oracle that always plays the best arm) of $\mathcal{O}(\sqrt{K T \log T})$ for
$K$ arms and $T$ rounds.
\textbf{Thompson Sampling}~\cite{Thompson1933} maintains a posterior distribution over arm reward
rates and selects the arm with the highest sampled value at each step, achieving
similar regret bounds with often better empirical performance.

\textbf{Contextual bandits} extend the above by conditioning the action choice
on a context vector $x_t$ that encodes the current situation.
\textbf{LinUCB}~\cite{Li2010LinUCB} assumes a linear reward model $r = x_t^\top \theta_a + \epsilon$
and maintains a ridge regression estimate of $\theta_a$ for each arm, with an
UCB-style confidence ellipsoid for exploration.
For non-linear reward models, neural contextual bandits~\cite{Zhou2020NeuralUCB}
replace the linear head with a neural network while preserving the UCB
structure.

\textbf{Semi-bandits}~\cite{Audibert2014Combinatorial} provide richer feedback: the agent selects a composite
action~(e.g., a subset of arms) and observes the reward for each component
individually.
This additional signal enables faster learning when the action space is
structured.
Bandit algorithms are used in HEP data quality monitoring~(DQM,
Section~\ref{sec:dqm}), where each monitoring decision, such as which detector
channels to observe, is treated as stateless and the semi-bandit structure
captures the multi-channel nature of detector monitoring.
The key distinction from full RL is that bandit methods assume no state
transitions: successive monitoring decisions in a run are treated as
independent.
This is a modeling choice rather than a fundamental constraint; as DQM systems
mature, incorporating run-level temporal correlations will likely motivate a
transition to full sequential RL formulations.

\subsubsection{Other RL Paradigms}
\label{sec:rl:paradigms}

Several RL paradigms beyond the single-task setting recur in the open problems
of the following sections.
\textbf{Meta-RL}~\cite{finn2017model, beck2025tutorial} learns, from a
distribution of related tasks, a policy or a learning procedure that adapts to a
new task from a small amount of experience; model-agnostic
meta-learning~(MAML)~\cite{finn2017model} does so by learning an initialization
from which a few gradient steps suffice.
\textbf{Multi-task and transfer RL}~\cite{Zhang2021MTLSurvey, Teh2017Distral}
share representations or policies across tasks, so that a policy trained on one
task can be reused on another with little additional data.
\textbf{Continual RL}~\cite{Khetarpal2022ContinualRL} learns from a sequence of
changing tasks without catastrophic forgetting of the earlier ones.
\textbf{Hierarchical RL}~\cite{Sutton1999Options, Dayan1993Feudal} decomposes a
task into temporally extended sub-tasks: in the options framework, an option
$\langle \mathcal{I}, \pi, \beta \rangle$ comprises an initiation set
$\mathcal{I}$ of states in which it is available, an intra-option policy $\pi$,
and a termination condition $\beta$, and a high-level policy that selects
options acts on a slower timescale over a semi-Markov decision process.
\textbf{Bayesian RL}~\cite{Ghavamzadeh2015BayesRL} maintains a posterior
distribution over the environment dynamics or the reward model, rather than
over returns as in distributional RL, and uses it to balance exploration and
exploitation.
\textbf{Multi-objective RL}~\cite{Liu2015MORL, Yang2019MOPolicy} optimizes a
vector of rewards and returns a set of Pareto-optimal policies, or a single
policy conditioned on a preference vector, instead of a single scalarized
trade-off; scalarized multi-objective rewards are discussed in
Section~\ref{sec:rl:reward}, together with curriculum learning.

% -----------------------------------------------------------------------------
\subsection{Practical Considerations}
\label{sec:rl:practical}

Deploying RL to real physical systems introduces a set of challenges that are
largely absent from simulation benchmarks.
Dulac-Arnold et al.~\cite{dulac2021challenges} identified nine such
challenges empirically: limited real-system samples, actuation and sensing
delays, high-dimensional continuous spaces, hard constraints, partial
observability, multi-objective rewards, real-time inference, offline log data,
and policy explainability.
HEP applications encounter most of these simultaneously, making the practical
considerations below not just theoretical guidance but design requirements.

\subsubsection{Reward Design}
\label{sec:rl:reward}

The reward function is the most critical and most difficult design choice in any
RL application.
A poorly specified reward leads to \textbf{reward hacking}: the agent finds
high-reward behaviors that satisfy the letter of the objective but violate its
intent.
The classic example is a boat-racing agent that achieves high reward by
collecting bonus items in circles rather than finishing the
race~\cite{Clark2016FaultyRewards}; analogous failure modes arise in HEP when a
beam-tuning agent optimizes a transmission proxy without respecting emittance
constraints, or when a jet grooming policy removes constituents that satisfy the
reward criterion but destroy infrared and collinear~(IRC) safety.

\textbf{Reward shaping} adds auxiliary reward terms to guide early learning
without changing the optimal policy, provided the additional terms are
potential-based: $r'(s, a, s') = r(s, a, s') + \gamma\Phi(s') -
\Phi(s)$~\cite{Ng1999PotentialShaping}.
Any such shaped reward preserves the set of optimal policies of the original
MDP.
In accelerator control, intermediate rewards based on partial beam transmission
or upstream orbit deviation serve this role, providing dense feedback during a
tuning sequence whose true outcome is only known at the end.

\textbf{Sparse rewards} arise naturally in several HEP problems: the resolution
of a detector design is known only after full simulation of the complete
design~\cite{Qasim2025DetectorRL}, and the consistency of a BSM model or a
string compactification can be fully assessed only once the construction is
complete.
Most reviewed works nevertheless shape these objectives into dense per-step
rewards, such as step-wise improvements of beam quality or of a consistency
score~(Sections~\ref{sec:accelerators} and~\ref{sec:bsm}).
Three complementary approaches address sparsity: reward shaping~(above);
\textbf{curriculum learning}~\cite{Bengio2009Curriculum}, which starts the agent from easier sub-problems
and progressively increases difficulty, used in RL-ABC~\cite{Ibrahim2026RLABC}
as stage learning; and \textbf{hierarchical RL}~\cite{Sutton1999Options,
Dayan1993Feudal}~(Section~\ref{sec:rl:paradigms}), which decomposes the problem
into a high-level policy selecting sub-goals and a low-level policy achieving
them, providing denser intermediate rewards naturally.

\textbf{Multi-objective rewards} are the norm rather than the exception in HEP.
Accelerator operation must optimize transmission while satisfying emittance and
safety bounds; detector design must balance resolution, cost, and material budget.
Naive scalarization of multiple objectives into a weighted sum is sensitive to
weight choice and can miss Pareto-optimal solutions.
Distributional RL methods~\cite{Dabney2018IQN} allow direct optimization of
risk-aware objectives such as the Conditional Value at Risk~(CVaR), which is
relevant when avoiding worst-case outcomes~(e.g., beam loss events) matters as
much as maximizing expected return.
When a reward is genuinely difficult to specify, IRL and GAIL~(Section
\ref{sec:rl:irl}) offer alternatives by inferring the reward from expert
demonstrations.

\subsubsection{Sim-to-Real Transfer}
\label{sec:rl:sim2real}

Many HEP RL policies, in particular for accelerator control, are trained in
simulation and then deployed on real systems.
The \textbf{sim-to-real gap} has two sources: \textit{model error}~(the
simulator does not perfectly reproduce real physics) and \textit{distributional
shift}~(the real system drifts over time due to temperature, radiation damage,
or mechanical settling, while the simulator remains static).
Both can degrade performance and, in safety-critical settings, cause dangerous
behavior.

\textbf{Domain randomization}~\cite{Tobin2017DomainRand} varies simulator
parameters~(magnet strengths, beam emittance, noise levels) over a wide range
during training, forcing the policy to learn behaviors that are robust across
the distribution of possible conditions rather than optimized for a single
nominal model.
Kaiser et al.~\cite{Kaiser2022ICML} and Awal et al.~\cite{Awal2024Injection}
both apply domain randomization to achieve zero-shot or near-zero-shot transfer
from simulation to real accelerators.
\textbf{System identification} fits simulator parameters to match real
observations before or during deployment, narrowing the gap at the cost of
additional measurement campaigns.
\textbf{Domain adaptation} fine-tunes the policy on a small number of real
transitions after simulation training, which is particularly effective when
real data collection is feasible but expensive.
In HEP settings where interactions are severely limited~(e.g., an operational
collider where beam time is allocated in shifts), a fast simulator is typically
used as the training environment, and few or no real correction steps are needed
after deployment; when no reliable simulator exists, a surrogate learned from
operational data can play the same role, as in model-based
RL~(Section~\ref{sec:rl:modelbased}).

\subsubsection{Safety in RL}
\label{sec:rl:safety}

Many HEP control tasks carry hard operational safety constraints.
In accelerator operation, beam loss must not exceed activation thresholds;
quench limits on superconducting magnets impose instantaneous current change
bounds; the bias currents of cryogenic detectors must remain within narrow
windows or risk damaging the sensor.
These constraints cannot be treated as soft reward penalties because a single
violation may cause irreversible damage.

\textbf{Constrained MDPs}~(CMDPs)~\cite{Altman1999CMDP} formalize this by
augmenting the standard MDP with auxiliary cost functions $c_k(s, a)$ and
requiring the policy to satisfy $\mathbb{E}_\pi[\sum_t c_k(s_t, a_t)] \leq
d_k$ for each constraint $k$.
Lagrangian relaxation converts this into an unconstrained saddle-point problem:
the constraint multipliers $\lambda_k$ are adapted online to maintain
feasibility.
\textbf{Action projection} is an alternative: after computing the unconstrained
policy action, it is projected onto the feasible set defined by the constraints,
using a lightweight quadratic program at each step.
\textbf{Control barrier functions}~(CBFs)~\cite{Cheng2019CBF} provide formal
forward-invariance guarantees: if the system state satisfies the safety
condition at time $t$, the CBF filter on the policy action guarantees it will
satisfy it at $t+1$.
In HEP, safety-constrained RL has so far been handled primarily through reward
penalties and conservative action clipping rather than formal CMDP methods; this
gap is one of the key open problems identified in
Section~\ref{sec:accelerators:future}.

\subsubsection{Sample Efficiency}
\label{sec:rl:sampleeff}

RL algorithms can require millions of environment interactions to converge.
In simulation, this is acceptable; on real physical systems, it is prohibitive.
A full accelerator simulation may take seconds per step; a
Geant4 detector simulation takes hours; and every step on a real machine
consumes scarce beam time.
Sample efficiency is therefore a first-order design requirement in all HEP RL
applications, not merely a performance metric.

The main strategies for improving it are: \textbf{replay buffers}~(DQN, SAC,
TD3), which allow each real transition to be reused for multiple gradient
updates; \textbf{model-based planning}~(Section~\ref{sec:rl:modelbased}), which
substitutes real interactions with cheaper simulated ones; \textbf{transfer
learning}, which initializes policy or value networks from pre-trained
representations~(e.g., from a related beamline or a lower-fidelity simulator)
to reduce the number of new interactions required; and \textbf{data
augmentation}, which synthetically expands the observed transition distribution
by applying state-space transformations that preserve physical validity.
\subsubsection{Real-Time Inference and Deployment}
\label{sec:rl:realtime}

Control systems in HEP operate at fixed frequencies: for example, the LHC orbit
feedback runs at tens of hertz, detector front-end readout operates in
nanosecond windows, and Level-1 trigger systems must decide within
microseconds.
For RL policies to be deployed in these loops, inference latency must match the
control frequency.

Standard neural network inference on CPUs achieves millisecond latency,
sufficient for slow accelerator tuning tasks~(seconds per correction step) but
not for fast feedback loops.
Deployment on field-programmable gate arrays~(FPGAs) via tools such as
hls4ml~\cite{schulte2026hls4ml,Duarte2018hls4ml} compiles trained
neural networks to a synthesizable hardware description, achieving
nanosecond-to-microsecond latency.
St.\ John et al.~\cite{StJohn2021Booster} demonstrated this for the Fermilab
Booster, compiling a trained RL controller to an FPGA and testing it on a
bench setup.
As RL is adopted in faster HEP control loops, hardware-aware neural architecture
design and quantization-aware training will become standard requirements.

\subsubsection{Physics-Informed RL}
\label{sec:rl:physicsinformed}

A recurring theme across all HEP RL applications is the incorporation of domain
knowledge directly into the RL formulation: in many HEP settings, physics
constraints \textit{define} what a valid solution looks like, and embedding them
into the RL problem is the only way to guarantee that learned policies respect
them. Physics knowledge manifests at four levels of the RL problem, in particular:

\begin{itemize}
    \item \textbf{State representation.} Beam optics parameters~(Twiss
          functions, dispersion) or beam statistics can provide a compact,
          physics-grounded state for accelerator agents, as in the
          Elegant-based environments of
          Section~\ref{sec:accelerators:steering}, although most deployed
          agents act directly on beam position monitor~(BPM) or screen
          readings.
          Graph-based hit structures encode the geometric relationships between
          detector layers for tracking agents.
          Particle four-momenta and energy flow representations capture the
          relevant degrees of freedom for jet agents.

    \item \textbf{Reward function.} The Coulomb scattering probability serves as
          a label-free physics reward for tracking~(Section~\ref{sec:tracking}),
          replacing hit-to-particle truth associations with a physical
          consistency criterion.
          In jet grooming, the reward favors the removal of soft wide-angle
          radiation, a qualitative proxy for the IRC-safety considerations
          behind analytic groomers~(Section~\ref{sec:jets}).
          Consistency with observed quark mass ratios and mixing angles provides
          the reward signal for BSM model-building
          agents~(Section~\ref{sec:bsm}).

    \item \textbf{Action space.} Restricting actions to physically plausible
          parameter ranges, such as bounded magnet current increments, reduces
          the effective search space, focusing exploration on regions where the
          physics is well-defined.
          Masks can also remove infeasible actions entirely: grammar masks
          admit only symmetry-allowed couplings in BSM model
          building~(Section~\ref{sec:bsm:graph}), and a safety layer~\cite{Dalal2018SafetyLayer} enforces
          unique hit assignments in multi-agent
          tracking~(Section~\ref{sec:tracking:marl}).

    \item \textbf{Environment model.} Fast differentiable simulators such as
          Cheetah~\cite{Kaiser2024Cheetah} for linear accelerator optics provide
          physically accurate dynamics models for MBRL without the cost of
          full particle tracking codes.
          Learned surrogates can also be constrained to a physically
          interpretable form: the sparse dictionary surrogate of Colen et
          al.~\cite{Colen2026ExplainableRL} recovers the linear energy-gain law
          of the CEBAF linacs and constrains the policy through it.
\end{itemize}

\begin{figure}[tp]
    \centering
    \input{src/2/rl_map.tex}
    \caption{Modular view of an RL system, labeled with the subsections that
    cover each component.
    The policy of the agent acts on the environment through an optional safety
    layer and receives observations; an optional learned model supports
    planning.
    The reward is either designed or learned from demonstrations by inverse
    RL, while imitation learning maps demonstrations to the policy directly.
    The environment is a simulator or the real system, bridged by sim-to-real
    transfer, and experience comes either from online interaction through a
    replay buffer or from a fixed dataset in offline RL.
    The R badges depict that the reward is computed from environment
    transitions; the P badges indicate where physics knowledge enters in
    physics-informed RL.}
    \label{fig:rl_map}
\end{figure}
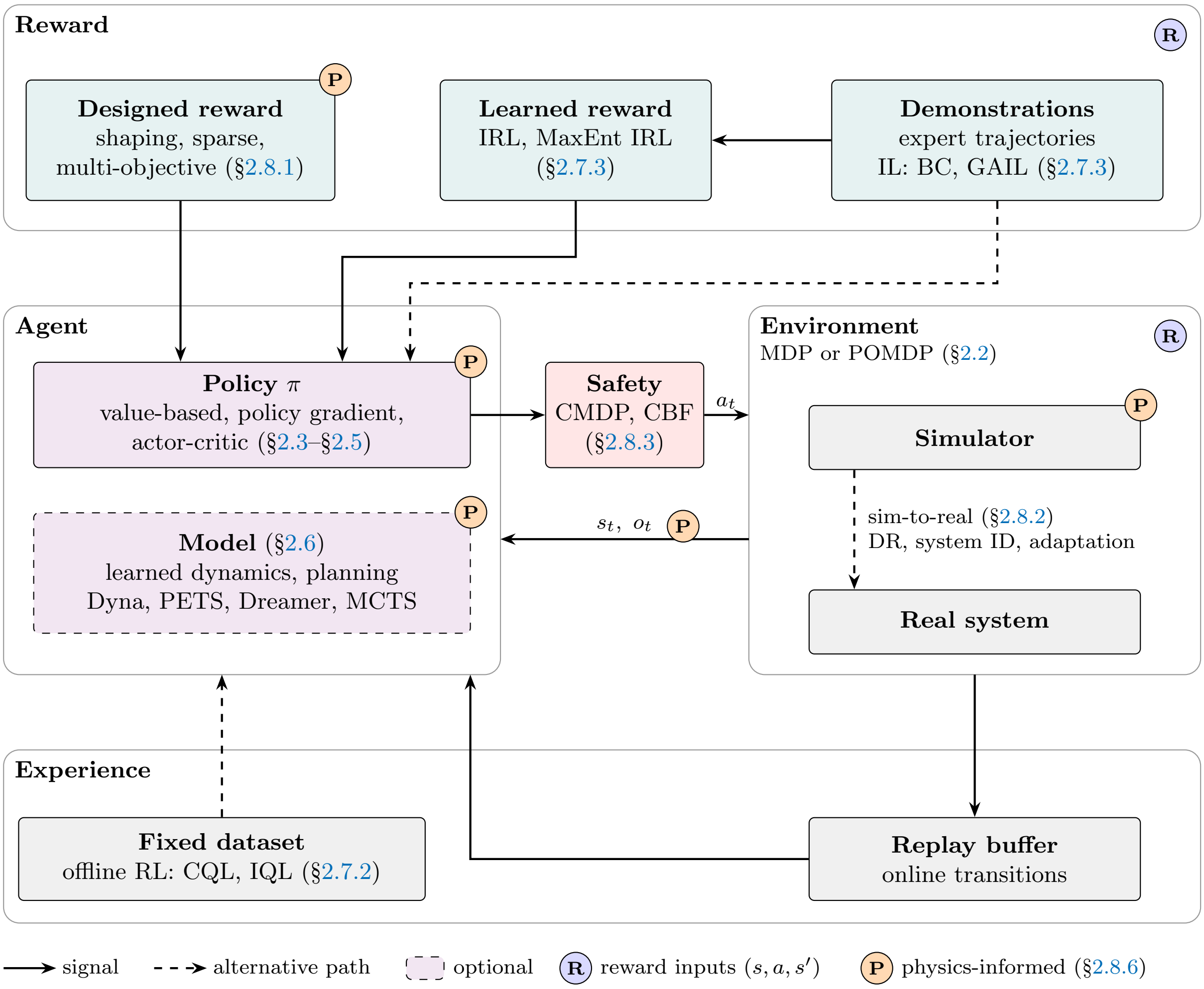

% -----------------------------------------------------------------------------
The methods introduced in this section are organized as modular components of a generic RL system, as shown in
Figure~\ref{fig:rl_map}: this also serves
as a common frame for the HEP applications reviewed in the following sections,
since each of them fixes one choice per component. In addition, these applications provide concrete opportunities to embed physics domain knowledge in RL agents. For a comprehensive discussion of general methods to enable physics-informed RL, we suggest the reader consult the survey of Banerjee et al.~\cite{Banerjee2025PIRL}.

%% file: src/2/rl_map.tex
% Modular map of the RL background (Section 2).
% Requires: \usepackage{tikz}
%           \usetikzlibrary{arrows.meta, positioning, fit, calc, backgrounds}
\begin{tikzpicture}[
    font=\small,
    >={Stealth[length=2mm]},
    block/.style={draw, rounded corners=2pt, align=center, inner sep=3pt},
    rew/.style={block, fill=teal!10, minimum height=1.6cm},
    agent/.style={block, fill=violet!10, minimum width=5.8cm},
    safe/.style={block, fill=red!10, minimum width=2.1cm, minimum height=1.4cm},
    env/.style={block, fill=black!6, minimum width=4.4cm, minimum height=0.85cm},
    data/.style={block, fill=black!6, minimum height=1.1cm},
    frame/.style={draw=black!40, rounded corners=6pt},
    framelabel/.style={anchor=north west, font=\small\bfseries, inner sep=4pt},
    optional/.style={dashed},
    loop/.style={->, thick},
    flow/.style={->, thick, dashed},
    lbl/.style={font=\footnotesize, inner sep=2pt},
    badge/.style={circle, draw, inner sep=0pt, minimum size=4.2mm,
                  font=\scriptsize\bfseries},
    badgeR/.style={badge, fill=blue!15},
    badgeP/.style={badge, fill=orange!30},
]

% --- Reward module -------------------------------------------------------
\draw[frame] (0, 0) rectangle (15.9, -3.0);
\node[framelabel] at (0, 0) {Reward};
\node[badgeR] at (15.5, -0.4) {R};
\node[rew, minimum width=4.1cm] (des) at (2.35, -1.8)
    {\textbf{Designed reward}\\shaping, sparse,\\multi-objective (\S\ref{sec:rl:reward})};
\node[rew, minimum width=3.6cm] (irl) at (7.6, -1.8)
    {\textbf{Learned reward}\\IRL, MaxEnt IRL\\(\S\ref{sec:rl:irl})};
\node[rew, minimum width=4.4cm] (demo) at (13.2, -1.8)
    {\textbf{Demonstrations}\\expert trajectories\\IL: BC, GAIL (\S\ref{sec:rl:irl})};
\node[badgeP] at (des.north east) {P};

% --- Agent module --------------------------------------------------------
\draw[frame] (0, -4.0) rectangle (6.6, -8.9);
\node[framelabel] at (0, -4.0) {Agent};
\node[agent, minimum height=1.4cm] (pol) at (3.3, -5.45)
    {\textbf{Policy} $\pi$\\value-based, policy gradient,\\actor-critic
     (\S\ref{sec:rl:valuebased}--\S\ref{sec:rl:continuous})};
\node[agent, optional, minimum height=1.6cm] (model) at (3.3, -7.55)
    {\textbf{Model} (\S\ref{sec:rl:modelbased})\\learned dynamics, planning\\Dyna, PETS, Dreamer, MCTS};
\node[badgeP] at (pol.north east) {P};
\node[badgeP] at (model.north east) {P};

% --- Safety layer --------------------------------------------------------
\node[safe] (safe) at (8.25, -5.45) {\textbf{Safety}\\CMDP, CBF\\(\S\ref{sec:rl:safety})};

% --- Environment module --------------------------------------------------
\draw[frame] (9.9, -4.0) rectangle (15.9, -8.9);
\node[framelabel, align=left] at (9.9, -4.0)
    {Environment\\[-1pt]{\footnotesize\normalfont MDP or POMDP (\S\ref{sec:rl:mdp})}};
\node[badgeR] at (15.5, -4.4) {R};
\node[env] (sim)  at (12.9, -5.75) {\textbf{Simulator}};
\node[env] (real) at (12.9, -8.2)  {\textbf{Real system}};
\node[badgeP] at (sim.north east) {P};
\draw[flow] ([xshift=-1.6cm]sim.south) -- node[lbl, right, align=left, xshift=1mm]
    {sim-to-real (\S\ref{sec:rl:sim2real})\\DR, system ID, adaptation} ([xshift=-1.6cm]real.north);

% --- Experience module ---------------------------------------------------
\draw[frame] (0, -9.9) rectangle (15.9, -12.2);
\node[framelabel] at (0, -9.9) {Experience};
\node[data, minimum width=5.4cm] (fixed)  at (2.9, -11.35)
    {\textbf{Fixed dataset}\\offline RL: CQL, IQL (\S\ref{sec:rl:offline})};
\node[data, minimum width=4.4cm] (replay) at (12.9, -11.35)
    {\textbf{Replay buffer}\\online transitions};

% --- Signals (solid) -----------------------------------------------------
\draw[loop] (des.south) -- (des.south |- pol.north);
\draw[loop] (irl.south) -- (irl.south |- 0, -3.35) -| ([xshift=1.2cm]pol.north);
\draw[loop] (demo.west) -- (irl.east);
\draw[loop] (pol) -- (safe);
\draw[loop] (safe.east) -- (9.9, -5.45) node[lbl, above, pos=0.5] {$a_t$};
\draw[loop] (9.9, -7.1) -- node[lbl, above] (obs) {$s_t,\ o_t$} (6.6, -7.1);
\node[badgeP, right=1mm of obs] {P};
\draw[loop] (real.south |- 0, -8.9) -- (replay.north);
\draw[loop] (replay.west) -- (6.2, -11.35) -- (6.2, -8.9);

% --- Alternative paths (dashed) ------------------------------------------
\draw[flow] (demo.south) -- (demo.south |- 0, -3.7) -| ([xshift=2.1cm]pol.north);
\draw[flow] (fixed.north) -- (fixed.north |- 0, -8.9);

% --- Legend --------------------------------------------------------------
\begin{scope}[shift={(0, -12.8)}]
    \draw[loop] (0, 0) -- (0.7, 0) node[right, lbl] {signal};
    \draw[flow] (2.0, 0) -- (2.7, 0) node[right, lbl] {alternative path};
    \node[block, fill=violet!10, optional, minimum width=5mm, minimum height=3mm,
          inner sep=0pt] at (5.6, 0) {};
    \node[lbl, anchor=west] at (5.9, 0) {optional};
    \node[badgeR] at (7.6, 0) {R};
    \node[lbl, anchor=west] at (7.85, 0) {reward inputs $(s, a, s')$};
    \node[badgeP] at (11.6, 0) {P};
    \node[lbl, anchor=west] at (11.85, 0) {physics-informed (\S\ref{sec:rl:physicsinformed})};
\end{scope}
\end{tikzpicture}

%% file: sections/3_accelerators.tex
% =============================================================================
%  3. Accelerator Control and Beam Optimization
% =============================================================================

% -----------------------------------------------------------------------------
\subsection{Background: The Problem, Traditional Methods, and the Case for RL}
\label{sec:accelerators:background}

Particle accelerators are among the most complex control systems in science.
A modern synchrotron or linear accelerator contains hundreds to thousands of
independently controlled elements: dipole and quadrupole magnets, radio-frequency~(RF) cavities,
collimators, and correction coils, all of which must be jointly tuned to steer,
focus, and accelerate a particle beam to the required energy, intensity, and
emittance.
The control problem is further complicated by time-varying disturbances (ground
motion, power supply ripple, temperature drifts, radiation damage), beam loading
effects, and hard safety constraints: excessive beam loss can quench
superconducting magnets, damage vacuum components, or activate surrounding
material.

In practice, a modular control system is built to manage this complexity.
Such a system combines low-level hardware parameters with higher-level
accelerator physics parameters, whose physics-to-hardware translation is stored
in databases and software rules derived by human domain experts.
For correction and tuning, low-level feedback loops and high-level physics
algorithms adjust beam parameters based on instrumentation observables.
The critical limitation of this approach is portability: a control module is
designed and tuned for a specific accelerator, so each newly commissioned machine
requires brand-new tools for tuning and performance optimization.
This is both costly and time-consuming; even a marginal improvement in the
control pipeline translates into improved scientific yield across the many domains
that rely on particle accelerators: fundamental physics research, medical
treatments, vaccine development, and materials science.

\paragraph{Traditional approaches.}
Accelerator control has historically relied on a combination of optics models,
hand-crafted feedback controllers, response matrix inversion, and extensive
expert manual tuning.
\textbf{Model-based control} using measured or simulated transfer matrices is
effective when the machine is well characterized but degrades as the model
becomes stale due to thermal or mechanical drift.
The \textbf{proportional-integral-derivative}~(PID) controller~\cite{Astrom2006PID} applies a correction
proportional to the instantaneous error, its integral over time, and its
derivative: it is reliable and well-understood but must be re-tuned for each
machine configuration and cannot handle non-linear dynamics or multi-element
coupling.
\textbf{Numerical optimization methods} are widely used to minimize an objective
function over the parameter space of the accelerator without requiring an explicit
model.
The \textbf{Nelder--Mead simplex}~\cite{Nelder1965Simplex} is a widely used
gradient-free method in accelerator operation, applied to routine tuning of
injection efficiency, orbit, and beam intensity; it is robust and simple to
implement but converges to local minima and scales poorly with the number of
parameters.
\textbf{Robust Conjugate Direction Search}~(RCDS)~\cite{Huang2013RCDS} improves
over the simplex by exploiting approximate curvature information and is widely
deployed for orbit and optics correction at synchrotron light sources.
\textbf{Extremum seeking}~(ES)~\cite{Scheinker2024ES} is a model-free,
gradient-estimation technique suitable for real-time optimization of drifting
systems; it has been applied to free-electron laser~(FEL) pulse energy
maximization, beam trajectory
control, and multi-objective emittance optimization at LCLS and AWAKE.
The \textbf{covariance matrix adaptation evolution strategy}~(CMA-ES)~\cite{Hansen2003CMAES} and other evolutionary strategies
maintain a population of candidate solutions and update a covariance matrix to
adapt the search direction, offering better global search capability than simplex
methods at the cost of more function evaluations.
\textbf{Bayesian optimization}~(BO)~\cite{Roussel2024BOReview} treats the
accelerator as an expensive black-box function, builds a Gaussian process~(GP)
surrogate of the objective, and selects the next query point by maximizing an
acquisition function~(e.g., Expected Improvement or Upper Confidence Bound).
BO is sample-efficient for low-to-moderate-dimensional problems and can
incorporate safety constraints~\cite{Kirschner2022SafeBO} and prior physics
knowledge into the GP model~\cite{Hanuka2021PhysicsGP}; a further extension by
Boltz et al.~\cite{Boltz2025NNPrior} uses neural network surrogate models trained
on beam dynamics simulations or historical data as informative prior mean
functions, substantially accelerating convergence in high-dimensional spaces even
when the surrogate only partially matches experimental conditions.
The GP surrogate nevertheless scales as $\mathcal{O}(n^3)$ in the number of
observations, and standard kernels lose accuracy beyond a few tens of
parameters, which motivates sparse or structured GP approximations.
\textbf{Genetic algorithms}~(GA) and particle swarm
optimization~\cite{Kennedy1995PSO} have been
applied to multi-objective beam optics design~\cite{Emery2003GA} but require
large numbers of objective evaluations and produce a static solution that does
not adapt online.
Most of these methods treat each tuning task as an independent optimization run
and do not learn a reusable mapping from machine state to corrective action.

\paragraph{Why RL?}
Accelerator tuning is inherently a sequential decision problem: adjusting one
magnet changes the beam state seen by downstream elements, which may then require
further corrections.
This state-dependent, sequential structure makes it a natural MDP. The MDP formulation for accelerator control is typically:
\textit{state}: a vector of beam diagnostics (beam position monitor readings,
current transformer measurements, emittance estimates);
\textit{action}: incremental changes to magnet currents or RF phases;
\textit{reward}: a scalar encoding beam quality, such as the negative
root-mean-square error~(RMSE) between measured and reference trajectory,
transmission fraction, or emittance.

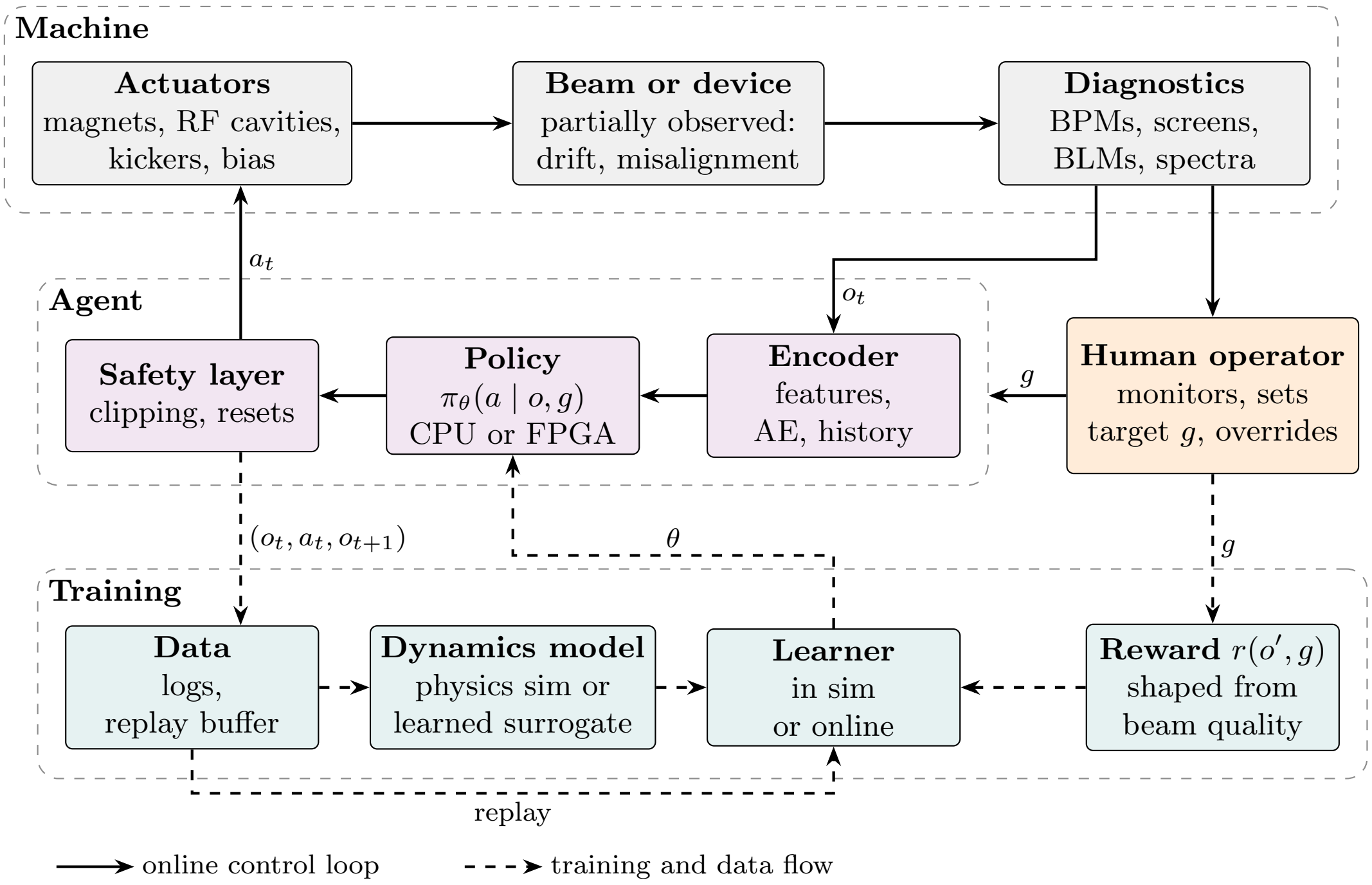
\begin{figure}[t]
    \centering
    \input{src/3/rl_loop.tex}
    \caption{Generalized RL control loop for accelerators and physics
    equipment.
    Solid arrows: online control loop.
    The encoder maps the diagnostics to an observation $o_t$, the policy
    $\pi_\theta$~(deployed on a CPU or FPGA) selects actuator changes $a_t$
    given the operator target $g$, and a safety layer clips actions or resets
    the episode.
    The human operator monitors the machine, sets the target, and can
    override the agent.
    Dashed arrows: training.
    The learner updates $\theta$ from rollouts of a dynamics model~(a physics
    simulator, or a surrogate fitted to logged data), from replayed real
    transitions, or both, under a reward shaped from beam quality.}
    \label{fig:accelerator_rl}
\end{figure}

Once trained, an RL agent computes each corrective action from the current
accelerator state in a single forward pass~(Figure~\ref{fig:accelerator_rl}) and
typically converges within a few steps, whereas numerical optimizers must query
the machine iteratively for each new task.
RL offers the following structural advantages over traditional approaches:
\begin{itemize}
    \item It operates \textit{online} and, when trained over a distribution of
          machine conditions, can follow moderate drifts without retraining.
    \item It optimizes directly against non-differentiable physical objectives:
          transmission efficiency, emittance, or luminosity.
    \item With a fast simulator or a learned surrogate, the training cost is
          moved offline, so that the number of real-machine interactions at
          deployment can be much smaller than for online optimizers such as
          BO~\cite{Kaiser2024RLOvsBO}.
    \item It can incorporate safety constraints through reward penalties, safe
          action filtering, or, in principle, constrained
          MDPs~(Section~\ref{sec:accelerators:future}).
    \item A single trained policy can, in principle, generalize across machine
          configurations and tasks, unlike control modules designed for a single
          deployment.
\end{itemize}
The accelerator control problem embeds nearly all of the real-world RL challenges
discussed in Section~\ref{sec:rl:practical}: data efficiency, large and partially
observable state and action spaces, hard safety constraints, multi-objective
rewards, real-time inference, and the need for policy explainability.

% -----------------------------------------------------------------------------
\subsection{Beam Steering, Tuning, and Trajectory Control}
\label{sec:accelerators:steering}

The earliest and most developed strand of RL for accelerator physics concerns
beam steering and tuning: the problem of adjusting corrector magnets to minimize
orbit distortion or maximize beam transmission along a beamline.
Table~\ref{tab:accelerators} summarizes the MDP formulations of representative
works reviewed in this section, in terms of observation, action, reward, and training
and transfer strategy, following the notation of
Figure~\ref{fig:accelerator_rl}; their baselines and outcomes are collected in
the cross-domain Table~\ref{tab:summary}.

\begin{table}[t!]
\centering
\caption{MDP formulations of representative reinforcement learning approaches
reviewed in Section~\ref{sec:accelerators}; baselines and outcomes are reported in
Table~\ref{tab:summary}.
Symbols follow Figure~\ref{fig:accelerator_rl}: observation $o$, action $a$,
and reward $r$; all action spaces are continuous except for the Fermilab
Booster, where the current is discretized.
\textit{Training and transfer}: environment used for policy training (physics
simulator, learned or empirical surrogate, or real machine) and strategy used to
bridge it to the real machine.
DR: domain randomization; FT: fine-tuning on the target lattice;
RMS: root mean square; RMSE: root mean squared error; MAE: mean absolute
error; $\Delta f$: step-wise
improvement in beam quality metric; RLO: reinforcement learning-trained
optimization; DDRL: deep differentiable reinforcement learning; TES:
transition-edge sensor.}
\label{tab:accelerators}
\small
\setlength{\tabcolsep}{4pt}
\renewcommand{\arraystretch}{1.2}
\begin{tabular}{@{}p{3.3cm} p{2.7cm} p{2.6cm} p{3.0cm} p{3.0cm}@{}}
\toprule
\textbf{Facility; task; algorithm} &
\textbf{Observation $o$} &
\textbf{Action $a$} &
\textbf{Reward $r$} &
\textbf{Training and transfer} \\
\midrule

CERN AWAKE \& LINAC4; trajectory correction; PER-NAF~\cite{Kain2020AWAKE}
  & BPM readings (orbit vector)
  & Corrector current increments
  & $-\mathrm{RMSE}(\mathbf{x}_\mathrm{meas},\,\mathbf{x}_\mathrm{ref})$
  & Online on the real machine; separate sim-trained agent deployed zero-shot \\

Fermilab Booster; magnet regulation; DQN~\cite{StJohn2021Booster}
  & LSTM surrogate output
  & Discretized 1-D current
  & $-|e_t|$ (regulation error)
  & LSTM surrogate learned from real logs \\

KARA; betatron oscillation damping; PPO~\cite{Scomparin2026KARA}
  & Last 8 BPM samples
  & Stripline kicker signal
  & L2 penalty on beam position
  & Online on the real machine; \SI{2.8}{\micro\second} inference \\

LHC; tune feedback; NAF, PPO~\cite{Grech2022LHCTune}
  & Horizontal \& vertical tune errors
  & 16 quadrupole current changes
  & Negative mean squared tune error
  & Gym sim of the feedback loop; sim only \\

CAFe~II; orbit correction; trend-based SAC~\cite{Chen2023TBSAC}
  & BPM readings $+$ two-step trend
  & Corrector increments
  & $-\mathrm{RMS}$ orbit $+$ per-BPM trend term
  & Physics sim; zero-shot \\

DESY ARES; beam focusing under partial observability; TD3 (RLO)~\cite{Kaiser2022ICML}
  & Actuator values $+$ beam params $+$ target $b'$
  & $\Delta$ actuator setpoints
  & Signed log-MAE improvement; asymmetric penalty
  & Cheetah; DR over $I$, $M$; zero-shot \\

DESY ARES; beam focusing \& steering; TD3 (RLO)~\cite{Kaiser2024RLOvsBO}
  & Screen beam parameters $+$ actuator values $+$ target
  & $\Delta$ actuator setpoints
  & $\Delta f$ (step improvement)
  & Cheetah; DR; zero-shot \\

FLUTE and ARES lattices; transverse tuning; SAC~\cite{Xu2023LatticeAgnostic}
  & Screen beam params $+$ target $+$ magnet settings
  & Magnet setpoints
  & $-$normalized MAE to target
  & Cheetah; DR over lattices; evaluated in sim; optional FT \\

COSY J\"ulich; injection optimization; SAC~\cite{Awal2024Injection}
  & Transverse beam profile
  & Beam preparation element settings
  & Cross-section match to target
  & Physics sim; DR; live evaluation \\

VEPP-5; beamline transmission; SAC $\to$ DDPG~\cite{Ibrahim2025RL, Ibrahim2026RLABC}
  & Beam statistics $+$ aperture margins
  & Quadrupole \& dipole currents
  & Multi-term particle transmission loss
  & Elegant; stage learning (curriculum) \\

FERMI FEL (Elettra); intensity optimization; MBRL~\cite{Hirlaender2020FERMIFEL}
  & Ensemble model state
  & Undulator \& chicane settings
  & Photon pulse intensity
  & Learned neural network ensemble; uncertainty-based stopping \\

CEBAF (JLab); heat load \& trip rate; DDRL~\cite{Rajput2025DDRL}
  & Cavity and beam state
  & Cavity settings
  & Pareto-weighted heat load $+$ trip rate
  & Differentiable surrogate learned from real data; not deployed \\

CEBAF (JLab); heat load \& trip rate under energy constraint; LC-TD3~\cite{Colen2026ExplainableRL}
  & Cavity gradients
  & Cavity gradients
  & $-$heat, $-$trip, or both; penalty on surrogate-predicted energy
  & Empirical surrogate from real data; sim only \\

Gran Sasso (CRESST); TES bias tuning; SAC~\cite{CRESST2023RL}
  & Detector response curve
  & TES temperature \& bias current
  & Energy resolution proxy
  & Detector response sim; live underground training \\

\bottomrule
\end{tabular}
\end{table}

\paragraph{Kain et al.~\cite{Kain2020AWAKE}} reported the first online training of
model-free RL agents for trajectory correction at CERN, on two facilities:
the AWAKE plasma-wakefield experiment and the LINAC4 linear accelerator.
Because the action space is continuous~(incremental magnet current setpoints),
the authors employ a variant of DQN known as the \textbf{Normalized Advantage
Function}~(NAF)~\cite{GuLSL16NAF}, which extends Q-learning to continuous actions
by parameterizing the advantage as a negative definite quadratic:
\begin{equation}
    A(s, a) = -\frac{1}{2}
    \bigl(a - \mu(s)\bigr)^\top P(s)\,\bigl(a - \mu(s)\bigr),
    \label{eq:naf}
\end{equation}
where $P(s)$ is a state-dependent positive-definite matrix and $\mu(s)$ is the
greedy action, both parameterized as neural network outputs.
With this form, $a^\star = \mu(s)$ recovers the optimal action analytically
without a separate actor network.
The Q-function is therefore $Q(s, a) = V(s) + A(s, a)$, with the greedy policy
recovered in closed form.
To improve sample efficiency, the agent uses a Prioritized Experience Replay
buffer~(PER)~\cite{Schaul2015PER}, yielding the full \textbf{PER-NAF} agent.
The reward is defined as the negative root mean squared deviation between the
measured and reference beam trajectories:
\begin{equation}
    r_t = -\sqrt{\frac{1}{N}\sum_{i=1}^{N}
    \bigl(x_i^{\mathrm{meas}} - x_i^{\mathrm{ref}}\bigr)^2},
    \label{eq:acc_reward}
\end{equation}
where $N$ is the number of beam position monitors and $x_i$ are their readings.
On both machines, the agents are trained online from scratch: at AWAKE, the
training for eleven degrees of freedom takes about 350 iterations, or roughly
half an hour of beam time.
In a separate test, an agent trained only on a simulation of the AWAKE line is
deployed on the real machine without retraining and corrects the trajectory to
better than \SI{2}{\milli\metre} RMS within one or two iterations, as well as
the agent trained online.
Safety is managed implicitly: the learning episode is reset whenever the actions
of the agent violate operational bounds, providing the agent with knowledge of system
constraints through experience rather than explicit constraint enforcement.
This does not guarantee complete safety during online operation, a limitation
common to most existing accelerator RL deployments and discussed further in
Section~\ref{sec:accelerators:future}.

\paragraph{O'Shea, Bruchon, and Gaio~\cite{OShea2020FERMIPG}} applied model-free
policy gradient RL directly on the FERMI free-electron laser at Elettra
Sincrotrone Trieste, without any pre-training in simulation.
The algorithm is first characterized on a simple number-guessing environment to
select its hyperparameters, then trained in situ on two tasks: maximizing the
FEL output intensity and optimizing an accelerator-based terahertz source.
Using up to five control parameters and no domain knowledge, the agent optimizes
the target signals within a few hundred steps and shows a modest tolerance to
machine faults without dedicated preparation.
An earlier study by the same group~\cite{Bruchon2020FERMIBasicRL} applied basic
model-free RL techniques to the alignment of the seed laser that controls the
FERMI intensity, motivated by the limitations of the gradient ascent and
extremum seeking methods used in operation: episodic Q-learning with linear
function approximation reaches the optimal working point, and a natural policy
gradient~(NPG)~\cite{Kakade2001NPG} version of REINFORCE~\cite{Williams1992REINFORCE}
recovers performance after drifts; both were first
tested on the laser alignment of the electro-optical sampling station and then
on the FEL itself.
Together with the model-based study of Hirlaender and
Bruchon~(Section~\ref{sec:accelerators:modelbased}), these works make FERMI one
of the few facilities where model-free and model-based RL have been compared on
the same machine.

\paragraph{Kaiser, Stein, and Eichler~\cite{Kaiser2022ICML}} carried out the first
systematic study of RL for accelerator optimization under partial observability
and without any real-machine training time, demonstrated on the ARES experimental
area at DESY.
The formulation explicitly acknowledges that the ARES state is a POMDP: neither
the incoming beam $I$ nor the magnet misalignments $M$ are measurable, so the
agent observes only the current actuator values $x_t$, the measured beam
parameters $b_t$, and the operator-specified target $b'$.
To handle this partial observability, both $I$ and $M$ are treated as uniformly
distributed random variables over physically motivated ranges and sampled fresh at
the start of each training episode, forcing the policy to generalize over all
plausible machine states rather than memorize a single configuration.
Actions are defined as continuous delta adjustments $\Delta x_t$ rather than
absolute setpoints, which naturally accommodates magnet hysteresis.
The reward is defined as the signed improvement in the logarithm of a weighted
mean absolute error between observed and target beam parameters:
\begin{equation}
    r_t = \begin{cases}
        \hat{r}_t & \text{if } \hat{r}_t > 0, \\
        2\hat{r}_t & \text{otherwise},
    \end{cases}
    \qquad
    \hat{r}_t = \ln\!\sum_p w_p |p_t - p'| - \ln\!\sum_p w_p |p_{t+1} - p'|,
    \label{eq:ares_reward}
\end{equation}
where $p$ indexes beam parameters~($\mu_x, \mu_y, \sigma_x, \sigma_y$) and the
asymmetric penalty~(doubling negative rewards) discourages the agent from
traversing high-error states even if it can subsequently recover.
Three TD3 agents~\cite{Fujimoto2018TD3} trained for millions of steps in the
Cheetah simulator~\cite{Kaiser2024Cheetah, Stein2022Cheetah} are deployed
zero-shot on the real ARES machine.
In simulation, the agents outperform all classical optimizers tested except the
Nelder--Mead simplex, which matches their final error but needs an order of
magnitude more steps to converge.
On the real accelerator, the agents reach a smaller median beam deviation than
Nelder--Mead, which, however, was given fewer steps there because of limited
beam time.
In a direct comparison with two experienced human operators, the operators reach
a slightly smaller final error, but take about an order of magnitude longer than
the agents.
The same group, together with the Karlsruhe Institute of Technology~(KIT),
subsequently trained lattice-agnostic
policies with randomized magnet lattice configurations: a SAC agent trained on
a dummy lattice tunes the beam on simulated models of the FLUTE and ARES
lattice sections without retraining, and fine-tuning for about 2\% of the
original training steps further improves it~\cite{Xu2023LatticeAgnostic}.

Zero-shot transfer has also been reported on a superconducting linac.
\textbf{Chen et al.}~\cite{Chen2023TBSAC} propose a trend-based variant of SAC
whose agents are trained in simulation and applied directly to the real machine.
On the CAFe~II linac, orbit correction over three cryomodules takes about one
tenth of the time required by human experts, with the RMS of the corrected orbit
below \SI{1}{\milli\metre}; on a light-particle injector, the agent raises the
radio-frequency quadrupole~(RFQ) transmission above 85\% within two minutes.
Earlier, Pang, Thulasidasan, and Rybarcyk~\cite{Pang2020LANSCE} trained a deep
RL controller against a high-fidelity simulator of a section of the LANSCE
linac, reporting better-than-human beam current and distribution in simulation.

The same group subsequently benchmarked this RL-trained optimization approach
against Bayesian optimization on the real machine~\cite{Kaiser2024RLOvsBO}; the
comparison is discussed in Section~\ref{sec:accelerators:modelbased}.

\paragraph{Awal, Hetzel, Gebel, and Pretz~\cite{Awal2024Injection}} applied a SAC
agent to injection optimization at the Cooler Synchrotron COSY at
Forschungszentrum J\"ulich.
The task is to shape the transverse beam cross-section at the injection point to
match the acceptance of the ring, optimizing over the settings of several beam
preparation elements.
Training is performed in a simulated environment with domain randomization over
beam parameters to ensure robustness.
The agent is then evaluated on live COSY beam shifts and successfully matches
human-operator level performance while completing the optimization in
substantially less time.
A systematic ablation study validates the contribution of domain randomization
and the neural network architecture to achieving a robust and generalizable
policy, providing useful empirical guidance for future accelerator RL deployments.

\paragraph{Ibrahim et al.~\cite{Ibrahim2025RL, Ibrahim2026RLABC}} develop, in a
line of work by HSE University and the Budker Institute of Nuclear
Physics~(BINP), an RL framework integrated
with the Elegant simulator~\cite{Borland2000Elegant} for transmission
optimization on the VEPP-5 injection complex positron transport beamline at the
BINP in Novosibirsk.
Their initial work~\cite{Ibrahim2025RL} introduces the Elegant-based RL
environment and demonstrates proof-of-concept results: a SAC agent optimizing
the currents of seven quadrupole magnets achieves 100\% transmission
efficiency for a nominal-emittance beam and 67.6\% for a beam with emittance
sixteen times higher, without any real-machine interaction.
The state is constructed from beam dynamics parameters~($\beta_x, \beta_y,
\eta_x, \eta_y$), transverse beam sizes, aperture limits, coupling factors, and
particle count at each longitudinal position, chosen so that the state is
approximately Markovian.
The reward is a multi-term function designed to penalize particle losses at
intermediate steps, total losses along the beamline, and the distance at which
all particles are lost before reaching the target.

Their follow-up research~\cite{Ibrahim2026RLABC} introduces \textbf{RL-ABC}~(Reinforcement
Learning for Accelerator Beamline Control), a full open-source framework
that automatically transforms standard Elegant lattice files into
Gymnasium-compatible \cite{Towers2025Gymnasium} RL environments\footnote{Gymnasium (formerly known as Gym \cite{BrockmanCPSSTZ16GYM}) provides a standard and abstract interface to implement reinforcement learning environments easily.}.
The framework pre-processes the lattice file to insert diagnostic watch points,
constructs a 57-dimensional observation from the particle distribution at each
watch point~(robust percentile statistics of the transverse positions and
momenta, a coarse transverse histogram, the surviving-particle fractions, the
correlation coefficients of the transverse coordinates, and the aperture
settings), and exposes a configurable transmission-based reward function
compatible with popular RL frameworks.
On a test beamline derived from the VEPP-5 positron transport channel, DDPG
agents reach about 70\% particle transmission.
The best configurations match those of differential
evolution~\cite{Storn1997DE}, but the DDPG
runs have a higher mean and a much smaller spread across independent runs,
whereas Bayesian optimization performs far worse under the same budgets.
A further design feature is \textbf{stage learning}: training starts on a
subset of the beamline elements, or of the control variables, and the trained
weights and replay buffer warm-start the next stage, forming a curriculum of
manageable sub-problems; on the test beamline, this warm start speeds up early
training but does not change the final result.

% -----------------------------------------------------------------------------
\subsection{Model-based, Hybrid, and Comparative Studies}
\label{sec:accelerators:modelbased}

A second line of work improves sample efficiency by learning or incorporating
an explicit model of the accelerator dynamics, enabling the agent to plan or
train without costly real machine interactions.

\paragraph{Hirlaender and Bruchon~\cite{Hirlaender2020FERMIFEL}} applied
model-based RL with Bayesian neural network~(BNN) ensembles to intensity
optimization at the FERMI FEL at Elettra Sincrotrone Trieste.
The key finding is that learning an explicit dynamics model provides higher
representational power, improved sample efficiency, and increased robustness to
noise compared to model-free baselines, though the asymptotic performance of
model-free agents remains superior once sufficient data are available.
Model ensembling is used to estimate the epistemic uncertainty of the learned
dynamics model; the authors use it to decide when to stop training the policy on
model-generated data and to check its performance on the real machine, which
limits over-fitting to model errors.
This uncertainty-aware criterion is important for MBRL on a physical system,
where exploiting model errors can produce harmful actions.

\paragraph{Dong, Tian, and Sun~\cite{Dong2026NSLSII}} applied model-based RL to
orbit feedback in the NSLS-II storage ring, where the standard approach inverts
the orbit response matrix by singular value decomposition~(SVD).
Their framework combines two optimization loops: trajectory optimization of the
expected return within a differentiable dynamics model, and online optimization
of that model from the ongoing agent-machine interaction, so that the learned
dynamics follow the drift of the electron ring.
In simulation, the online model updates let the agent keep reducing the orbit
error over long runs, overtaking the SVD-based correction; on the real storage
ring, where the model was refit once to the current machine state, the method
reaches an RMS alignment error of about \SI{1}{\micro\metre}, against about
\SI{5}{\micro\metre} for the SVD-based feedback.
Adapting the dynamics model during operation, rather than fixing it after
training, has so far been demonstrated over long runs only in simulation.

\paragraph{Learned state representations.}
Several accelerator diagnostics are high-dimensional signals, such as screen
images or frequency spectra, which are poorly suited as direct policy inputs.
Two CERN studies address this by learning a compact state with autoencoders.
\textbf{Velotti et al.}~\cite{Velotti2023AWAKEElectron} automate the setup of
the \SI{18}{\mega\electronvolt} AWAKE electron line: model-free RL agents, with
TD3 performing best, are trained on a state encoding learned by autoencoders
rather than on hand-selected observables, and a variational
autoencoder~(VAE)~\cite{kingma2013auto} provides a synthetic model of the
beamline used to choose the agent hyperparameters.
On the real line, numerical optimizers need up to a few hundred iterations to
reach a suitable configuration, whereas the trained agents need only a handful
of steps; however, agents tested weeks or months after training need more
steps or fail to converge, because the mapping between settings and beam has
drifted.
\textbf{Madysa et al.}~\cite{Madysa2022LEIR} apply the same idea to intensity
optimization in the Low Energy Ion Ring~(LEIR), where the agent tunes electron
cooler and Linac3 settings to maximize the intensity at the end of the injection
plateau.
A VAE compresses longitudinal Schottky spectra into a latent state, and the
agent is pretrained on a surrogate of the electron cooling dynamics learned from
the same data.
In both cases, representation learning and surrogate modeling share the same
data, so the cost of building the environment is paid once.

\paragraph{Meier et al.~\cite{Meier2022SRFGun}} optimized the beam properties
of the superconducting radio-frequency~(SRF) photoinjector developed at HZB as the
electron source for the SeaLab accelerator.
The costly simulation of the gun is replaced by a neural network approximation,
and the RL procedure exploits the cheap derivative of this approximation,
requiring fewer function evaluations than common optimization methods to reach a
given accuracy.
This work anticipates the differentiable surrogate idea that DDRL develops
further.

\paragraph{Rajput et al.~\cite{Rajput2025DDRL}} introduced \textbf{Deep
Differentiable Reinforcement Learning}~(DDRL), an approach that combines a
differentiable physics surrogate with RL by back-propagating gradients of the
objective directly through the surrogate model and into the policy parameters.
The method is applied to simultaneous multi-objective optimization of heat load
and cavity trip rates at the Continuous Electron Beam Accelerator
Facility~(CEBAF) at Jefferson Lab, a superconducting linac operating under strict
per-element and cumulative energy constraints.
A physics-based differentiable surrogate is built from historical operational data and
used both as the RL environment and as a differentiable path for gradient
computation; this eliminates the need to sample rollouts for policy gradient
estimation and instead allows direct optimization of the policy parameters via
backpropagation through the model.
The results are evaluated as a Pareto front over the two objectives, in an
offline optimization task of increasing dimension, from a single cryomodule to
the 200-dimensional CEBAF North linac.
On the smaller problems, DDRL, model-free RL, Bayesian optimization, and a
multi-objective GA reach similar fronts, and the GA converges fastest; on the
full linac, DDRL converges fastest, while multi-objective Bayesian optimization
and model-free RL do not converge within the time limit.
Unlike DDRL, the GA yields a static set of solutions rather than a control
policy, and must be re-run when the linac changes, for example when a cavity
goes offline.
This work clarifies when differentiability confers an advantage: when a
high-quality differentiable surrogate is available, the objective landscape is
smooth, and the problem is high-dimensional, DDRL can be far more efficient
than model-free RL; when the objective is
discontinuous, the surrogate is inaccurate, or online adaptability to
non-stationary conditions is required, model-free RL remains the appropriate
choice.

\paragraph{RLO vs.\ Bayesian optimization.}
The most rigorous existing comparison of RL and Bayesian optimization for online
accelerator tuning is provided by \textbf{Kaiser et al.}~\cite{Kaiser2024RLOvsBO},
conducted on the ARES linac at DESY using both simulation~(300 trials) and real
machine~(22 live trials).
The study frames RL in the paradigm of \textit{reinforcement learning-trained
optimization}~(RLO): the policy is trained offline in simulation to act as a
general-purpose optimizer for the tuning task, then deployed zero-shot on the
real machine.
The key quantitative findings are the following.
In simulation, RLO reaches a median final beam deviation about an order of
magnitude smaller than BO and converges in several times fewer steps.
On the real machine, the gap narrows: RLO outperforms BO in a majority of the
live trials, with notably smoother actuator trajectories; erratic magnet
movements cause hysteresis and power supply glitches on real hardware, so
smoothness is operationally significant beyond pure accuracy.
RLO is also orders of magnitude faster at inference than BO, since it requires a
single forward pass of the policy network rather than acquisition function
optimization at each step, making it suitable for real-time control loops where
the latency of BO is prohibitive.
The practical trade-off is equally clear: BO requires no simulation, no training,
and transfers to new lattices or working points without re-engineering; RLO
requires a high-speed simulation, substantial reward-engineering effort, and a
separate training phase for each new task.
The authors state that BO is the right
choice when a fast simulation is unavailable, the lattice is subject to frequent
modification, or a quick deployment is needed; RLO is preferable when tuning
quality and inference speed are paramount, and the engineering investment can be
amortized over many deployment cycles.
For large machines such as the High-Luminosity LHC~(HL-LHC), where many
correctors must be tuned, conditions evolve over fills, and real-time feedback
loops are required, this analysis suggests that RLO could become an attractive
option, provided the necessary simulation infrastructure is in place; a first simulation study of
an LHC feedback loop is discussed in Section~\ref{sec:accelerators:feedback}.

% -----------------------------------------------------------------------------
\subsection{Real-Time Feedback and Stabilization}
\label{sec:accelerators:feedback}

The works above are \textit{tuning} problems: the agent drives the machine
toward a target configuration within an episode of tens of steps, after which
the settings are held fixed.
A distinct class of problems is \textit{feedback}: the agent acts continuously
at the rate of the control loop to reject disturbances, so latency constraints
of micro- to milliseconds dominate the design, and the policy must be executed
on dedicated hardware close to the actuators.

\paragraph{St.\ John, Herwig, Kafkes et al.~\cite{StJohn2021Booster}} addressed
gradient magnet power supply~(GMPS) regulation at the Fermilab Booster
synchrotron, where the task is to track a time-varying current reference with
high accuracy and low latency.
Because building an accurate physics simulator of the Booster is difficult, the
authors construct a surrogate model: an LSTM network trained on real accelerator
operational data to emulate the dynamical response of the Booster.
This surrogate serves as a safe and cost-free training environment for a double
DQN agent that selects among a small set of discrete adjustments of the
regulator.
The agent is trained to maximize the negative absolute regulation error,
$r_t = -|e_t|$, where $e_t$ is the tracking error, and improves the reward by
about a factor of two over the existing PID controller on the surrogate.
The key engineering contribution is the implementation of the trained network on
an FPGA, with an inference latency of a few microseconds, well within the loop
period of tens of milliseconds of the Booster regulation; the FPGA controller
was tested on a bench setup rather than in the operational loop.
A preceding study by Kafkes and Schram~\cite{Kafkes2021BoosterTwin} validated the
LSTM digital twin of the GMPS and showed that a DQN policy trained on it
stabilizes the regulation under realistic time-varying disturbances, with a
preliminary two- to fourfold improvement over the PID baseline; the authors
motivate offline training on the twin by the risk of training online on the
operational system.

\paragraph{Scomparin et al.~\cite{Scomparin2026KARA}} performed RL feedback
control on the Karlsruhe Research Accelerator~(KARA) storage ring, damping
induced horizontal betatron oscillations in real time.
The observation is the last eight horizontal beam position samples from a beam
position monitor at the revolution frequency of about \SI{2.7}{\mega\hertz},
and the action is the signal sent to a stripline kicker.
A compact PPO actor with a single hidden layer runs on the AI Engine array of an
AMD-Xilinx Versal device, with a full inference loop of \SI{2.8}{\micro\second}.
Training is split between the device and the control room: an experience
accumulator on the device collects transitions, the policy update runs on a
control room computer, and the updated weights are loaded back onto the
hardware.
The agents are trained from scratch on the real machine and are compared with
the commercial bunch-by-bunch feedback installed at KARA, which is based on
finite impulse response~(FIR) filters: the authors summarize the performance as
comparable, and every trained configuration, across reward definitions and
network sizes, reaches a higher damping rate than the FIR controller.
This is a direct experimental answer to the latency question raised in
Section~\ref{sec:accelerators:future}: online RL with microsecond inference is
feasible on current hardware.
The work builds on an earlier KARA framework by Wang et al.~\cite{Wang2021KARAFPGA},
which targeted inference latencies below \SI{100}{\micro\second} on a Xilinx
Zynq UltraScale+ device for longitudinal beam dynamics feedback and validated
it on cart-pole and pendulum control tasks.

\paragraph{Grech, Valentino, Alves, and Hirlaender~\cite{Grech2022LHCTune}}
studied RL for the tune feedback of the LHC, the component of the beam-based
feedback system that corrects the horizontal and vertical betatron tunes.
In a dedicated Gym simulation of the feedback loop, the state is the pair of
normalized tune errors and the action sets the current changes of 16 tuning
quadrupoles.
Five algorithms are tested~(NAF, PPO, SAC, TD3, and the model-based AE-DYNA),
and the best NAF and PPO agents reach the target in about 9 steps on average,
compared to about 20 for the reimplemented proportional-integral~(PI)
controller, which inverts the tune response matrix.
When one or three actuators are stuck at a random setting, both agents degrade
more gracefully than the PI controller: with one stuck actuator, the PI
controller fails in all test episodes, whereas the NAF agent still succeeds in
one third of them, and with three, the agents end closer to the target.
The study is simulation-only, but it targets an operational LHC feedback loop
directly, with the same actuators and observables as the existing controller.

% -----------------------------------------------------------------------------
\subsection{Cryogenic and Specialized Equipment Control}
\label{sec:accelerators:cryogenic}

A striking application of RL to physics instrumentation control beyond the
accelerator domain is the work of the CRESST collaboration~\cite{CRESST2023RL}.
CRESST operates cryogenic phonon detectors with transition-edge sensors~(TES) at
millikelvin temperatures in the Gran Sasso underground laboratory, targeting
direct dark matter-nucleus scattering.
Optimal detector sensitivity requires careful and simultaneous tuning of the TES
temperature and the bias current in its readout circuit: small deviations from
the optimal operating point degrade energy resolution significantly.
This tuning has historically been performed manually by a trained physicist, a
process that can take up to several hours per detector and does not scale to the large detector
arrays envisioned for next-generation searches.

The authors automated this procedure with RL in two settings: first, training on
a simulation of the CRESST detector response used as a virtual RL environment;
second, training \textit{live} on three physical CRESST detectors operating in
the Gran Sasso underground setup.
In both cases, SAC agents optimized a standard detector as fast as human
experts and with comparable results.
This work carries two important implications for the field.
First, it shows that an RL agent can be trained and operated live inside a deep
underground physics experiment, interfaced with the experiment control
software, in an environment where manual intervention is costly.
Second, autonomous detector tuning is a prerequisite for the large detector
arrays of future experiments: the planned CRESST upgrade foresees several tens
to hundreds of detector modules, and because the agents need no manual
interaction, the procedure can be run in parallel for all of them.

% -----------------------------------------------------------------------------
\subsection{Open Problems and Future Research Directions}
\label{sec:accelerators:future}

\paragraph{Safe RL and operational constraints.}
All existing deployments handle safety through reward penalties, episode resets
on constraint violation, or conservative action clipping.
None provides formal safety guarantees.
Constrained MDPs with Lagrangian relaxation and control barrier
functions~(Section~\ref{sec:rl:safety}) offer principled alternatives: the
former would encode beam loss, quench, and actuator limits as explicit cost
constraints, while the latter would certify that the machine state remains in
the safe set at every step.

An additional challenge specific to model-based accelerator RL is propagating
safety constraints through the learned surrogate model: an action that the
surrogate predicts to be safe may still violate constraints on the real machine
due to model error.
Ensemble-based uncertainty estimates~\cite{Hirlaender2020FERMIFEL} are a first
step toward addressing this, but robust constraint satisfaction under surrogate model uncertainty
remains an open theoretical problem with direct operational consequences.
Safe model-based RL~\cite{Berkenkamp2017SafeMBRL} offers a more principled
path: by maintaining a Lyapunov function over the state space and using a
Gaussian process model of the dynamics, the agent can certify a growing safe
region of the parameter space and guarantee that exploration never leaves it.
The GP dynamics model provides calibrated uncertainty estimates that tighten
the safe region where the surrogate is reliable and expand it only as new data
confirms safety, providing exactly the uncertainty-aware constraint propagation
that ensemble methods approximate heuristically.
Adapting this framework to the accelerator setting requires replacing the GP
with a more scalable model, such as a neural network ensemble, and
incorporating the known physics of beam optics as a structured mean function;
both remain open research directions.
The key open question is whether the Lyapunov stability certificates can be
maintained when the dynamics model is a high-dimensional neural network rather
than a GP, where coverage guarantees are harder to establish formally.

\paragraph{Generalizable policies across facilities.}
Current policies are trained for a specific beamline and machine configuration.
Building policies that generalize across lattice configurations, facilities, or
beam energies would dramatically reduce the engineering effort for new
deployments.
The lattice-agnostic approach of~\cite{Xu2023LatticeAgnostic} is an early step,
demonstrating zero-shot generalization, in simulation, to lattice sections of
two different linacs through domain randomization over lattice parameters.

The cross-facility generalization problem is structurally equivalent to
sim-to-real transfer~(Section~\ref{sec:rl:sim2real}): the target facility is an
out-of-distribution environment relative to the source, and the gap has the same
two components, namely model error in the physics simulation and distributional
shift in operating conditions.
The methods developed for sim-to-real transfer therefore apply directly:
\begin{itemize}
    \item \textbf{Domain randomization} over lattice parameters, magnet misalignment, and
beam emittance during training forces the policy to discover behaviors that are
robust across a family of machines rather than optimal for a single
nominal configuration~\cite{Tobin2017DomainRand}.

    \item \textbf{System identification}~\cite{Kaiser2024Cheetah} can narrow the
remaining gap by fitting the simulation parameters of a new facility to a small
number of real measurements before deployment, reducing the effective
distribution shift to a residual that the built-in robustness of the policy can
tolerate.

    \item \textbf{Meta-RL}~(Section~\ref{sec:rl:paradigms}) treats each facility as a distinct task within a shared latent space of accelerator configurations~\cite{beck2025tutorial}, so that the policy can specialize to a new facility from a handful of interactions.
First steps exist at the AWAKE electron line: Hirlaender et al.~\cite{Hirlaender2024MetaRL} use MAML to adapt a policy within 3 to 6 interaction steps to a POMDP whose hidden parameters drift over time, and a Gaussian process model-based variant addresses few-shot training when no simulation is available~\cite{Hirlaender2024FewShot}.

    \item \textbf{Model-based RL with online planning}~\cite{Chua2018PETS} offers a
complementary route to generalization.
If the world model~\cite{Ha2018WorldModels} is expressive enough to represent the dynamics of a family
of accelerators, the agent can adapt to a new facility by updating its world
model from a small number of real transitions and re-planning with the updated
model, without retraining the policy network.
Planning algorithms such as MCTS or model-predictive
control~\cite{Koller2018SafeMPC} can then generate facility-specific action
sequences at deployment time by rolling out the updated model, combining the
data efficiency of model-based methods with the flexibility of online
re-planning.

    \item \textbf{Offline RL} from archived operational logs of multiple
facilities~(Section~\ref{sec:rl:offline}) could pretrain a broad prior over
accelerator dynamics without any real machine time, which a lightweight online
fine-tuning step could then specialize to a new target.
\end{itemize}

In principle, a truly general RL agent could optimize the full operational
workflow of a particle accelerator, generalizing across equipment
types~(linear, circular) and tasks~(magnet tuning, beam regulation, injection);
recent progress in zero-shot sim-to-real transfer~\cite{Kaiser2022ICML,
Awal2024Injection} and lattice-agnostic policies~\cite{Xu2023LatticeAgnostic}
suggests that steps toward this goal are feasible.

\paragraph{Real-time deployment on FPGAs and edge hardware.}
The Fermilab Booster~\cite{StJohn2021Booster} and KARA~\cite{Scomparin2026KARA}
works~(Section~\ref{sec:accelerators:feedback}) demonstrated hardware inference
with microsecond latency for single feedback loops, the latter with online
training on the real machine.
Extending this to full-machine control at LHC or HL-LHC frequencies requires
co-design of RL policy architectures with hardware constraints: policies must be
small enough to fit on the target FPGA device, execute within the timing budget
of the control system, and remain robust to the fixed-point arithmetic of
hardware inference.

Three complementary techniques reduce the computational footprint of a trained
policy without retraining from scratch:
\begin{itemize}
    \item \textbf{Quantization-aware training}~\cite{Krishnan2022QuaRL} reduces the
precision of network weights and activations during training to match hardware
arithmetic; the work of Kresse and Lampert~\cite{Kresse2025QuantizedRL}
demonstrates that continuous-control RL policies can be quantized to as few as
2--3 bits per weight with negligible performance loss on MuJoCo benchmarks \cite{todorov2012mujoco} and
deployed to an Artix-7 FPGA with microsecond inference latency and micro-joule
energy consumption per action.
Crucially, the same work finds that input quantization must be treated
separately: the precision of the input state is generally the bottleneck,
because quantization error in the first layer prevents the policy from
distinguishing similar sensor states that matter for fine control, whereas
deeper layers, where the information is spread over a more redundant latent
space, tolerate much lower bitwidths.

 \item \textbf{Pruning}~\cite{Han2016DeepCompression} removes redundant parameters to reduce model size,
and \textbf{hls4ml}~\cite{Duarte2018hls4ml, schulte2026hls4ml} compiles trained
neural networks directly to synthesizable hardware description language, enabling
sub-microsecond FPGA inference for neural network classifiers in the LHC L1
trigger system.

    \item \textbf{Policy distillation}~\cite{Rusu2016PolicyDistillation} is a
complementary approach that is particularly well suited to the FPGA deployment
problem: a large, expressive teacher policy is trained first without hardware
constraints, and a compact student policy is then trained via supervised
regression on the action distribution of the teacher rather than by direct RL.
The student inherits the behavior of the teacher while being dramatically
smaller and faster; the compression ratio can be adjusted freely to meet the
FPGA resource budget, unlike quantization, which is constrained by the
fixed-point arithmetic of the hardware.
\end{itemize}

Hardware-aware RL training, where the quantization and latency constraints are
incorporated as penalties during policy optimization, remains an open research
direction.
The recurrent networks usually employed to handle partial
observability~(Section~\ref{sec:rl:mdp}) are particularly expensive in
hardware: LSTM~\cite{Hochreiter1997LSTM} and gated recurrent unit~(GRU)~\cite{Cho2014GRU} cells require sequential state
updates that cannot be fully parallelized on an FPGA, motivating the development
of hardware-friendly alternatives such as linear recurrent
units~\cite{Orvieto2023LRU} or S4-based state-space
models~\cite{Gu2022S4} for RL policy networks: the latter requires only a fixed-size state vector multiplication step during update, with no gating operations, making it significantly cheaper on an FPGA than an LSTM of comparable capacity.

\paragraph{Explainable and verifiable policies.}
Operators will hand control of a machine to a policy only if they can check
that its decisions follow from the physics of the beamline, yet none of the
deployments reviewed above evaluates an explanation of the learned policy.
Colen et al.~\cite{Colen2026ExplainableRL} take a first step by making the
constraints that shape the policy inspectable, rather than the policy itself.
In TD3 agents with learnable constraints~(LC-TD3) for surrogate environments of
the CEBAF linacs, calibrated on
historical data and ranging from a single cryomodule to the 197 cavities of the
North Linac, a learned surrogate predicts the energy gain from the state and
action, and a penalty on the predicted energy enters the actor objective, so
that the policy minimizes heat load and trip rate within the target energy band.
When the surrogate is a sparse dictionary model in the spirit of sparse
identification of nonlinear dynamics~\cite{Brunton2016SINDy}, it yields a closed-form expression for the
energy gain, linear in the cavity gradients, whose coefficients agree with the
ground-truth cavity parameters of the environment: operators can therefore
verify the physics behind the constraint the agent obeys.
On the North Linac, the constrained agents converge to the target energy band in
88\% to 100\% of the trials for every objective, whereas TD3 with the energy
penalty in the reward converges in at most 50\%; in the multi-objective setting,
the sparse variant achieves the largest Pareto front coverage, while the neural
surrogate variant fails to satisfy the energy constraint.
The approach does not require a differentiable environment, but it converges
more slowly and to worse configurations than DDRL~\cite{Rajput2025DDRL}, which
starts from an accurate differentiable surrogate, and it has not yet been tested
on the real machine.

The explanation, however, covers the constraint and not the policy.
Policy distillation, introduced above for FPGA deployment, offers a route to the
latter: the student can be chosen for readability rather than size, for example
a soft decision tree or a program in a domain-specific
language~(Section~\ref{sec:bsm:future}), or, for steering and orbit correction,
a low-order map from BPM readings to corrector settings that operators can
compare directly with the inverted orbit response matrix.
A single compact student could then serve both purposes, fitting the latency
budget of the control loop and remaining readable by operators.
The open questions are how much fidelity to the teacher such a student retains
in the nonlinear regimes where RL outperforms classical controllers, and whether
its explanations remain valid after transfer to the real machine.

\paragraph{Bent crystal setup.}
Bent crystals are used at CERN in two settings whose setup is a sequential
alignment task, and both are candidates for RL.
At the SPS, a thin bent crystal placed upstream of the electrostatic septum
shadows the septum blade and reduces slow extraction losses; the crystal is
currently set up with numerical optimizers, and RL agents have been proposed to
shorten the setup time~\cite{Velotti2022CrystalRL}, but no RL results on this
task have been reported so far.
At the LHC, crystal collimator alignment is currently addressed with supervised
models trained on beam loss monitor signals~\cite{Ricci2024CrystalML}.
Both problems share the same structure: the crystal angle is changed in small
increments, and beam loss monitor readings provide feedback after each move,
which maps onto an episodic MDP with a learned, possibly zero-shot transferred,
policy in place of a scan.
The open question is whether such a policy reduces setup time or improves
robustness relative to the optimizers and supervised models already in
operation, which requires the same kind of direct comparison that
Kaiser et al.~\cite{Kaiser2024RLOvsBO} carried out for beam tuning at ARES.

%% file: src/3/rl_loop.tex
% Generalized RL control loop for accelerators and physics equipment (Section 3).
% Requires: \usepackage{tikz}
%           \usetikzlibrary{arrows.meta, positioning, fit, calc, backgrounds}
\begin{tikzpicture}[
    font=\small,
    >={Stealth[length=2mm]},
    block/.style={draw, rounded corners=2pt, align=center, minimum height=1.15cm,
                  inner sep=3pt},
    machine/.style={block, fill=black!6, minimum width=3.2cm},
    agent/.style={block, fill=violet!10, minimum width=2.6cm},
    train/.style={block, fill=teal!10, minimum width=2.6cm},
    human/.style={block, fill=orange!15, minimum width=3.0cm, minimum height=1.6cm},
    tier/.style={draw=black!45, dashed, rounded corners=6pt, inner sep=8pt},
    tierlabel/.style={anchor=north west, font=\small\bfseries, inner sep=3pt},
    loop/.style={->, thick},
    flow/.style={->, thick, dashed},
    lbl/.style={font=\footnotesize, inner sep=2pt},
]

% --- Machine tier --------------------------------------------------------
\node[machine] (act)  at (1.7, 0)  {\textbf{Actuators}\\magnets, RF cavities,\\kickers, bias};
\node[machine] (beam) at (6.6, 0)  {\textbf{Beam or device}\\partially observed:\\drift, misalignment};
\node[machine] (diag) at (11.6, 0) {\textbf{Diagnostics}\\BPMs, screens,\\BLMs, spectra};

% --- Agent tier ----------------------------------------------------------
\node[agent] (safe) at (1.7, -2.8) {\textbf{Safety layer}\\clipping, resets};
\node[agent] (pol)  at (5.0, -2.8) {\textbf{Policy}\\$\pi_\theta(a \mid o, g)$\\CPU or FPGA};
\node[agent] (enc)  at (8.3, -2.8) {\textbf{Encoder}\\features,\\AE, history};
\node[human] (op)   at (12.2, -2.8) {\textbf{Human operator}\\monitors, sets\\target $g$, overrides};

% --- Training tier -------------------------------------------------------
\node[train] (data)  at (1.7, -5.8)  {\textbf{Data}\\logs,\\replay buffer};
\node[train] (dyn)   at (5.0, -5.8)  {\textbf{Dynamics model}\\physics sim or\\learned surrogate};
\node[train] (learn) at (8.3, -5.8)  {\textbf{Learner}\\in sim\\or online};
\node[train] (rew)   at (12.2, -5.8) {\textbf{Reward} $r(o', g)$\\shaped from\\beam quality};

% --- Tier frames ---------------------------------------------------------
\begin{scope}[on background layer]
    \node[tier, fit=(act)(diag), inner ysep=12pt, yshift=4pt] (Tm) {};
    \node[tier, fit=(safe)(enc), inner ysep=12pt, yshift=4pt] (Ta) {};
    \node[tier, fit=(data)(rew), inner ysep=12pt, yshift=4pt] (Tt) {};
\end{scope}
\node[tierlabel] at (Tm.north west) {Machine};
\node[tierlabel] at (Ta.north west) {Agent};
\node[tierlabel] at (Tt.north west) {Training};

% --- Online control loop (solid) -----------------------------------------
\draw[loop] (act) -- (beam);
\draw[loop] (beam) -- (diag);
\draw[loop] ([xshift=-6mm]diag.south) |- ($(diag.south)!0.5!(enc.north)$) -| node[lbl, pos=0.75, right] {$o_t$} (enc.north);
\draw[loop] (enc) -- (pol);
\draw[loop] (pol) -- (safe);
\draw[loop] ([xshift=5mm]safe.north) -- node[lbl, right] {$a_t$} ([xshift=5mm]act.south);
\draw[loop] ([xshift=6mm]diag.south) -- ([xshift=6mm]diag.south |- op.north);
\draw[loop] (op.west) -- node[lbl, above] {$g$} (op.west -| Ta.east);

% --- Training and data flow (dashed) -------------------------------------
\draw[flow] ([xshift=5mm]safe.south) -- node[lbl, right] {$(o_t, a_t, o_{t+1})$} ([xshift=5mm]data.north);
\draw[flow] (data) -- (dyn);
\draw[flow] (dyn) -- (learn);
\draw[flow] (rew) -- (learn);
\draw[flow] (op) -- node[lbl, right] {$g$} (rew);
\draw[flow] (data.south) -- ++(0, -0.45) -| node[lbl, pos=0.25, below] {replay} (learn.south);
\draw[flow] (learn.north) -- ++(0, 0.75) -| node[lbl, pos=0.25, above] {$\theta$} (pol.south);

% --- Legend --------------------------------------------------------------
\draw[loop] ($(Tt.south west) + (0.2, -0.9)$) -- ++(0.8, 0)
    node[right, lbl] {online control loop};
\draw[flow] ($(Tt.south west) + (4.4, -0.9)$) -- ++(0.8, 0)
    node[right, lbl] {training and data flow};
\end{tikzpicture}

%% file: sections/4_dqm.tex
% =============================================================================
%  4. Data Quality Monitoring
% =============================================================================

% -----------------------------------------------------------------------------
\subsection{Background: The Problem, Traditional Methods, and the Case for RL}
\label{sec:dqm:background}

Large-scale particle physics experiments such as ATLAS \cite{ATLAS2008}, CMS \cite{CMS2008}, LHCb \cite{LHCb2008Detector}, and ALICE \cite{ALICE2008}
operate detectors with of order $10^8$ readout channels, sampled every
25\,ns during LHC bunch crossings.
The resulting data rates reach tens of terabytes per hour, and before any physics
analysis can proceed, human operators must validate the recorded data; detector subsystems
must be operating correctly, and runs or luminosity blocks affected by
malfunctions like noise bursts, high-voltage trips, dead channels, or timing
misalignment, must be flagged and excluded.
This process is called \textbf{data quality monitoring}~(DQM) and is of
paramount importance: undetected detector malfunctions introduce biases that can
invalidate entire physics analyses, while overly conservative flagging discards
good data and reduces statistical power.

\paragraph{Online and offline DQM.}
In a typical LHC experiment, DQM is performed at two levels.
The \textit{online} stage occurs synchronously with data-taking, at fixed time
intervals of several minutes: the time needed to accumulate enough counts for
statistically meaningful histogram evaluation.
When anomalies are detected, they must be corrected before data collection
continues, so speed of detection takes priority over depth of analysis.
The \textit{offline} stage runs after data recording is complete; the absence of
time constraints allows evaluation of high-level distributions requiring larger
datasets and more time to reconstruct, catching anomalies that the online pass
missed and flagging the corresponding data for exclusion.
Both stages currently rely heavily on expert shifters who inspect hundreds of
diagnostic histograms per run and issue binary good/bad certifications,
amounting to thousands of expert decisions per year across the LHC experiments.

\paragraph{Traditional approaches.}
Automated DQM has been pursued through threshold-based alarms and reference
histogram comparison, using statistical tests such as the Kolmogorov--Smirnov~(KS)
test and the $\chi^2$ test against reference histograms accumulated from good
runs.
More recently, supervised classifiers and autoencoder-based anomaly detectors
have been trained on reference data from good runs, with or without
labels~\cite{Pol2019DQM, AutoDQM2025CMS}.
A representative recent example is the work of Asres, Omlin, Wang et al.~\cite{Asres2025TL},
who apply a semi-supervised hybrid autoencoder, trained on healthy data and
incorporating convolutional, graph, and recurrent neural networks, to
spatio-temporal anomaly detection in the CMS Hadron Calorimeter~(HCAL).
The authors show that transfer learning from a model pre-trained on a different
section of the calorimeter reduces the number of trainable parameters and the
amount of training data needed for a new section, and improves robustness to
anomalies contaminating the training data, addressing a key practical bottleneck
in DQM automation.
MEDIC~\cite{Bassa2026MEDIC}~(Monitoring for Event Data Integrity and Consistency), a simulation-driven
framework that uses a modified version of the Delphes fast detector simulation \cite{delphes}
to generate controlled fault scenarios, enables the study and development of DQM
methodologies without requiring real experimental data: the neural network learns
to both identify anomalies and localize the specific malfunctioning detector
components responsible, going beyond binary good/bad classification.
Supervised approaches achieve good accuracy when labeled data are abundant, but
suffer in low-label regimes: expert annotation is expensive, labels for rare or
novel failure modes are scarce, and the distribution of detector conditions
shifts over time as the machine and detector age.
Conversely, unsupervised and transfer-learning approaches mitigate the labeling burden but
remain static once trained and cannot incorporate expert feedback to correct
systematic mistakes.
None of the existing automated approaches actively solicit human expertise in a
principled way: the human shifter is either in the loop for every decision or out
of it entirely.

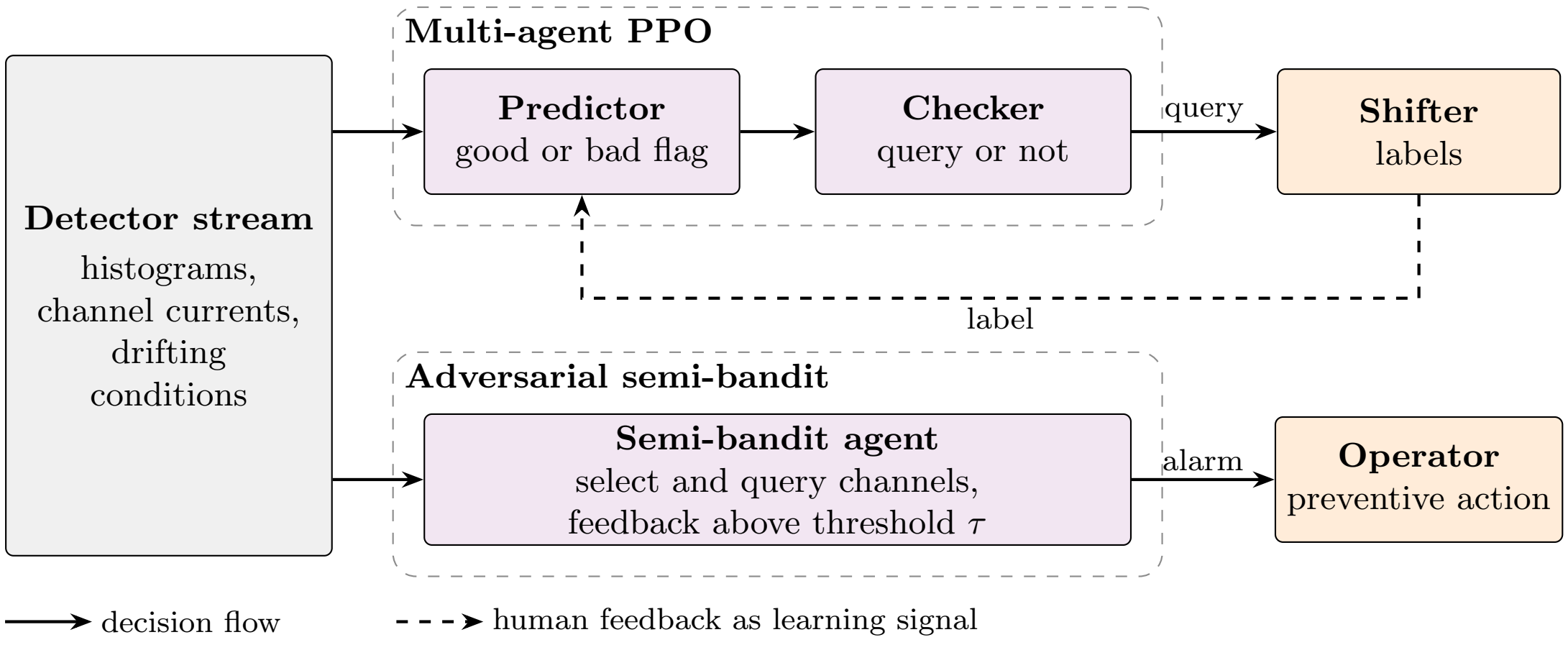
\begin{figure}[t]
    \centering
    \input{src/4/dqm_hitl.tex}
    \caption{Human-in-the-loop and bandit formulations of data quality
    monitoring.
    Top: the multi-agent PPO system of Jullian Parra et
    al.~\cite{jullian2025human}, in which a predictor flags each histogram as
    good or bad and a checker decides whether to query the human shifter, whose
    label updates the predictor.
    Bottom: the adversarial thresholding semi-bandit of
    Bower~\cite{Bower2020ATSB}, which selects and queries channels of the
    ALICE Transition Radiation Detector~(TRD),
    receives feedback only above the threshold $\tau$, and raises early alarms
    to the operator.
    In the first case the human supplies the learning signal; in the second,
    the agent supplies alerts to the human.}
    \label{fig:dqm_hitl}
\end{figure}

\paragraph{Why RL?}
The DQM problem can be formalized as a sequential decision-making task with an adaptive human in
the loop.
An RL~(or bandit) agent observes diagnostic histograms, decides which ones to
flag for expert review, and updates its policy based on the feedback of the expert.
This framing has three structural advantages over static supervised approaches.
First, it naturally minimizes annotation effort by actively selecting the most
informative queries: requesting human intervention only when the agent is
uncertain, rather than for every histogram or for none.
Second, it adapts online to concept drift: as detector conditions change, the
policy updates from the new feedback without requiring a full retraining cycle.
Third, in the bandit formulation, it provides theoretical regret guarantees on
the number of missed anomalies over time, which static classifiers cannot offer.
Bandit algorithms are particularly well suited to the stateless, high-throughput
DQM setting where each certification decision is largely independent. Instead, a full RL formulation
with state transitions becomes appropriate when sequential decisions within a
run are correlated~(e.g., when the status of one subsystem affects the expected
behavior of a downstream subsystem).
Figure~\ref{fig:dqm_hitl} contrasts the two formulations reviewed below.

% -----------------------------------------------------------------------------
\subsection{Human-in-the-Loop RL for DQM}
\label{sec:dqm:hitl}

\textbf{Jullian Parra et al.}~\cite{jullian2025human} provide the first
proof-of-concept for applying human-in-the-loop RL to automate the DQM process
in particle physics experiments.
The setting is motivated by DQM at the LHC experiments and is developed on a
simplified synthetic dataset that captures the essential structure of the
problem: a stream of histograms, each representing a detector observable, must
be classified as good or bad with as few expert queries as possible.

The system implements a \textbf{multi-agent PPO} architecture with two
cooperating agents.
The \textit{predictor} observes the current histogram and emits a binary
good/bad classification.
The \textit{checker} observes both the histogram and the decision of the
predictor and decides whether to query the human shifter; if queried, the label
given by the shifter is used to update the predictor.
The objective of the predictor is to match the judgment of the human; the
objective of the checker is to call the shifter only when the predictor is uncertain, minimizing
total annotation effort while maintaining accuracy.
The state space of the DQM environment comprises the current histogram, the
operating mode of the detector, and the human label if the shifter has been
called.
To handle scarce labeled data, the authors propose data augmentation: artificially
generated histograms are inserted into the training stream to accelerate learning
and improve generalization to new operating conditions.

Three key results emerge.
First, the system learns to request human intervention selectively, primarily at
the start of training or after operating conditions change, and rarely calls the
shifter once the predictor is accurate.
Second, random and unbiased noise in human classification can be reduced: a
predictor trained on shifter labels flipped at random in 30\% of the cases
exceeds the 70\% accuracy of those labels, and the accuracy improves further
when the shifter can consult the outputs of the predictor before deciding,
so that algorithm and shifter improve through mutual feedback.
Third, the framework adapts online to changing detector conditions without
retraining, a property that static supervised classifiers lack by construction.
A preliminary test on a real offline dataset from the LHCb experiment supports
the synthetic setup as a proxy: on 2\,300 shifter-labeled data points, an agent
trained only on augmented synthetic data reaches 91\% accuracy on nominal and
75\% on anomalous points, although the shifter labels may themselves be noisy
and the dataset cannot be released.
The main open challenge, according to the authors, is robustness against
\textit{shifter bias}: a mismatch between the average knowledge of the shifters
and that of the detector experts introduces correlated label noise that the
current system does not address; the authors suggest reference histograms
built from the decisions of the shifters and periodically checked by detector
experts.

% -----------------------------------------------------------------------------
\subsection{Bandit-based Monitoring}
\label{sec:dqm:bandits}

An alternative approach treats DQM monitoring as a bandit problem, without
maintaining state across rounds: at each round, the agent selects which detector
channels to observe and receives feedback only on the observed channels.

\textbf{Bower}~\cite{Bower2020ATSB} introduces the \textbf{Adversarial
Thresholding Semi-Bandit}~(ATSB) formalism and applies it to real-time monitoring
of the anode currents of the ALICE TRD~\cite{ALICE:2017ymw}.
The operational challenge is as follows: the TRD automatically switches off a
device channel when the current exceeds a safety limit, but doing so reduces the
data-collection capacity of the experiment.
A monitoring agent must therefore selectively observe channels whose current
exceeds or deviates from a threshold, intervening before the automatic shutdown
triggers and allowing the human operator to take preventive action.

Formally, the ATSB problem extends the classical non-stochastic multi-armed
bandit~(MAB) to a semi-bandit setting with thresholding feedback.
In each round $t \in \{1, \ldots, T\}$, the player selects a subset of the $K$
arms~(detector channels) $i \in \{1, \ldots, K\}$.
In the semi-bandit setting, the player observes individual arm rewards rather
than only the total reward of the selected subset, enabling fine-grained credit
assignment across a large number of detector channels.
The adversarial~(non-stochastic) formulation is critical: because detector
conditions can change arbitrarily over time, for example due to beam instabilities,
radiation damage, or environmental fluctuations, no stationary reward
distribution can be assumed.
Let $x_{t,i} \in [0, 1]$ be the feedback of arm $i$ in round $t$, $\tau \in
(0, 1)$ a threshold, which may be static or adaptive, $I_t$ the subset of arms
pulled by the player, with $k_t = |I_t|$, and $I^\star_t$ the subset of arms
whose feedback exceeds the threshold, with $q_t = |I^\star_t| = \sum_{i=1}^{K}
\mathbf{1}[x_{t,i} > \tau]$.
The objective of the player is twofold: to pull only arms whose feedback exceeds
the threshold, and to accumulate as much of their combined feedback as possible.
Performance is therefore measured through the \textbf{ATSB regret}:
\begin{equation}
    \mathcal{R}_T = \mathbb{E}\!\left[\sum_{t=1}^{T} \sum_{i \in I^\star_t} x_{t,i}
    - \sum_{t=1}^{T} \frac{1}{\gamma_t} \sum_{i \in I_t} x_{t,i}\,
    \mathbf{1}[x_{t,i} > \tau]\right],
    \qquad \gamma_t = |k_t - q_t| + 1,
    \label{eq:bandit_regret}
\end{equation}
where the expectation is taken over the randomness of the arm selection: pulled
arms contribute only when their feedback exceeds the threshold, and the factor
$1/\gamma_t$ penalizes pulling more or fewer arms than those above threshold,
so that the regret vanishes in a round only when $I_t = I^\star_t$.

The ATSB decision process is decomposed into two sub-decisions per round.
\textit{Selection} decides which arms to pull; \textit{querying} decides whether
to observe none, some, or all selected arms for feedback, managing the cost of
observation.

The proposed algorithm, \textbf{AliceBandit}, comprises three modules: the first
maintains a list of arms that exceeded the threshold in the current round and
pulls all of them; the second queries pulled arms according to Bernoulli random
variables that are updated during play; the third ensures that any arm not pulled
for at least $t_{\max}$ rounds is pulled regardless, preventing any channel from
being permanently ignored.
AliceBandit solves the monitoring task for both fixed-threshold over-current
detection and threshold-interval current-deviation detection.
In the latter setting, which is the use case of major practical interest, AliceBandit identified
deviations in TRD anode current channels \textit{hours} before a trip event,
enabling the human operator to take preventive action before the automatic
shutdown triggered.

The two RL-based DQM approaches reviewed here occupy opposite ends of the
decision-making spectrum: ATSB~\cite{Bower2020ATSB} treats each
monitoring decision as stateless and provides formal regret guarantees, while
the multi-agent PPO system~\cite{jullian2025human} maintains an explicit agent
state and adapts its query behavior as detector conditions change.
Table~\ref{tab:dqm} places both against traditional automated DQM methods in
terms of feedback, label requirements, online adaptation, guarantees, and
validation on real data.
This contrast reflects a genuine open question: whether the sequential
structure of a data-taking run is rich enough to warrant full RL, or whether
the marginal gain over bandit methods justifies the added complexity.
Answering it will require validation on real experiment data across multiple
runs and operating conditions; so far, only a preliminary offline test on LHCb
data exists for the PPO system~\cite{jullian2025human}.

\begin{table}[t]
\centering
\caption{Comparison of data quality monitoring approaches discussed in
Section~\ref{sec:dqm}, spanning traditional automated methods and
reinforcement learning-based systems; the outcomes of the RL-based systems are
also reported in Table~\ref{tab:summary}.
\textit{Supervision}: source of the supervisory signal during operation
(passive labels, active queries or rewards, or none) and the labels required for
deployment (full: fully labeled dataset; weak: weakly labeled or reference
histograms only).
\textit{Online adaptation}: whether the system updates its model during
operation without full retraining, and whether formal guarantees are provided.
\textit{Validation}: data on which the method was evaluated, and the experiment
or subsystem.}
\label{tab:dqm}
\small
\setlength{\tabcolsep}{4pt}
\renewcommand{\arraystretch}{1.3}
\begin{tabular}{@{}p{2.8cm} p{2.8cm} p{3.4cm} p{2.4cm} p{3.6cm}@{}}
\toprule
\textbf{System} &
\textbf{Method} &
\textbf{Supervision} &
\textbf{Online adaptation} &
\textbf{Validation} \\
\midrule

Threshold / $\chi^2$ / KS tests
  & Rule-based
  & None
  & No
  & Real data; all LHC experiments \\

Pol et al.~\cite{Pol2019DQM}
  & Supervised $+$ autoencoder
  & Passive; full labels (known faults)
  & No
  & Real data; CMS muon detectors \\

AutoDQM~\cite{AutoDQM2025CMS}
  & Unsupervised autoencoder
  & None
  & No
  & Real data; CMS Run~3 \\

Asres et al.~\cite{Asres2025TL}
  & Transfer-learning autoencoder
  & Passive; weak labels (source domain)
  & No
  & Real data; CMS HCAL \\

MEDIC~\cite{Bassa2026MEDIC}
  & Supervised neural network, simulation-driven
  & Passive; full labels (simulated)
  & No
  & Simulation; generic collider detector \\

Bower~\cite{Bower2020ATSB}
  & Adversarial semi-bandit
  & Active; per-channel reward, no labels
  & Yes, with regret bound
  & Real data; ALICE TRD \\

Jullian Parra et al.~\cite{jullian2025human}
  & Multi-agent PPO
  & Active; weak labels from online expert queries
  & Yes
  & Synthetic data; offline LHCb test \\

\bottomrule
\end{tabular}
\end{table}

% -----------------------------------------------------------------------------
\subsection{Open Problems and Future Research Directions}
\label{sec:dqm:future}

\paragraph{Active learning and uncertainty quantification.}
Current systems query experts reactively, based on the uncertainty of the agent
about the current histogram.
Combining RL with principled uncertainty quantification, using calibrated
predictive uncertainty~(e.g., conformal prediction~\cite{Vovk2005Conformal} or
Bayesian deep learning~\cite{Gal2016BayesianDL}) to identify histograms where annotation is most
valuable, would improve sample efficiency further.
A natural extension is to model the attention budget of the expert explicitly: for example, the
agent should not only ask ``is this histogram anomalous?'' but also ``is this
the right moment to ask, given that the shifter has already been queried five
times in the last hour?''

Two families of RL methods are particularly well-suited to this setting.
The first is \textbf{distributional RL}~(Section~\ref{sec:rl:valuebased}).
The spread of its return distribution reflects the intrinsic~(aleatoric)
variability of the outcome of a certification decision, not the epistemic
uncertainty of the agent, which requires ensembles or Bayesian
methods~(below).
In the DQM context, a wide return distribution for a given histogram signals an
ambiguous case, making it a natural trigger for an expert query; conversely,
narrow distributions can be acted upon autonomously without annotation.
This connects directly to the risk-sensitive behavior enabled by distributional
RL: the agent can be configured to be systematically conservative~(preferring
to query when uncertain) by targeting a low quantile of the return distribution,
rather than its mean, for decision-making.
\textbf{Bayesian RL}~(Section~\ref{sec:rl:paradigms}) is complementary: in the
DQM setting, the posterior becomes a posterior over detector malfunction rates
and anomaly distributions that is updated as expert labels arrive;
information-theoretic criteria, such as the expected information gain, then
guide the query decisions of the agent, selecting the histogram that would most
reduce uncertainty about the current detector state.
Both approaches address a key limitation of the existing systems: the queries of the discussed
PPO-based DQM agent~\cite{jullian2025human} are based on a point estimate
of its own prediction confidence, which does not quantify how much new information
each query would provide relative to the total annotation budget of the expert.

\paragraph{Real-world deployment and validation on experiment data.}
The PPO-based DQM system has been validated mainly on synthetic data, with only a
preliminary offline test on real LHCb data~\cite{jullian2025human}.
The critical next step is systematic validation on real experiment data: real
detector
histograms exhibit long-tailed failure distributions, correlated multi-subsystem
anomalies, and label noise patterns that synthetic datasets cannot fully capture.
Deployment in a shadow mode, where the RL agent runs alongside the existing DQM
system without affecting decisions, would allow controlled comparison and
identification of failure modes before operational adoption.

Four RL sub-fields offer complementary tools for bridging this gap.
\textbf{Offline RL}~(Section~\ref{sec:rl:offline}) is the most directly
applicable: years of archived DQM decisions and expert certifications constitute
a large fixed dataset from which a policy can be learned without any new
interaction with the real experiment.
Conservative methods such as CQL are critical here, because the behavioral
policy, i.e., the human shifter, covers only a subset of possible detector
conditions.
\textbf{Inverse RL}~(Section~\ref{sec:rl:irl}) complements offline RL by
inferring from certification histories the implicit reward that experienced
shifters have been optimizing, rather than requiring an explicit reward to be
hand-designed.
This is valuable because expert DQM judgment encodes subtle heuristics~(e.g.,
tolerating minor calorimeter noise during early fills but flagging it at
peak luminosity) that are difficult to articulate as a scalar reward but are
readily recoverable from demonstration data via maximum entropy
IRL~\cite{Ziebart2008MaxEnt}.
\textbf{Meta-RL}~(Section~\ref{sec:rl:paradigms}) addresses the rapid
adaptation problem:
detector conditions shift between LHC runs, fills, and years, so a policy
trained on Run~2 data may underperform on Run~3 without retraining.
A meta-learned policy that treats each run period as a separate task can adapt
to new conditions from a handful of expert labels, rather than requiring a full
offline training campaign each time.
\textbf{Model-based RL}~(Section~\ref{sec:rl:modelbased}) is useful in the shadow-mode
deployment phase: a learned model of the anomaly dynamics of the detector can
generate synthetic failure scenarios from the small number of real anomalies
observed in shadow operation, augmenting the scarce real data and accelerating
policy improvement before full deployment.

\paragraph{Multi-subsystem and multi-experiment DQM.}
Each RL-based DQM application reviewed here targets a single detector subsystem.
Building a shared RL policy across multiple subsystems, or even multiple
experiments, via transfer learning or multi-task RL would reduce the engineering
overhead for new deployments and exploit cross-subsystem correlations.
For example, a fault in a shared clock or trigger system can appear
simultaneously in several subsystems; a shared policy could detect such
correlated signatures faster than independent single-subsystem agents.

\textbf{Multi-agent RL}~(Section~\ref{sec:rl:marl}) is the most natural
framework for multi-subsystem DQM: each subsystem is assigned a dedicated agent
that observes its local histograms and issues local certification decisions,
while under centralized training with decentralized execution the critic has
access to all subsystem observations simultaneously.
This allows the joint policy to capture cross-subsystem correlations, such as
the correlated signatures above, without requiring any single
agent to process the full set of experiment histograms at inference time.
Competitive dynamics between agents can further be used to model the resource
allocation problem: when the total annotation budget is limited, agents must
implicitly negotiate the uncertain histograms of which subsystem are most valuable
to present to the shifter.

\textbf{Transfer learning and multi-task RL}~(Section~\ref{sec:rl:paradigms}) enable policies trained on one
subsystem or experiment to be reused elsewhere with minimal additional data.
A policy pre-trained on CMS tracker histograms shares structural similarities
with one needed for ATLAS inner detector monitoring: both involve spatial hit
occupancy maps, timing distributions, and noise rate histograms, differing
mainly in granularity and geometry.
Multi-task RL with a shared encoder and task-specific heads would allow a single
model to serve multiple subsystems simultaneously, while transfer learning via
fine-tuning on a small number of labeled histograms from the new subsystem
would dramatically reduce the annotation cost of deploying DQM RL in a new
experimental context.

\paragraph{Concept drift and long-run adaptation.}
Detector conditions change on timescales of years due to radiation damage,
hardware replacements, and operating condition changes.
\textbf{Continual RL}~(Section~\ref{sec:rl:paradigms}) is a natural fit for the
long-horizon DQM problem: as the detector evolves, the agent must retain
knowledge of historical failure modes while incorporating new ones, without
resetting its policy from scratch at each run period.
Standard replay buffers, elastic weight consolidation~\cite{Kirkpatrick2017EWC},
and progressive neural network architectures~\cite{Rusu2016Progressive} are candidate mechanisms for
preserving past knowledge while permitting new learning.
A recent perspective on deployable continual learning for particle
accelerators~\cite{Rajput2025ContinualAccel} identifies the accelerator
applications in which continual learning could maintain model performance under
distribution drift; an analogous framework for
detector DQM would need to handle the additional complexity of non-stationary
anomaly distributions and evolving expert labeling criteria over multi-year run
periods.

%% file: src/4/dqm_hitl.tex
% Human-in-the-loop and bandit formulations of DQM (Section 4).
% Requires: \usepackage{tikz}
%           \usetikzlibrary{arrows.meta, positioning, fit, calc, backgrounds}
\begin{tikzpicture}[
    font=\small,
    >={Stealth[length=2mm]},
    block/.style={draw, rounded corners=2pt, align=center, minimum height=1.15cm,
                  inner sep=3pt},
    stream/.style={block, fill=black!6, minimum width=3.0cm, minimum height=4.6cm},
    agent/.style={block, fill=violet!10, minimum width=2.9cm},
    human/.style={block, fill=orange!15, minimum width=2.6cm},
    tier/.style={draw=black!45, dashed, rounded corners=6pt, inner sep=8pt},
    tierlabel/.style={anchor=north west, font=\small\bfseries, inner sep=3pt},
    loop/.style={->, thick},
    flow/.style={->, thick, dashed},
    lbl/.style={font=\footnotesize, inner sep=2pt},
]

% --- Nodes ---------------------------------------------------------------
\node[stream] (stream) at (1.5, -1.6) {\textbf{Detector stream}\\[2pt]histograms,\\channel currents,\\drifting\\conditions};

\node[agent] (pred)  at (5.3, 0)    {\textbf{Predictor}\\good or bad flag};
\node[agent] (check) at (8.9, 0)    {\textbf{Checker}\\query or not};
\node[human] (shift) at (13.0, 0)   {\textbf{Shifter}\\labels};

\node[agent, minimum width=6.5cm] (bandit) at (7.1, -3.2)
    {\textbf{Semi-bandit agent}\\select and query channels,\\feedback above threshold $\tau$};
\node[human] (oper)  at (13.0, -3.2) {\textbf{Operator}\\preventive action};

% --- Frames --------------------------------------------------------------
\begin{scope}[on background layer]
    \node[tier, fit=(pred)(check), inner ysep=12pt, yshift=4pt] (Tp) {};
    \node[tier, fit=(bandit), inner ysep=12pt, yshift=4pt] (Tb) {};
\end{scope}
\node[tierlabel] at (Tp.north west) {Multi-agent PPO};
\node[tierlabel] at (Tb.north west) {Adversarial semi-bandit};

% --- Decision flow (solid) -----------------------------------------------
\draw[loop] (stream.east |- pred) -- (pred);
\draw[loop] (pred) -- (check);
\draw[loop] (check) -- node[lbl, above] {query} (shift);
\draw[loop] (stream.east |- bandit) -- (bandit);
\draw[loop] (bandit) -- node[lbl, above] {alarm} (oper);

% --- Human feedback (dashed) ---------------------------------------------
\draw[flow] (shift.south) -- ++(0, -0.95) -| node[lbl, pos=0.25, below] {label} (pred.south);

% --- Legend --------------------------------------------------------------
\draw[loop] ($(stream.south west) + (0.0, -0.6)$) -- ++(0.8, 0)
    node[right, lbl] {decision flow};
\draw[flow] ($(stream.south west) + (3.6, -0.6)$) -- ++(0.8, 0)
    node[right, lbl] {human feedback as learning signal};
\end{tikzpicture}

%% file: sections/5_jets.tex
% =============================================================================
%  5. Jet Physics
% =============================================================================

% -----------------------------------------------------------------------------
\subsection{Background: The Problem, Traditional Methods, and the Case for RL}
\label{sec:jets:background}

Jets are collimated sprays of hadrons produced by the fragmentation of quarks and
gluons created in high-energy collisions.
They are the most abundant objects in LHC events and are central to a vast range
of physics analyses, from precision quantum chromodynamics~(QCD) measurements to
searches for new physics.
Working with jets requires a series of algorithmic decisions: how to cluster
detector deposits into a jet, how to remove contaminating soft radiation
(\textit{grooming}), and how to order or represent jets and their constituents
as inputs to downstream ML tasks.
Each of these is a discrete or sequential decision problem, making jet physics
a natural domain for RL.

\paragraph{Traditional approaches.}
Standard jet clustering algorithms such as anti-$k_T$~\cite{Cacciari2008AntiKt},
$k_T$~\cite{Catani1993kT}, and Cambridge\slash Aachen~(C/A)~\cite{Dokshitzer1997CA,
Wobisch1999CA} define a distance metric between
pseudo-particles and merge the closest pair iteratively until all pairs exceed a
jet radius threshold $R$.
The anti-$k_T$ algorithm is the standard jet reconstruction algorithm at the LHC: its clustering sequence produces geometrically
regular, cone-like jets that are straightforward to calibrate, but its merging
criterion is entirely fixed and insensitive to the internal structure or physics
goal of the analysis.
Grooming algorithms such as Soft Drop~\cite{Larkoski2014SoftDrop} post-process
the clustered jet by recursively undoing the groupings in the C/A tree and removing the
softer branch whenever the Soft Drop condition
$z > z_{\mathrm{cut}} (\Delta R_{ab} / R)^\beta$ is not satisfied, where $z$ is
the momentum fraction and $\Delta R_{ab}$ is the angular separation between the
two branches.
Soft Drop is IRC-safe by construction and widely used for jet mass measurements
and boosted object tagging, but its parameters $z_{\mathrm{cut}}$ and $\beta$ are
hand-tuned and constant across all jets, regardless of their kinematic
configuration.
Input ordering for sequence-based classifiers~(LSTM, transformer~\cite{Vaswani2017Attention}) that take
lists of jets or jet constituents is typically chosen by $p_T$ ordering or
angular ordering, without optimization for any specific downstream task.

\paragraph{Why RL?}
All three problems (clustering, grooming, and ordering) share the structure of
a sequential decision process in which the objective~(e.g., mass resolution,
tagging performance, or IRC safety) depends on the full sequence of decisions,
not just any individual step.
This delayed, holistic reward structure is precisely what RL is designed to
optimize.
RL offers two concrete advantages over hand-crafted algorithms.
First, a learned policy can adapt its decisions to the specific phase-space
configuration of each jet: a grooming agent can be more aggressive on jets with
large soft contamination and more conservative on others, whereas Soft Drop
applies the same criterion uniformly.
Second, RL can optimize directly against a physics observable, such as a
discriminant score, a mass resolution, or a likelihood, that has no closed-form
gradient with respect to the algorithmic decisions, making standard
backpropagation inapplicable.

\begin{figure}[t]
    \centering
    \input{src/5/construction.tex}
    \caption{Emergent construction pattern in RL for HEP: an agent builds a
    physics object edit by edit, as in several of the applications of
    Sections~\ref{sec:jets} to~\ref{sec:analysis}.
    The agent, a policy followed by a feasibility mask, applies an edit $a$ to
    the partial object $s$; the edited object becomes the next state $s'$ and is
    scored by an evaluator, whose reward trains the policy.
    An optional search procedure or prior guides the policy, and the complete
    object is returned as output.
    Subtitles give examples from the reviewed applications.}
    \label{fig:construction}
\end{figure}
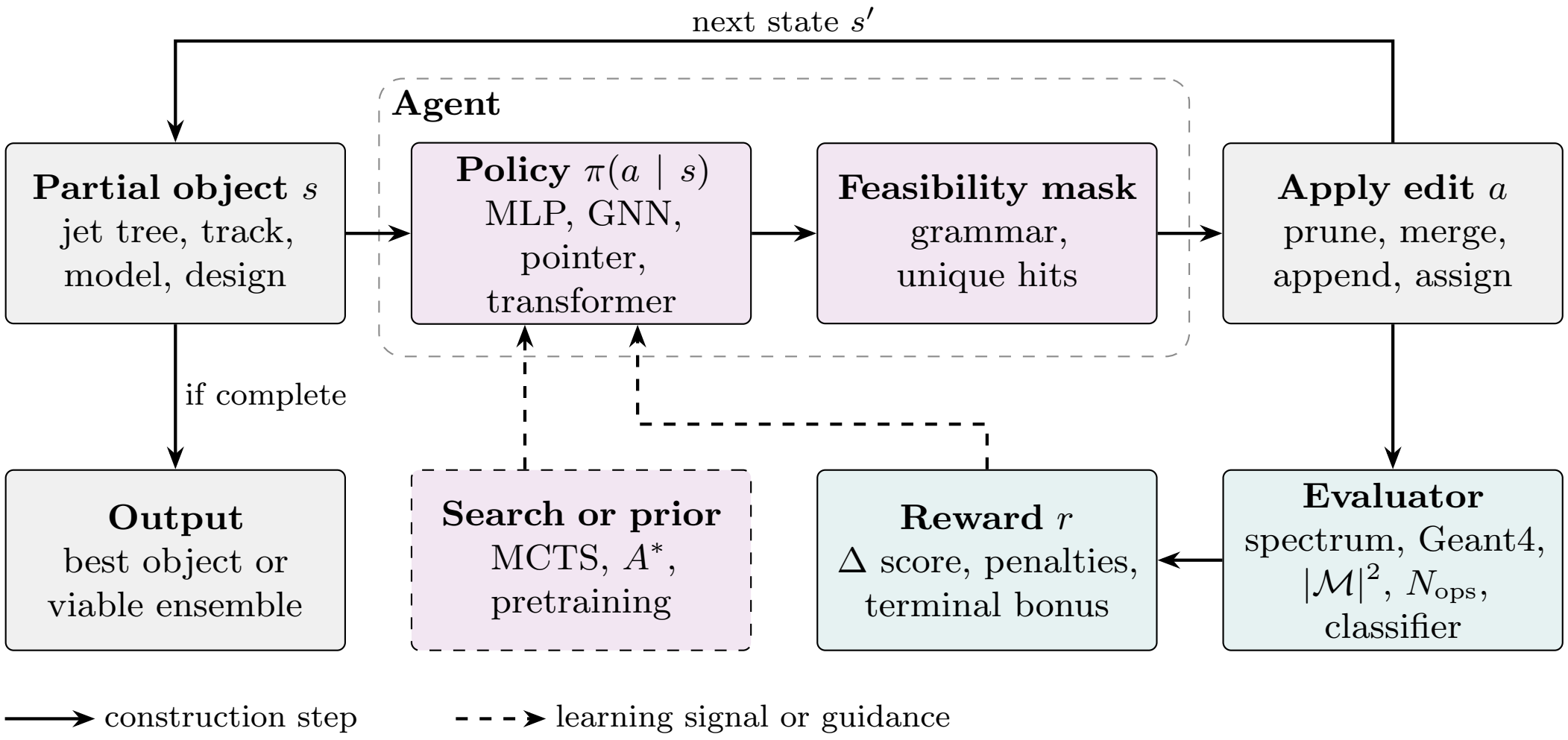

These problems are instances of a pattern that recurs across several of the
applications reviewed in this survey: the agent builds or refines a partial
object one edit at a time, a feasibility mask restricts the admissible edits,
and an evaluator scores the result~(Figure~\ref{fig:construction}).
The partial object can be a jet clustering tree or an input
ordering~(this section), a track candidate on a hit
graph~(Section~\ref{sec:tracking}), a BSM model encoded as a charge matrix, a
field graph, or a token sequence~(Section~\ref{sec:bsm}), a detector
configuration~(Section~\ref{sec:detectors}), or an assignment of reconstructed
objects, a system of integral identities, or a symbolic
expression~(Section~\ref{sec:analysis}).
The domains differ mainly in the evaluator, which ranges from analytic physics
models to full detector simulation and learned classifiers.

% -----------------------------------------------------------------------------
\subsection{Jet Grooming}
\label{sec:jets:grooming}

\textbf{Carrazza and Dreyer}~\cite{Carrazza2019GroomRL} reformulated jet
grooming as a sequential decision process and trained a DQN agent to make
step-by-step separate-and-remove decisions on the C/A clustering tree.
A jet is represented as a binary tree in which each internal node contains
kinematic information about the two child branches.
The children $a$ and $b$ at each node are ordered by transverse momentum such
that $p_{T,a} > p_{T,b}$.
The state at each node is:
\begin{equation}
    s = \bigl(z,\; \Delta_{ab},\; \psi,\; m,\; k_t \;\big|\; a, b\bigr),
    \label{eq:jet_state}
\end{equation}
where $z = p_{T,b}/(p_{T,a} + p_{T,b})$ is the momentum fraction of the softer
branch, $\Delta_{ab}$ is the angular distance between the two branches, $\psi$ is
the azimuthal angle, $m$ is the jet mass, and $k_t = p_{T,b}\,\Delta_{ab}$ is
the transverse momentum of $b$ relative to $a$.
The binary action of the agent at each node is to keep or remove the softer branch $b$.
The reward function comprises two terms.
The first, $R_M$, compares the current jet mass to a target mass $m_{\rm target}$
using a Cauchy-shaped profile:
\begin{equation}
    R_M = \frac{\Gamma^2}{\pi\bigl(|m - m_{\rm target}|^2 + \Gamma^2\bigr)},
    \label{eq:jet_rm}
\end{equation}
peaking at $m = m_{\rm target}$ with width $\Gamma$.
For background events where no reference mass exists, this term is replaced by:
\begin{equation}
    R_M^{\rm bkg} = \frac{m}{\Gamma_{\rm bkg}}\exp\!\left(-\frac{m}{\Gamma_{\rm bkg}}\right).
    \label{eq:jet_rm_bkg}
\end{equation}
The full reward at each node combines $R_M$ with a term that rewards removal
of soft wide-angle radiation and penalizes removal of hard-collinear emissions:
\begin{equation}
    R(m, a_t, \Delta, z) = R_M +
    \frac{a_t\min\!\bigl(1,\,e^{g(\alpha_1,\beta_1,z_1)}\bigr)
    + (1-a_t)\max\!\bigl(0,\,1 - e^{g(\alpha_2,\beta_2,z_2)}\bigr)}{N_{\rm SD}},
    \label{eq:jet_reward}
\end{equation}
where $g(\alpha_i, \beta_i, z_i) = -\alpha_i \ln(1/\Delta) + \beta_i \ln(z_i/z)$
and $\Gamma, N_{\rm SD}, \alpha_i, \beta_i, z_i$ are hyperparameters.
Unlike Soft Drop, the learned groomer carries no IRC-safety guarantee.
A grooming procedure is IRC-safe if its output is insensitive to the addition of
arbitrarily soft or collinear emissions; violations of IRC safety make the
associated cross-sections ill-defined in perturbation theory.
In GroomRL, this requirement enters only indirectly, through the reward term
that favors the removal of soft wide-angle radiation: no constraint is placed on
the action space, and the theoretical calculability of the learned groomer is
discussed only qualitatively.
The resulting agent performs favorably against Recursive Soft Drop~\cite{Dreyer2018RecursiveSD} in jet mass
reconstruction, reduces background contamination qualitatively, and shows a
moderate sensitivity to hadronization, while adapting its aggressiveness to the
kinematic configuration of each jet rather than applying a fixed criterion
uniformly.
The trained policy is released as the open-source \textbf{GroomRL}
framework~\cite{Carrazza2019LibGroomRL}.

% -----------------------------------------------------------------------------
\subsection{Jet Clustering and Hierarchical Reconstruction}
\label{sec:jets:clustering}

\textbf{Brehmer, Macaluso, Pappadopulo, and Cranmer}~\cite{Brehmer2021MCTS}
applied Monte Carlo Tree Search (MCTS) \cite{Coulom2006MCTS} with a learned neural policy and value function
to the problem of hierarchical jet clustering.
The goal is to reconstruct the most plausible binary tree of parton splittings
that explains the observed final-state momenta.
The MDP is formulated as follows.
The state at time $t$ is the set of particles not yet merged:
$s_t = \{p_{t,1}, \ldots, p_{t,n_t}\}$, where the initial state $s_N$ contains
all observed final-state momenta and the episode terminates when a single
particle remains.
The action is the choice of which pair $(i, j)$ to merge, replacing them with
a parent $p_{t-1,i} = p_{t,i} + p_{t,j}$.
Each transition reduces the particle count by one, so the episode has exactly
$N-1$ steps for $N$ initial particles.
The reward at each step is the log splitting probability under a parton shower
model: $r(s_t, a_t, s_{t-1}) = \log p_s(s_t \mid s_{t-1}, a_t)$, making the
cumulative return the log-likelihood of the full clustering tree.

Since the transition model is known analytically, model-based planning via MCTS
is applied, playing the role of the search component in
Figure~\ref{fig:construction}.
At each state, the agent selects the merge action that maximizes the following
PUCT criterion~(predictor plus upper confidence bounds applied to
trees)~\cite{Rosin2011PUCT, Silver2016Go}:
\begin{equation}
    U(s, a) = Q(s, a) + c\,\pi(a \mid s)\,
    \frac{\sqrt{N(s)}}{1 + N(s, a)},
    \label{eq:puct}
\end{equation}
where $Q(s, a)$ is the mean normalized reward, $N(s)$ and $N(s, a)$ are the
state and action visitation counts, $\pi(a \mid s)$ is the neural network prior
policy, and $c$ balances exploration and exploitation.
As in AlphaZero~\cite{Silver2018AlphaZero}, the policy network is trained to agree with the MCTS decisions
and in turn guides the search, which is still run at inference time; a policy
trained by behavioral cloning alone performs much worse.
The MCTS agent generates clustering trees of higher log-likelihood than both
greedy and beam search baselines, with the advantage growing with the number of
final-state particles.

A direct successor to the MCTS approach is the work of
\textbf{Greenberg, Macaluso, Monath et al.}~\cite{Greenberg2021Astar}, who
replace MCTS with $A^*$ search~\cite{Hart1968Astar} over a novel cluster trellis data structure.
The $A^*$ algorithm is best-first search guided by $f(n) = g(n) + h(n)$, where
$g(n)$ is the cost of the clustering history up to state $n$ and $h(n)$ is an
admissible heuristic upper-bounding the remaining log-likelihood.
The key contribution is a compact trellis encoding of the clustering state space
that reduces the otherwise super-exponential search space to at most exponential
in the number of jet constituents, making exact maximum a posteriori~(MAP)
clustering tractable for jets
with up to $\sim$15 constituents and approximate clustering competitive for
datasets exceeding $10^{1000}$ possible trees.
On the Ginkgo jet simulator benchmark~\cite{Cranmer2019Ginkgo}, $A^*$ improves
on MCTS, which the authors describe as a strong baseline only for small jets
where it is feasible to run, and the approximate variant remains applicable to
much larger jets.
This work makes $A^*$ the stronger maximum-likelihood jet clustering algorithm
among those reviewed, and connects the jet clustering problem to the broader
literature on hierarchical clustering in computer science.

% -----------------------------------------------------------------------------
\subsection{Jet Input Ordering and Representation}
\label{sec:jets:ordering}

The performance of classifiers that operate on ordered lists of physics
objects, whether the jets of an event or the constituents of a jet, depends on
the order in which the objects are presented, but the optimal ordering is not
known a priori and varies by downstream task.
Standard choices~($p_T$ ordering, angular ordering, anti-$k_T$ sequence) are
physically motivated but not optimal for any specific classifier.

\textbf{Erdmann, Fischer, and Noll}~\cite{Erdmann2020ReinforcedSorting} frame
the ordering of the jets in an event as an RL problem and design a two-stage
system: a \textit{sorting network}~(the agent) and an \textit{analysis
network}~(the environment), which in the terms of Figure~\ref{fig:construction}
act as the policy and the evaluator, respectively.
The case study is the separation of $t\bar{t}H$~($H \to b\bar{b}$) from
$t\bar{t}+b\bar{b}$ events, whose final state contains six jets, a charged
lepton, and a neutrino; the analysis network is a Lorentz Boost Network~\cite{Erdmann2019LBN}, whose
performance depends on the order of its inputs.
The sorting network receives the $p_T$-sorted event and assigns a position index
to each jet, which specifies how the jets are reordered before being passed,
together with the lepton and the neutrino, to the analysis network.
Training alternates between the two networks: given a current ordering $S_i(e)$,
all pairwise swaps of jets~(plus the identity) are evaluated on the analysis
network, the top-ranking permutation is used as the learning target for the
sorting network, and the analysis network is then retrained with the improved
ordering.
Both networks are first pretrained, the sorting network to reproduce the $p_T$
ordering, to avoid premature convergence before switching to RL-driven
optimization.
During training, the classification accuracy rises from that obtained with
$p_T$-sorted jets toward the upper benchmark obtained with an ordering based on
generator information.

This paper illustrates a general and transferable principle: RL can optimize
discrete preprocessing decisions that have no closed-form gradient with respect
to the downstream loss, simply by using the downstream performance to select the
training targets of the ordering policy.
The same principle applies to any sequential representation choice in particle
physics such as constituent selection, variable ordering, and feature masking, 
where the optimal choice depends on the downstream task.

% -----------------------------------------------------------------------------
\subsection{Open Problems and Future Research Directions}
\label{sec:jets:future}

\paragraph{End-to-end joint optimization.}
Current work optimizes grooming, clustering, and ordering independently.
A unified RL policy trained jointly on all three decisions and coupled end-to-end
to a downstream physics analysis could discover synergies between them that
sequential optimization misses.
The main challenge is defining a reward that reflects the full analysis goal~(e.g.,
signal significance or exclusion limit) without intermediate surrogates, as the
gradient of such an objective through discrete decisions is not available.

This setting is naturally framed as a multi-objective RL
problem~(Section~\ref{sec:rl:paradigms}): grooming quality~(mass resolution),
clustering quality~(parton shower log-likelihood), ordering quality~(downstream
classifier performance), and IRC safety are four distinct objectives that cannot
in general be simultaneously maximized, and a scalarized reward would fix an
arbitrary priority among them.
A preference-conditioned policy~\cite{Yang2019MOPolicy} would instead let the
physicist select the operating point at inference time, for example preferring
mass resolution for a boosted $W$ search and tagging performance for a top
tagger, with one trained policy shared across analyses.

\paragraph{Graph-based RL for variable-multiplicity jets.}
Jet constituent lists have variable length and permutation symmetry, making them
natural candidates for graph neural network~(GNN) representations \cite{battaglia2018relational}.
The implications of adopting a GNN policy backbone go beyond a better state
encoder and affect the RL algorithm itself in three ways.
First, a permutation-equivariant GNN policy~\cite{Munikoti2024GraphRL} constrains
the policy class to respect the physical symmetry of the jet: the action
distribution is invariant to any reordering of the constituents, which tightens
sample complexity and prevents the agent from exploiting arbitrary input orderings.
Second, each message-passing layer propagates information across the constituent
graph before emitting an action, providing a form of implicit look-ahead over the
$k$-hop neighborhood that is baked into the forward pass rather than achieved by
explicit tree search.
Third, when actions are defined as edge additions or removals on the constituent
graph, the action space changes at every step as merges are applied, requiring
modifications to standard policy gradient estimators: the policy must be
evaluated over the current edge set via a masked softmax or auto-regressive edge
sampling scheme, and advantage estimation must account for credit flowing through
both the temporal and graph-topological dimensions simultaneously.

\paragraph{Foundation models for jet physics.}
The emergence of cross-task foundation models for jet physics, such as
OmniJet-$\alpha$~\cite{Birk2024OmniJet}, raises the question of whether an RL
policy pretrained on diverse jet samples could generalize across grooming,
clustering, and ordering tasks simultaneously.
A pretrained jet representation used as a shared state encoder across all three
RL problems would reduce the engineering overhead of per-task policy design, but
the benefits extend further than architectural convenience.

First, in a \textit{transfer learning} scenario, the foundation model backbone
provides a rich prior over jet kinematics that the RL policy can exploit from
the first training step rather than learning from scratch.
Because the encoder already captures the relevant structure of the jet
constituent space, the policy head only needs to learn the task-specific
decision logic on top of a fixed or lightly fine-tuned representation.
This translates directly into higher \textit{sample efficiency}: fewer
environment interactions are needed to reach a given level of performance,
which is practically significant given that training RL policies on realistic
jet samples from full Geant4~\cite{Agostinelli2003GEANT4} simulation is computationally expensive.

Second, foundation model backbones offer a principled path toward
\textit{sim-to-real transfer} for jet RL policies.
A common failure mode of simulation-trained policies is that the state
representation is sensitive to low-level features that differ between
simulation and real detector data: calorimeter granularity, noise thresholds,
or particle flow calibrations.
A foundation model pretrained on both simulated and real jets, or fine-tuned
with domain adaptation techniques on real data, would produce a state encoder
whose representation is robust to these low-level distributional shifts.
The RL policy operating on top of this encoder would then inherit that
robustness, requiring only a lightweight fine-tuning step on real data rather
than a full retraining cycle.
This two-stage approach, pretraining a general encoder followed by RL
fine-tuning of the decision head, mirrors the paradigm that has proven
effective in natural language processing~(pretrained language models fine-tuned with RL from human
feedback~(RLHF)~\cite{Ouyang2022RLHF}) and, with supervised rather than RL
fine-tuning, in robotics~(self-supervised sensorimotor pretraining followed by
task-specific fine-tuning~\cite{Radosavovic2023RPT}), and represents a promising
but as yet unexplored direction for jet physics.

\paragraph{IRC safety as a hard constraint.}
None of the RL jet algorithms reviewed here enforces IRC safety: the grooming
policy of~\cite{Carrazza2019GroomRL} addresses it only through the reward.
A general treatment of IRC safety as a constrained RL problem, applicable to
clustering, grooming, and ordering, has not yet been developed.
Connecting constrained RL theory~(Section~\ref{sec:rl:safety}) to the formal definitions
of IRC safety~\cite{Sterman1977IRC} in perturbative QCD~\cite{Ellis1996QCD} is
an open theoretical problem with direct implications for the use of learned jet
algorithms in precision measurements.

The safe RL literature offers two complementary tools for this problem.
The first is Lyapunov-based safe model-based RL~\cite{Berkenkamp2017SafeMBRL},
which maintains a certified safe region of the state space and proves that
data collection can expand this region without ever leaving it.
Translated to jet grooming, the constraint cannot be a property of a single
partial sequence, since IRC safety is a property of the algorithm as a whole; it
would instead be expressed as conditions on the decision rule, for example that
the removal decision be insensitive to the addition of soft or collinear
branches, and the agent would be certified to respect them at every
de-clustering step.
The second is model predictive control~(MPC) for safe exploration~\cite{Koller2018SafeMPC},
which uses a learned GP model of the dynamics to construct confidence intervals
over future trajectories and guarantees that the trajectory satisfies safety
constraints at every step via a terminal set constraint.
In the jet context, the dynamics model would predict the kinematic consequences
of a grooming or clustering decision, and the MPC layer would reject actions
whose predicted trajectories violate the IRC safety condition before they are
applied.

Both approaches also highlight a broader benefit of model-based RL
for IRC-safe jet algorithms.
A model-free agent with a per-step action mask can only exclude inadmissible
actions at the current step; it cannot reason about whether a currently
admissible action leads to a state from which no admissible completion exists.
A model-based agent with look-ahead can verify, before committing to a merge or
removal, that at least one admissible continuation exists from the resulting
state, analogously to how MPC guarantees the existence of a feasible return
trajectory at every timestep~\cite{Koller2018SafeMPC}.
This look-ahead capability would be especially valuable in clustering, where the
properties of the final tree depend on the full sequence of merges and a single
early decision can preclude admissible completions.
Planning with MCTS or $A^*$~\cite{Greenberg2021Astar} already provides this
look-ahead property implicitly; embedding IRC safety as an explicit constraint
into the tree search is a natural and well-motivated extension.

%% file: src/5/construction.tex
% Emergent construction pattern in RL for HEP (Section 5, referenced in Sections 6-9).
% Requires: \usepackage{tikz}
%           \usetikzlibrary{arrows.meta, positioning, fit, calc, backgrounds}
\begin{tikzpicture}[
    font=\small,
    >={Stealth[length=2mm]},
    block/.style={draw, rounded corners=2pt, align=center, minimum height=1.6cm,
                  text width=2.8cm, inner sep=3pt},
    object/.style={block, fill=black!6},
    agent/.style={block, fill=violet!10},
    score/.style={block, fill=teal!10},
    tier/.style={draw=black!45, dashed, rounded corners=6pt, inner sep=8pt},
    tierlabel/.style={anchor=north west, font=\small\bfseries, inner sep=3pt},
    loop/.style={->, thick},
    flow/.style={->, thick, dashed},
    lbl/.style={font=\footnotesize, inner sep=2pt},
]

% --- Construction step (top row) -----------------------------------------
\node[object] (obj)  at (1.6, 0)  {\textbf{Partial object} $s$\\jet tree, track,\\model, design};
\node[agent]  (pol)  at (5.2, 0)  {\textbf{Policy} $\pi(a \mid s)$\\MLP, GNN, pointer,\\transformer};
\node[agent]  (mask) at (8.8, 0)  {\textbf{Feasibility mask}\\grammar,\\unique hits};
\node[object] (edit) at (12.4, 0) {\textbf{Apply edit} $a$\\prune, merge,\\append, assign};

% --- Output, guidance, and scoring (bottom row) --------------------------
\node[object] (out)   at (1.6, -2.9)  {\textbf{Output}\\best object or\\viable ensemble};
\node[agent, dashed] (prior) at (5.2, -2.9) {\textbf{Search or prior}\\MCTS, $A^*$,\\pretraining};
\node[score]  (rew)   at (8.8, -2.9)  {\textbf{Reward} $r$\\$\Delta$ score, penalties,\\terminal bonus};
\node[score]  (eval)  at (12.4, -2.9) {\textbf{Evaluator}\\spectrum, Geant4,\\$|\mathcal{M}|^2$, $N_{\rm ops}$, classifier};

% --- Agent frame ---------------------------------------------------------
\begin{scope}[on background layer]
    \node[tier, fit=(pol)(mask), inner ysep=12pt, yshift=4pt] (Ta) {};
\end{scope}
\node[tierlabel] at (Ta.north west) {Agent};

% --- Construction step (solid) -------------------------------------------
\draw[loop] (obj) -- (pol);
\draw[loop] (pol) -- (mask);
\draw[loop] (mask) -- (edit);
\draw[loop] (edit.north) -- ++(0, 0.9) -| node[lbl, pos=0.25, above] {next state $s'$} (obj.north);
\draw[loop] (edit) -- (eval);
\draw[loop] (eval) -- (rew);
\draw[loop] (obj) -- node[lbl, right] {if complete} (out);

% --- Learning signal and guidance (dashed) -------------------------------
\draw[flow] (rew.north) -- ++(0, 0.4) -| ([xshift=5mm]pol.south);
\draw[flow] ([xshift=-5mm]prior.north) -- ([xshift=-5mm]pol.south);

% --- Legend --------------------------------------------------------------
\draw[loop] ($(out.south west) + (0.0, -0.6)$) -- ++(0.8, 0)
    node[right, lbl] {construction step};
\draw[flow] ($(out.south west) + (4.0, -0.6)$) -- ++(0.8, 0)
    node[right, lbl] {learning signal or guidance};
\end{tikzpicture}

%% file: sections/6_tracking.tex
% =============================================================================
%  6. Charged Particle Tracking
% =============================================================================

% -----------------------------------------------------------------------------
\subsection{Background: The Problem, Traditional Methods, and the Case for RL}
\label{sec:tracking:background}

Charged particle tracking is the reconstruction of particle trajectories from
discrete position measurements~(\textit{hits}) left in a tracking detector.
A tracker consists of multiple layers of detector cells; when a charged particle
crosses a cell, it deposits charge in one or more adjacent pixels, producing a
hit whose centroid estimates the true particle position.
In a typical LHC detector, each bunch crossing contains tens of overlapping
soft-collision vertices~(\textit{pileup}) on top of the hard interaction of
interest, resulting in tens of thousands of hits across of order ten detector
layers.
Associating the correct hits to each particle trajectory is a combinatorial
assignment problem whose complexity grows super-linearly with pileup multiplicity.
At the HL-LHC, with up to 200 pileup interactions per bunch crossing, thousands
of charged particles, and of order $10^5$ space-points per event, tracking will
be one of the dominant computational
bottlenecks in offline reconstruction~\cite{Ju2021ExaTrkX}.

\paragraph{Traditional approaches.}
Tracking traditionally involves four sequential steps: seed generation, track
finding, track fitting, and track selection.
\textit{Seed generation} constructs an initial estimate of the five helix
parameters $(q/p,\, \lambda,\, \phi,\, x_\perp,\, y_\perp)$ by connecting
hits in adjacent detector layers.
\textit{Track finding} extends each seed through subsequent layers, branching
when more than one compatible hit is found.
\textit{Track fitting} uses the Kalman filter~\cite{Fruhwirth1987Kalman}
to estimate the particle state from the sequence of hit measurements, propagating
the state between layers and incorporating material effects via the estimated
radiation length.
\textit{Track selection} removes fake candidates that are not associated with a
real charged particle.
The Kalman filter is the standard algorithm for LHC tracking and provides
theoretically optimal state estimation given the assumed linear dynamics; its limitation is
the combinatorial explosion of track candidates as pileup increases, with
reconstruction time growing super-linearly with detector occupancy.
The required computing time is projected to exceed the available CPU budget
at HL-LHC conditions without significant algorithmic advances~\cite{Bocci2020HLLHC}.

More recently, graph neural networks~(GNNs) have been applied to tracking as
a graph edge classification problem, with hits as nodes and edges representing
candidate track segments~\cite{Ju2021ExaTrkX}.
The Exa.TrkX pipeline processes detector hits as a graph and predicts edge
labels~(true track segment vs.\ combinatorial background) using message-passing
GNNs, achieving competitive efficiency with significantly reduced computational
cost on GPU hardware.
GNN-based trackers are differentiable and scalable, but require labeled training
data: ground-truth hit-to-particle associations must be extracted from Monte Carlo
simulation, potentially introducing discrepancies between simulation and reality.

\paragraph{Why RL?}
RL offers a qualitatively different framing of the tracking problem.
Rather than learning a discriminator for pre-generated track candidates, an RL
agent \textit{actively navigates} the detector hit graph, deciding at each step
which hit to follow next.
The key advantage is that this formulation is naturally \textit{label-free}: the
reward can be defined in terms of physical consistency without requiring
hit-to-particle truth associations.
Specifically, the trajectory of a charged particle traversing a detector is
governed by multiple Coulomb scattering, which determines the expected angular
deflection between successive detector layers; this provides a physics-grounded
reward signal that distinguishes genuine tracks from combinatorial fakes purely
from the measured hit positions.
Furthermore, because the reward depends on the material and geometry of the
detector rather than on labels from a particular simulation, the formulation
could ease the transfer to new detector configurations, where supervised GNN
approaches are typically retrained; this potential has not yet been demonstrated.

% -----------------------------------------------------------------------------
\subsection{RL for Track Finding}
\label{sec:tracking:finding}

\textbf{Kortus et al.}~\cite{Kortus2023Tracking}
formulated charged particle tracking as a graph MDP and trained a PPO agent to
reconstruct proton tracks in the Bergen Digital Tracking Calorimeter~(DTC), a
high-granularity detector developed for proton computed tomography~(pCT).
The MDP is defined on a graph $G = (V, E)$ over the layered point cloud of
detector readouts: vertices are hit centroids, and directed edges connect
neighboring readouts across adjacent detector layers in the reverse direction,
establishing the possible states and valid transitions for the policy.
A full track candidate is represented as a trajectory $\tau = (s_0, a_0,
\ldots, a_{T-1}, s_T)$ where $T$ is the episode length.
Track finding thus instantiates the construction pattern of
Figure~\ref{fig:construction}: the partial object is a track candidate, each
edit appends a hit from the next layer, and the evaluator is the multiple
Coulomb scattering model introduced below.

The policy and value function are implemented using a shared encoder-decoder
Pointer Network~\cite{Vinyals2015Pointer} architecture with multi-head
attention, computing attention weights $\alpha_i^\pi$ and $\alpha_i^V$ for each
valid action $a_t^{(i)} \in \mathcal{A}_t$, i.e., for each neighboring hit of
the current vertex in the next layer:
\begin{equation}
    \pi_\theta\!\left(a_t^{(i)} \mid s_t\right) =
    \frac{\exp\!\left(\alpha_i^\pi\right)}
    {\sum_{j=1}^{|\mathcal{A}_t|} \exp\!\left(\alpha_j^\pi\right)},
    \qquad
    V_{\pi_\theta}(s_t) = \frac{1}{|\mathcal{A}_t|} \sum_{i=1}^{|\mathcal{A}_t|} \alpha_i^V,
    \label{eq:tracking_policy}
\end{equation}
so that the value estimate is the average of the value attention weights over
the same candidate hits.

The reward function exploits the physics of multiple Coulomb scattering.
When a particle traverses a detector layer, the angular deflection $\Delta\theta$
of its trajectory is distributed according to the Highland
formula~\cite{Navas2024PDG}, which depends on the particle momentum and on the
material thickness in units of the radiation length $X_0$; the momentum is
estimated from the residual range $R_0$ of the track.
The immediate reward at each step is the log-probability of the observed angular
deflection under this distribution:
\begin{equation}
    r_t = \log p\!\left(\Delta\theta(s_t : s_{t+1})
    \mid X_0(s_t : s_{t+1}),\, R_0(\tau)\right),
    \label{eq:tracking_reward}
\end{equation}
where $p(\Delta\theta \mid \cdot)$ is the probability of observing the angular
deflection between consecutive state transitions, $X_0(s_t : s_{t+1})$ is the
radiation length of the detector section enclosed by the transition, and
$R_0(\tau)$ is the estimated range of the current track candidate.
Maximizing this reward is equivalent to maximizing the log-likelihood of
observing the full track under the Coulomb scattering model, without any
labeled ground truth.

The agent is evaluated on unseen phantom geometries, varying particle densities,
and different beam spot positions, and is competitive with both a heuristic
search algorithm and a supervised model trained on ground truth.
Note that the phantom is the imaged object, not the detector: generalization is
shown across imaging conditions on a fixed detector, and transfer to a different
detector configuration remains untested.

% -----------------------------------------------------------------------------
\subsection{Multi-Agent RL for Tracking}
\label{sec:tracking:marl}

\textbf{Kortus et al.}~\cite{Kortus2026MARL} extended
the single-agent tracking formulation to a constrained multi-agent setting.
Multiple agents operate simultaneously on the hit graph, each responsible for
reconstructing a different proton track.
The agents compete for hits, since two agents cannot claim the same hit for
different tracks, and cooperate toward a joint objective, the total amount of
particle scattering over all reconstructed tracks in a readout frame.
Training is centralized and execution decentralized: the actors use only local
information, while global information shared between all agents is available
to a centralized critic during training.
The constrained optimization framework ensures that the learned joint policy
satisfies the combinatorial constraints of the assignment problem at each step,
rather than only in expectation.

In practice, uniqueness of the hit assignment is enforced by a safety layer that
solves a linear assignment problem for each joint action, and a cost-margin term
in the gradient estimation steers the policy toward lower total assignment cost,
which the authors relate to wider regions of high reconstruction performance and
reduced predictive instabilities.
In the terms of Figure~\ref{fig:construction}, several policies edit the same
hit graph, and the safety layer acts as a joint feasibility mask over their
actions.
On simulated data from the proton imaging detector, the constrained policies
outperform a conventional sequential track follower as well as single- and
multi-agent RL baselines, with the advantage of the constraints most visible at
high particle densities.
The formulation is relevant to the high-pileup regime of the HL-LHC, where many
tracks compete for the same hits, but its computational scaling and any speed-up
from parallel agents have not yet been evaluated.

% -----------------------------------------------------------------------------
\subsection{Open Problems and Future Research Directions}
\label{sec:tracking:future}

\paragraph{Scaling to full LHC occupancy.}
Existing RL tracking work uses the Bergen pCT DTC geometry with modest hit
multiplicities.
Scaling to the $\mathcal{O}(10^5)$ hits per event of a real LHC silicon strip
or pixel tracker at HL-LHC pileup, while maintaining reconstruction
efficiency and fake rate comparable to the Kalman filter, remains the central
open challenge.
A flat single-agent policy faces two compounding difficulties at LHC scale:
the action space at each step grows with the number of hits in the next layer,
and the episode length grows with the number of detector layers, making the
credit assignment problem progressively harder as the detector size increases.

\textbf{Hierarchical RL}~(Section~\ref{sec:rl:paradigms}) addresses both
difficulties.
A \textit{high-level} manager policy selects subgoals, such as a track seed, a
region of the detector to search, or an intermediate waypoint hit, and a
\textit{low-level} worker policy executes primitive hit-association actions to
reach each subgoal, receiving dense intrinsic rewards in addition to the
per-step scattering reward.
This decomposition is particularly natural for tracking: seeding~(identifying
promising hit triplets) and track extension~(growing the seed hit by hit) are
already treated as separate algorithmic stages in the Kalman filter pipeline,
so the hierarchical decomposition aligns with existing domain structure rather
than imposing it artificially.

\paragraph{Training on real detector data.}
Current RL tracking policies are trained exclusively in simulation.
Because the scattering reward is label-free, it can in principle be evaluated
directly on recorded hits: the hit graph of each recorded event is itself a
valid environment, and interacting with it requires no ground truth.
Training on recorded data would let the policy adapt to detector conditions
that simulators model imperfectly: dead channels, non-uniform efficiency, and
radiation damage.
The main obstacles are the absence of truth for validating the reconstructed
tracks and the cost of training on large volumes of recorded events.
Logged outputs of the standard reconstruction could additionally serve as
demonstrations for imitation learning or as a fixed dataset for offline
RL~(Section~\ref{sec:rl:offline}), where conservative methods such as
CQL~\cite{Kumar2020CQL} and IQL~\cite{Kostrikov2021IQL} limit extrapolation to
hit associations absent from the log.

\paragraph{Combining RL with Kalman filter refinement.}
A hybrid approach using RL for the combinatorial hit association step and the
Kalman filter for subsequent track fitting and smoothing could combine the
flexibility of RL with the theoretical optimality of the Kalman filter for
state estimation given a fixed hit assignment.
This two-stage pipeline aligns with how modern tracking is already structured
and would allow RL to be introduced incrementally into existing reconstruction
software frameworks such as ACTS~\cite{Acts2022}, without replacing the entire
pipeline.

%% file: sections/7_bsm.tex
% =============================================================================
%  7. Beyond the Standard Model: Model Building and Theory Exploration
% =============================================================================

% -----------------------------------------------------------------------------
\subsection{Background: The Problem, Traditional Methods, and the Case for RL}
\label{sec:bsm:background}

The Standard Model~(SM) of particle physics~\cite{Navas2024PDG}, despite its
extraordinary predictive success, leaves fundamental questions unanswered: the
origin of the fermion mass hierarchy~\cite{Feruglio2015FlavourPuzzle}, the
strong charge-parity~(CP) problem~\cite{Peccei2008StrongCP}, the nature of dark
matter~\cite{Bertone2018DMHistory}, and the matter-antimatter
asymmetry~\cite{Riotto1999Baryogenesis}.
Physicists address these by constructing theories \textit{beyond the Standard
Model}~(BSM)~\cite{CsakiTanedo2016BSM}: extensions of the SM gauge group,
matter content, or symmetry structure that predict new particles or
interactions while remaining consistent with all existing measurements.

The space of phenomenologically viable BSM models is vast, discrete, and largely
unexplored.
A theorist searching for models consistent with observed quark masses, lepton
mixing angles, and collider constraints must navigate a combinatorial space of
symmetry groups, field representations, and charge assignments.
The number of candidate models scales as $\sim 10^n$ where $n$ is of order ten
(e.g., of order $10^{14}$ charge assignments for the quark sector of a
single-$U(1)$ Froggatt--Nielsen model with charges between $-9$ and
$9$~\cite{Nishimura2023Flavor}), making
systematic enumeration infeasible even for the simplest symmetry extensions.
This search is typically performed by hand, guided by theoretical prejudice and
experience, or by exhaustive enumeration with automated consistency checks.
Both approaches scale poorly as the number of free parameters grows.

\paragraph{Traditional approaches.}
Manual model building relies on theoretical insight to restrict attention to
well-motivated corners of BSM space, such as minimal extensions or natural
solutions to the hierarchy problem.
Automated scans use random sampling or grid search over discrete parameter
spaces, combined with fast spectrum calculators~(e.g., SPheno~\cite{Porod2003SPheno, Porod2012SPheno31},
FlexibleSUSY~\cite{Athron2015FlexibleSUSY}) and comparison to experimental constraints
via tools such as CheckMATE~\cite{Drees2015CheckMATE} and data from the HEPData
repository~\cite{Maguire2017HEPData}.
Markov Chain Monte Carlo and nested sampling, in particular the MultiNest
algorithm~\cite{Feroz2009MultiNest}, have been used extensively for continuous
BSM parameter estimation, from early global fits of the constrained minimal
supersymmetric Standard Model~(CMSSM)~\cite{Trotta2008SuperBayeS} to the
modular GAMBIT framework~\cite{Athron2017GAMBIT}; however, these techniques
presuppose a fixed particle content and symmetry structure chosen in advance
and only sample couplings and masses within that fixed topology, rather than
searching over the discrete model-building choices themselves.
Bayesian optimization has been applied to continuous BSM parameter
spaces~\cite{Goodsell2023ActiveBO} but struggles with the discrete structure of symmetry
assignments; its Gaussian process surrogate is not well-suited to categorical
spaces.
Genetic algorithms have been used for string landscape
exploration~\cite{Cole2021GAandRL} but are not sample-efficient when the fitness
function requires an expensive spectrum calculation, and they do not learn a
generalizable policy; direct comparisons on string flux vacua and on heterotic
monad-bundle models found GA and RL complementary rather than one dominating the
other~(Section~\ref{sec:bsm:string}).

\paragraph{Why RL?}
BSM model building is a sequential decision process: a model is constructed step
by step by assigning charges, representations, or symmetry factors to fields.
The consistency of a model with observations can be fully evaluated only once
the model is specified; in practice, the reviewed works shape this objective
into per-step rewards based on improvements of a consistency score.
This combinatorial structure, together with the need for strategic exploration
rather than local search, maps naturally onto the RL framework.
In the terms of Figure~\ref{fig:construction}, the partial object is the model
under construction, the feasibility mask encodes the grammar or consistency
conditions on fields and couplings, and the evaluator is a spectrum calculation
or a likelihood; terminal states are collected as an ensemble of viable models.

This formulation places BSM model building directly within the domain of
\textbf{RL for combinatorial optimization}~(CO)~\cite{Mazyavkina2021COsurvey,
DarvariuHM24}, a field that has produced learned solvers for routing, graph
problems, and discrete assignment tasks by training policies to construct
solutions element by element.
The key insight shared between combinatorial optimization and BSM model building
is that hand-crafted heuristics, which encode domain knowledge about which
partial solutions are worth extending, can be replaced by a learned policy that
generalizes across problem instances.
In the CO literature, pointer networks~\cite{Vinyals2015Pointer} and graph
attention networks~\cite{Velickovic2018GAT, Kool2019AttentionRouting} have
proven effective as policy backbones for discrete construction problems; the
GNN-based policy of Wojcik et al.~\cite{Wojcik2025BSMGraphPRD}
follows the same architectural paradigm, with the BSM model grammar playing
the role of the problem-instance graph and the charge or coupling assignments
playing the role of the solution being constructed.

Three specific advantages of RL over the classical alternatives follow from this
framing.
Unlike genetic algorithms~\cite{Cole2021GAandRL}, RL learns a policy that
generalizes across related model-building tasks and can be guided by intermediate
physics-motivated reward shaping, rather than restarting the search for each
new task; in heterotic model building, agents trained on one
Calabi--Yau manifold remain more efficient than random walkers on new
ones~\cite{Larfors2020LineBundle}.
Unlike Bayesian optimization~\cite{Goodsell2023ActiveBO}, RL scales naturally
to discrete combinatorial spaces; GP surrogates are not well suited to
categorical variables without special kernels, and their $\mathcal{O}(n^3)$
cost in the number $n$ of evaluated models~\cite{Rasmussen2006GPML} limits them
to a number of evaluations that is a negligible fraction of model spaces with
$10^{10}$ or more configurations.
Unlike exhaustive enumeration, an RL agent can in principle concentrate its
sampling on the phenomenologically relevant corner of the model space rather
than visiting every configuration, analogously to how learned traveling
salesman problem~(TSP) solvers~\cite{Vinyals2015Pointer, Kool2019AttentionRouting}
find near-optimal tours without examining all $N!$ permutations.
This efficiency gain has not yet been benchmarked against classical scans for
field-theoretic BSM model building~(Section~\ref{sec:bsm:graph}), but in
heterotic string model building an RL agent reached viable models while
sampling only $\sim 10^{-8}$ of the state space, where $10^9$ random samples
typically found none~(Section~\ref{sec:bsm:string}).

% -----------------------------------------------------------------------------
\subsection{Flavor Physics and Froggatt--Nielsen Model Building}
\label{sec:bsm:flavor}

The \textbf{Froggatt--Nielsen~(FN) mechanism}~\cite{Froggatt1979FN} explains the
fermion mass hierarchy through a global $U(1)$ symmetry under which fields carry
integer charges.
The Yukawa coupling between two fields $i$ and $j$ is suppressed by a power of a
small parameter $\epsilon \sim \langle\phi\rangle/\Lambda$:
\begin{equation}
    y_{ij} \sim \epsilon^{|n_i + n_j|},
    \label{eq:fn_yukawa}
\end{equation}
where $\phi$ is a flavon field and $n_i, n_j$ are the FN charges of the two
fields.
The challenge is to find integer charge assignments $\{n_i\}$ for all SM fields
such that the resulting Yukawa textures reproduce the observed quark and lepton
masses and mixing angles.
Even restricting charges to the range $[-N, N]$ for moderate $N$, the number of
candidate charge matrices is $(2N+1)^{k}$ for $k$ fields, making enumeration
infeasible.

\textbf{Harvey and Lukas}~\cite{Harvey2021FNModels} applied RL to
field-theoretic particle physics model building, following earlier applications
to string vacua~(Section~\ref{sec:bsm:string}), using policy gradient methods
on the FN charge assignment problem.
The environment consists of the set $\{Q\}$ of all FN charge matrices, and each
action $a$ increases or decreases a single charge $Q_a$ by one while keeping the
remaining charges unchanged.
Harvey and Lukas leverage domain knowledge to design a domain-informed
consistency score $V(Q)$, computed directly from the predicted spectrum rather
than learned, that checks order-of-magnitude agreement between predicted and
observed mass ratios rather than measuring exact consistency:
\begin{equation}
    r(Q, a, Q') =
    \begin{cases}
        V(Q') - V(Q) & \text{if } V(Q') - V(Q) > 0, \\
        r_{\text{offset}} & \text{otherwise},
    \end{cases}
    \;+\; R_{\text{term}} \, \mathbb{1}[Q' \text{ terminal}]
    \label{eq:fn_reward}
\end{equation}
where $Q'$ is the state after applying action $a$, $r_{\text{offset}}$ is a
fixed negative value assigned when the action does not improve the consistency
score, and
$R_{\text{term}}$ is a fixed positive bonus awarded whenever $Q'$ is a terminal
state.
Through $V(Q)$, it is possible to identify phenomenologically promising terminal
states that give rise to approximately correct masses and mixings.
The score $V(Q)$ is evaluated at the flavon vacuum expectation values~(VEVs)
that maximize it, with the $\mathcal{O}(1)$ Yukawa coefficients held fixed during
training.
After training, a Monte Carlo search over these coefficients further optimizes the consistency score of
terminal states found during and after training, exploring only a small
fraction of the full search space.
This procedure yields 89 consistent models for a single $U(1)$ and 2\,019 for
two $U(1)$ factors, and the trained policy can recover specific models
previously proposed in the literature when initialized near them, though not
every known model is rediscovered during training.

\textbf{Nishimura, Miyao, and Otsuka}~\cite{Nishimura2023Flavor} extended this
line of work to the combined flavor structure of quarks and leptons, including
neutrino masses.
The richer reward function encodes consistency with the elements of the quark
Cabibbo--Kobayashi--Maskawa~(CKM) matrix, the lepton
Pontecorvo--Maki--Nakagawa--Sakata~(PMNS) mixing angles, and neutrino mass ratios
simultaneously.
A DQN agent navigates a substantially larger model space than the Harvey--Lukas
work and finds 21 models consistent with the measured masses and mixings,
favoring the normal ordering of neutrino masses when the ordering is not
specified in advance.
A key methodological contribution is extending the single-sector consistency
score of Harvey and Lukas to jointly cover the quark and lepton sectors within
one agent, with each observable entering as an unweighted log-ratio deviation
term alongside fixed shaping constants for non-improvement and termination.

\textbf{Nishimura, Miyao, and Otsuka}~\cite{Nishimura2024Axion} further applied
RL to the statistical search for an axion model from flavor, in which the
$U(1)$ flavor symmetry doubles as the Peccei--Quinn symmetry, so that its
Nambu--Goldstone boson is a QCD axion that solves the strong CP problem.
The reward is the same flavor consistency score as in the earlier work,
evaluated with masses renormalized at several high scales; the constraints
from dark matter, isocurvature perturbations, and inflation, and the
sensitivity of future axion searches, are analyzed afterwards for the charge
assignments found by the agents.
In six days of training, the agents find 156 such assignments, whereas the
authors estimate that conventional brute-force or random searches would need at
least 55 days to find the same number.

A dedicated short review of RL techniques for the flavor problem in particle
physics has been provided by Giarnetti and Meloni~\cite{Giarnetti2026FlavourRL},
covering Froggatt--Nielsen-based RL approaches in greater depth; our treatment
extends beyond flavor models
to graph-based BSM exploration and connects to the broader combinatorial search
literature.

% -----------------------------------------------------------------------------
\subsection{Graph- and Grammar-Based RL for BSM Model Space Exploration}
\label{sec:bsm:graph}

A more general approach to BSM model building represents theories as graphs,
where nodes are fields and edges encode symmetry or coupling relations.
This graph-based formulation accommodates a wider class of BSM models than
FN-style charge assignments: crucially, it does not require the user to specify
a fixed particle content in advance, allowing the agent to discover the number
and type of new particles as part of the search.

\textbf{Wojcik, Eu, and Everett}~\cite{Wojcik2024BSMGraph, Wojcik2025BSMGraphPRD}
developed a methodology for scanning BSM parameter spaces with RL using graph
neural networks, applied as a case study to a class of theories involving
vector-like leptons and a dark $U(1)$ gauge group inspired by the portal matter
paradigm~\cite{Rizzo2019PortalMatter, Wojcik2023LeptonPortal}.
A BSM theory is represented as a heterogeneous graph $G = (V, E)$ with two node
types: particle nodes for the field species and coupling nodes for the
interaction terms, connected by typed edges that specify how each field enters
a given coupling.
A graph grammar, built from a user-specified symmetry group and a finite list
of allowed field representations, defines the set of valid nodes and edges and
admits only interaction terms that contract to singlets under the preserved
symmetries; this provides a physics-informed feasibility constraint on the
action space, analogous to the safety layer that enforces unique hit
assignments in multi-agent tracking~(Section~\ref{sec:tracking:marl}).

The RL agent acts on the graph by sequentially adding or removing nodes and
edges, with each action modifying the particle content or coupling structure
of the model.
The environment also exposes continuous parameters~(particle masses and Yukawa
couplings on particle nodes, coupling constants on coupling nodes), so the
agent must navigate a mixed discrete-continuous parameter space simultaneously.
To handle this parameterized action space, the agent is trained with hybrid
proximal policy optimization~(H-PPO)~\cite{Fan2019HPPO}, which pairs parallel
sub-actors for the discrete and continuous action components with a single
critic.
Models are scored by a parsimony-penalized log-likelihood ratio relative to
the Standard Model:
\begin{equation}
    K(G) = \log \frac{\mathcal{L}(\text{data} \mid G)}{\mathcal{L}(\text{data} \mid \text{SM})}
    - k \, n_{\text{particles}},
    \label{eq:bsm_reward}
\end{equation}
where $\mathcal{L}(\text{data} \mid G)$ is the likelihood of the experimental
data under the model $G$, including the muon anomalous magnetic moment
$(g-2)_\mu$~\cite{Abi2021MuonG2} and precision electroweak and lepton flavor
violation constraints, $n_{\text{particles}}$ is the number of new particles,
and $k$ is a fixed parsimony weight~($k = 0.5$ in the case study).
Per-step rewards are based on improvements of $K$ over its running maximum
within the episode, and reaching a terminal state, defined by a threshold on
$K$, earns an additional fixed bonus.
A GNN-based actor-critic network processes the current graph state via message
passing and outputs a value estimate together with the action distributions:
log-probabilities over the discrete choices and a mean and variance for each
continuous parameter.

The trained agent identifies both known and previously unstudied vector-like
lepton models that accommodate the muon $(g-2)_\mu$ anomaly within a single
scan, exploring all six feasible sets of particle content and learning
hierarchical differences between model parameters to remain consistent with
lepton flavor violation observables.
Since then, the updated Standard Model prediction~\cite{Aliberti2025WP} and the
final Fermilab measurement~\cite{MuonG2Final2025} have removed the significant
$(g-2)_\mu$ tension, so this target is best read as a methodological benchmark.
The agent samples about 1.6 million models over the course of training; the
authors do not benchmark this against random, grid, or exhaustive scans, so
the efficiency gain over classical exploration remains to be quantified.
The authors inspected the resulting model ensembles, attributing the clustering
of discovered models in the target $(g-2)_\mu$ range directly to the reward
design. They caution that factors unrelated to the underlying physics also shape the sampled distribution, which need not match the prior of a theorist over the model space.

\textbf{Alexander, Bradley, Gouskos, and Niu}~\cite{Alexander2026Albert} replace
the graph representation with a token sequence.
In their framework, called Albert, a theory is a sequence of tokens for
symmetries, fields, and interactions generated under a rule-based grammar, and
a grammar mask sets the probability of inadmissible tokens to zero, playing the
same feasibility role as the graph grammar above.
The policy is a decoder-only transformer with about 25 million parameters,
pretrained by next-token prediction on 100\,000 theories sampled uniformly from
the grammar and then fine-tuned with GRPO~\cite{Shao2024GRPO}, the critic-free
policy gradient method also used for trigger control in
Section~\ref{sec:analysis:trigger}.
The reward has two stages: a consistency stage scores gauge anomaly
cancellation, perturbative unitarity, and the absence of light exotic states,
and an experimental stage computes predictions with a
SARAH~\cite{Staub2014SARAH} and SPheno pipeline, minimizes the $\chi^2$ against
the $W$ boson mass measured at the Large Electron--Positron collider~(LEP-II)
over the free
parameters by differential evolution~\cite{Storn1997DE}, and maps the result to a reward.
Starting from particle content known before 1990, the agent finds, within 320
candidate theories, six structurally distinct theories that all contain a
color-triplet fermion doublet and a right-handed singlet with the quantum
numbers of the top-bottom pair.
An ensemble of differential-evolution fits of the best theory predicts
$m_t = 178.9 \pm 5.0$~GeV and $m_H = 146.9 \pm 17.4$~GeV.
As for the graph-based agent, the paper does not compare against random or
exhaustive sampling, and the experimental stage uses a single observable.

% -----------------------------------------------------------------------------
\subsection{RL for String Compactifications}
\label{sec:bsm:string}

String compactifications provide a UV-complete setting for BSM model building in
which the four-dimensional particle content is fixed by discrete compactification
data, such as brane wrapping numbers or vector bundle data.
Consistency conditions and phenomenological requirements are checked directly on
these data, so the search for realistic vacua is a combinatorial construction
problem of the same type as in Sections~\ref{sec:bsm:flavor}
and~\ref{sec:bsm:graph}, and RL was applied to it before the field-theoretic
studies discussed above.

\textbf{Halverson, Nelson, and Ruehle}~\cite{Halverson2019BranesBrains} explored
type IIA compactifications with intersecting D6-branes on the
$T^6/(\mathbb{Z}_2\times\mathbb{Z}_2\times\mathbb{Z}_2)$ orientifold with an
asynchronous advantage actor-critic~(A3C) agent.
Actions add or remove branes from stacks or change their winding numbers, and
the reward combines the string consistency conditions, namely tadpole
cancellation, the K-theory constraint, and supersymmetry, with proximity to
Standard Model-like vacua.
In some tasks the agent finds $\mathcal{O}(200)$ times more consistent solutions
than a random walker.
It rediscovers the human-derived ``filler brane'' strategy for satisfying the
tadpole conditions and, where filler branes are unavailable, learns a strategy
that reaches the same goal twice as efficiently per unit time.
The agent also surfaced a structural result, that three untilted tori cannot
yield an odd number of chiral families, which the authors subsequently proved.
No exact Standard Model was found in the explored space, which is not guaranteed
to contain one.

\textbf{Larfors and Schneider}~\cite{Larfors2020LineBundle} applied RL to
heterotic model building, training A3C agents to construct $SU(5)$ grand
unified models from sums of line bundles over complete intersection Calabi--Yau
manifolds.
In the most favorable settings, the agents find unique consistent models up to
1\,700 times more efficiently than random exploration.
Agents trained on one manifold also outperform random walkers on new manifolds,
and the experiments scale well with $h^{1,1}$, which suggests that RL could make
model building tractable on manifolds with many K\"ahler moduli.
A recent successor, LB-Explorer~\cite{Yip2026LBExplorer}, replaces the A3C policy
with a transformer-based RL architecture for line bundle standard models with
$SU(5)\times S(U(1)^5)$ symmetry: the agent learns the $E_8$ embedding, anomaly
cancellation, poly-stability, chirality, and no-exotics conditions, while a
hybrid variant perturbs the solutions of the agent with CP-SAT constraint solvers to
impose some conditions exactly.

\textbf{Constantin, Harvey, and Lukas}~\cite{Constantin2022HeteroticRL} applied RL
to heterotic $E_8\times E_8$ compactifications, constructing $SO(10)$ grand
unified models from monad bundles on two Calabi--Yau threefolds with a freely
acting $\mathbb{Z}_3\times\mathbb{Z}_3$ symmetry: the bicubic~($h^{1,1}=2$) and
the triple trilinear~($h^{1,1}=3$).
The state is an integer matrix of line-bundle entries, and each action shifts one
entry by one unit together with a compensating shift that preserves the
first Chern class constraint.
An intrinsic value penalizes deviations from the index required for three
families after the quotient, anomaly violation, failure of the bundle and
equivariance conditions, and failed stability checks.
The reward is a variant of the value-improvement, offset, and terminal-bonus
structure of Harvey and Lukas~(Equation~\ref{eq:fn_reward}), with the
improvement raised to a power and additional per-step and boundary penalties,
and the agent is trained with an actor-critic method.
The trained policies reach phenomenologically promising states in nearly all
episodes within about 20 steps while sampling only $\sim 10^{-8}$ to
$\sim 10^{-10}$ (bicubic) and
$\sim 10^{-13}$ (triple trilinear) of the respective state spaces; by contrast,
$10^9$ random samples of the bicubic environment typically contain no terminal
state.
The search yields about 90 new candidate standard models on the bicubic, across
two choices of bundle ranks, and a
few hundred on the triple trilinear, the latter extrapolated from stability
checks on a random subsample.
In a follow-up on the same class of monad-bundle models, Abel, Constantin,
Harvey, and Lukas~\cite{Abel2022GAvsRL} showed that genetic algorithms also find
anomaly-free, supersymmetric three-family $SO(10)$ models, and that the two
methods have similar efficacy with partly complementary strengths.

\textbf{Cole, Krippendorf, Schachner, and Shiu}~\cite{Cole2021GAandRL} compared
a genetic algorithm with an A3C agent on flux vacua near a conifold limit,
searching over eight integer fluxes for configurations that reach a target
magnitude of the flux superpotential.
Both methods identify distinct vacua efficiently, and the authors do not single
out either as preferred.
Instead, they combine samples from the two qualitatively different search
strategies to reduce sampling bias; both reveal the same clustering in the flux
data, which the authors interpret as a hint of a previously unidentified
discrete symmetry.

\textbf{Krippendorf, Kroepsch, and Syvaeri}~\cite{Krippendorf2021FluxRL} applied
RL to type IIB flux vacua on conifold and symmetric torus backgrounds, targeting
vacua with prescribed values of the flux superpotential and the string coupling.
The agent finds such vacua significantly faster than Metropolis~\cite{Metropolis1953} and random
sampling.
More importantly for the interpretability question of
Section~\ref{sec:bsm:future}, the learned strategies can be read out and reveal
previously unknown correlations in the flux landscape.

% -----------------------------------------------------------------------------
\subsection{Open Problems and Future Research Directions}
\label{sec:bsm:future}

\paragraph{Extending beyond flavor models.}
Existing RL work on model building covers Froggatt--Nielsen flavor and axion
models~(Section~\ref{sec:bsm:flavor}), graph-based vector-like lepton models for
the muon $(g-2)_\mu$ anomaly and grammar-generated Lagrangians fitted to LEP
data~(Section~\ref{sec:bsm:graph}), and string
compactifications~(Section~\ref{sec:bsm:string}).
Extending these methods to supersymmetric~(SUSY) soft terms, extra dimensions, and composite
Higgs models would require new graph grammars, new consistency check pipelines,
and more expensive reward evaluations.
Connecting RL model builders to automated spectrum calculators has been
demonstrated for a single observable~\cite{Alexander2026Albert}; scaling this to
full collider constraint databases~(e.g., HEPData, CheckMATE) remains a
necessary engineering step.

The key challenge is that spectrum calculations can take seconds to minutes per
model, while RL requires millions of environment interactions for convergence
(about 1.6 million sampled models in Section~\ref{sec:bsm:graph}).
Two complementary strategies target this bottleneck:
\begin{itemize}
    \item\textbf{Model-based RL} with a learned surrogate for the spectrum
calculator~\cite{Janner2019MBPO} replaces the expensive simulator with a fast
neural network approximation, enabling millions of imagined model evaluations
between real spectrum calls; the approach mirrors what Hirlaender and
Bruchon~\cite{Hirlaender2020FERMIFEL} did for the FEL with a neural network
ensemble as the environment model.

    \item\textbf{Example-based reward learning.} Classical inverse
RL~(Section~\ref{sec:rl:irl}) infers a reward from expert demonstrations, but the BSM
literature records finished models rather than the sequences of decisions that
produced them.
Published models can instead serve as examples of successful outcomes, from
which a reward or value function is learned without
demonstrations~\cite{Fu2018VICE, Eysenbach2021RCE}.
Because such a learned reward is cheap to evaluate, it can pre-filter
candidates before expensive spectrum calls, and a policy trained against it
would explore the BSM space in a way that is consistent with human theoretical
intuition without being limited to it.
Two caveats apply: the literature provides positive examples only, and it is a
biased sample shaped by analytic tractability and by which model classes were
popular, so the learned reward inherits these biases unless they are corrected.
\end{itemize}

\paragraph{Reward design for multi-constraint models.}
As the number of simultaneous physical requirements grows, reward shaping becomes
more challenging: naive scalarization of constraints into a single reward can
lead the agent to sacrifice one constraint to optimize another, and the optimal
weight vector is unknown a priori. Multiple techniques can avoid this, such as:
\begin{itemize}
    \item\textbf{Constrained MDPs}~(Section~\ref{sec:rl:safety}) encode each
physical requirement as a separate cost signal with its own budget, and
primal-dual methods such as reward constrained policy
optimization~\cite{Tessler2019RCPO} keep the policy feasible.
The formulation fits requirements with a binary character, such as whether a
model reproduces the observed CKM matrix to within experimental uncertainty,
encoded as an indicator cost.
Because the constraint holds in expectation, individual models produced by the
policy can still violate it and must be filtered at evaluation time.

    \item\textbf{Multi-objective RL}~(Section~\ref{sec:rl:paradigms}) avoids
fixing the weight vector in advance: a preference-conditioned
policy~\cite{Yang2019MOPolicy}, conditioned on a preference vector
$\mathbf{w}$, can be varied at inference time to trace trade-offs between, for example, quark mass accuracy and
lepton mixing consistency, without retraining.
With linear preferences, such methods recover the convex coverage set, that is,
the convex part of the Pareto front; trade-offs in concave regions of the front
are not recovered.
Within this limit, the recovered front is physically informative: its shape
indicates which pairs of constraints are in tension and which can be jointly
satisfied without compromise.

    \item\textbf{Curriculum learning}~(Section~\ref{sec:rl:reward};~\cite{DBLP:conf/ijcai/PortelasCWHO20}) addresses the sparse
joint reward problem by introducing constraints progressively during training.
The natural curriculum in BSM model building is ordered by computational cost.
In the FN setting, fermion masses and the CKM and PMNS mixing follow from the
same diagonalization of the Yukawa textures, and axion properties such as the
domain wall number and the axion-photon coupling are analytic functions of the
charges~\cite{Nishimura2024Axion}, so these checks are cheap.
Fitting the $\mathcal{O}(1)$ coefficients and the flavon VEV requires a
numerical optimization per model, and observables such as the dark matter relic
density require external spectrum and relic-density codes.
Training the agent first against the cheap constraints and progressively
introducing the expensive ones prevents the agent from being overwhelmed by a
sparse joint reward and reduces the total number of expensive evaluations.

    \item \textbf{Reward decomposition}~\cite{VanSeijen2017HRA} learns separate value
functions for each physical constraint and combines them at decision time
via a linear sum, allowing the agent to disentangle which actions satisfy
which constraints.
\end{itemize}

\paragraph{Interpretability: what does the policy know?}
The sampling distribution of a trained BSM model-building policy can be read as
an implicit prior over the space of viable models, although
Section~\ref{sec:bsm:graph} shows that this distribution is also shaped by the
reward design and by factors unrelated to the physics.
Extracting interpretable physics from the policy, for example, identifying which
symmetry patterns it prefers and why, is an open problem at the intersection
of \textbf{explainable RL}~(XRL)~\cite{Heuillet2021XRL} and theoretical physics.
The string landscape studies of Section~\ref{sec:bsm:string} show that this is
possible in practice: learned strategies have exposed correlations in flux
vacua~\cite{Krippendorf2021FluxRL} and suggested a structural result on chiral
families that was then proven~\cite{Halverson2019BranesBrains}.

In attention-based GNN policies, such as graph attention
networks~\cite{Velickovic2018GAT}, the attention weights over graph edges
provide a natural first layer of attribution: they indicate how strongly each
coupling relation contributes to the decision of the policy at each step, and can be
inspected post hoc to identify recurring structural motifs in discovered models.
Such attention-based attribution has not yet been reported for the BSM policies
of Section~\ref{sec:bsm:graph}.
Attention weights are also not reliable explanations on their own: they often
disagree with gradient-based importance measures, and markedly different
attention distributions can yield the same
prediction~\cite{Jain2019AttentionNotExplanation}.
\textbf{Causal RL}~\cite{Deng2023CausalRL, Zeng2023CausalRLSurvey} offers a
stronger notion of interpretability by incorporating a structural causal
model~(SCM) of the state variables into the learning process, so that the agent
reasons about interventions rather than correlations across sampled
trajectories.
In BSM environments the reward is a known, deterministic function of the model
parameters, so reward-level interventions, such as changing a charge $n_i$ and
re-evaluating consistency, are directly computable without causal machinery.
The value of causal methods lies instead in explaining the policy: identifying
which features of the current partial model causally drive its decisions, as
opposed to features that merely correlate with them in the training
distribution.

\textbf{Policy distillation into soft decision trees}~\cite{Frosst2017SoftTree,
Coppens2019DistillTree} compresses the neural policy into a binary tree of soft
decisions: each inner node applies a learned sigmoid filter to the whole input,
and each leaf holds a fixed action distribution.
An explanation is the sequence of filters along the most probable path, which is
more transparent than a deep network but does not yield if-then rules over
individual features, since the splits are oblique rather than axis-aligned.
For FN models, where the state is a fixed-length charge vector, the learned
filters can be read as weighted combinations of charges that steer the
construction; variable-size graph states, as in Section~\ref{sec:bsm:graph},
would first require a fixed-length encoding.
\textbf{Programmatic policy distillation}~\cite{Verma2018PIRL} offers a
complementary path: a trained neural policy guides a search over programs in a
domain-specific language, yielding a policy that is directly readable and could,
in principle, be reused in automated theory-building pipelines.

%% file: sections/8_detectors.tex
% =============================================================================
%  8. Detector and Instrument Design
% =============================================================================

% -----------------------------------------------------------------------------
\subsection{Background: The Problem, Traditional Methods, and the Case for RL}
\label{sec:detectors:background}

The design of a particle physics detector is one of the most complex and
high-stakes engineering optimization problems in science.
A modern collider detector such as ATLAS~\cite{ATLAS2008} or CMS~\cite{CMS2008}
incorporates of order $10^8$ channels across multiple subsystems: tracking detectors, calorimeters, and muon
systems, each with hundreds of free geometric and material parameters.
Design decisions are strongly coupled; the longitudinal segmentation of a
calorimeter affects the energy-flow algorithm, which in turn affects jet
reconstruction and therefore all analyses using jets.
A single full simulation of a design candidate using
Geant4~\cite{Agostinelli2003GEANT4} typically takes hours to days of CPU time,
making exhaustive optimization infeasible.

A detector configuration can be formally represented as an ordered collection
of components $\mathcal{D} = \{C_1, C_2, \ldots, C_N\}$, where each $C_n$
corresponds to a sub-detector or functional module~\cite{Figalli2026DesigningAI}.
Each component is parameterized by a vector $C_n = [p_1, p_2, \ldots, p_{K_n}]$
encoding physical layout, material choices, and engineering properties.
These parameters may be continuous ($p_k \in \mathbb{R}$), discrete~(e.g., layer
counts or segmentation steps), or categorical~(e.g., sensor technology or
material type).
The design task is to maximize a scalar score function $S(\mathcal{D}) \in
\mathbb{R}$ quantifying the quality of the configuration, subject to physical,
mechanical, and engineering constraints:
\begin{equation}
    \mathcal{D}^\star = \arg\max_{\mathcal{D}}\; S(\mathcal{D}),
    \label{eq:detector_opt}
\end{equation}
where $S(\mathcal{D})$ depends on four components: simulation, reconstruction,
analysis, and cost.
The computational burden of evaluating $S(\mathcal{D})$ is the central
obstacle: each evaluation requires simulating large numbers of particle
interactions through the detector geometry, and the design space is large enough that
exhaustive search is infeasible.

\paragraph{Traditional approaches.}
Detector design has historically been driven by engineering judgment and
parametric studies: a small number of design parameters are varied manually while
the rest are fixed, and simulation is run at a grid of points.
More systematic optimization uses \textbf{Bayesian optimization}~(BO) with a
Gaussian process surrogate over the simulation
output~\cite{Cisbani2020AIDetector}, which is sample-efficient but limited to
low-to-moderate-dimensional continuous parameter spaces; the GP surrogate becomes
inaccurate and computationally expensive as dimensionality grows.
The \textbf{differentiable programming} approach, pioneered by the
\textit{MODE Collaboration}~\cite{Dorigo2023MODEWhitePaper, Aehle2025MODEProgress},
represents the detector geometry in a differentiable framework and
back-propagates gradients from physics observables to design parameters.
This is powerful for continuous parameters but requires the full simulation
pipeline to be differentiable, which excludes discrete design choices~(e.g., the
number of detector layers) and stochastic simulation steps.
Differentiable approximations using hybrid Monte Carlo gradient
estimators~\cite{Aehle2025MODEProgress} have been developed within MODE to
partially address this, but surrogate-based
gradients remain tied to reduced models and inherit their limitations.
A recent complementary approach uses diffusion model surrogates as fully
differentiable emulators of the simulation-reconstruction chain~\cite{Schmidt2025AIDO},
enabling gradient-based design exploration in both continuous and discrete
parameter spaces; discrete choices such as materials are handled by optimizing
a probability assigned to each category rather than the category itself.

\paragraph{Why RL?}
RL complements both BO and differentiable programming by naturally handling the
discrete and sequential nature of detector design: the number of layers, the
choice of material at each position, and the placement of sub-detectors are all
discrete decisions that can be made sequentially, building the configuration
one component at a time.
The reward is the score $S(\mathcal{D})$ evaluated by a full or fast simulation,
with no differentiability required.
Detector design therefore follows the construction pattern of
Figure~\ref{fig:construction}, with the partially built detector as the object,
layer placement as the edit, and a full or fast simulation as the evaluator;
unlike the previous cases, the agent observes only part of the design and
receives the score only at the end of the episode~(Section~\ref{sec:detectors:rl}).
A recent review by Figalli, Qasim, Owen, and Serra~\cite{Figalli2026DesigningAI}
provides a comparative perspective on all three methodologies and concludes that
RL is particularly well suited to combinatorial and mixed-discrete-continuous
design spaces, while BO excels for low-dimensional continuous problems and
differentiable programming is most powerful when large portions of the pipeline
can be made differentiable.

% -----------------------------------------------------------------------------
\subsection{RL for Calorimeter and Spectrometer Design}
\label{sec:detectors:rl}

\textbf{Qasim, Owen, and Serra}~\cite{Qasim2025DetectorRL} presented the first
systematic application of RL to physics instrument design.
They formulate detector design as a sequential MDP in which the agent places or
configures detector elements one at a time until the design is complete or the
resource budget is exhausted.
At each step, the agent predicts a continuous action to determine the position
of the next layer and a discrete action to select the physical properties of
that layer~(e.g., sensor type and granularity).
The environment is partially observable: the observation sent to the agent
contains only the current longitudinal location and the remaining resource
budget~(or the number of layers already placed), not the full geometry.
The reward $S(\mathcal{D})$ is the design score of the instrument quantifying its
resolution, awarded only at the final interaction step; the agent may also
receive a penalty for design violations~(e.g., two layers placed too close
together) at intermediate steps.

Two empirical studies are presented:
\begin{enumerate}
    \item \textbf{Calorimeter longitudinal segmentation}: the agent sequentially
          places active sensor layers along the calorimeter depth, choosing the
          position and the sensor type~(thickness) of each layer.
          The reward is derived from the energy resolution for electromagnetic
          and hadronic showers, evaluated with Geant4 simulations whose stored
          output is reused across designs, which reduces the cost of one design
          evaluation to a fraction of a second on a GPU.
          The PPO agent finds designs that improve the hadronic resolution
          substantially with respect to a hand-designed reference
          segmentation, while the electromagnetic resolution agrees with the
          reference within statistical uncertainty.

    \item \textbf{Spectrometer tracker placement}: the agent places tracking
          stations along the beam axis on both sides of a dipole magnet and
          chooses their transverse segmentation, with the reward based on the
          momentum resolution of reconstructed tracks and a penalty when the
          efficiency falls below 95\%.
          The resulting spectrometer has a significantly better momentum
          resolution than two baseline designs with three or four equidistant
          stations on each side of the magnet.
\end{enumerate}

The paper makes a strong methodological argument: RL should be considered
alongside gradient-based instrument optimization as a standard tool in the
detector design workflow, particularly for discrete and mixed-discrete-continuous
design spaces.
Figalli et al.~\cite{Figalli2026DesigningAI}, including the same authors,
extend this argument by proposing the Future Circular Collider~(FCC) detector
design challenge as a long-term benchmark for AI-assisted detector
optimization, and introduce distributional and
distributionally robust RL as principled frameworks for handling the two types
of uncertainty inherent in detector design: stochastic detector response~(handled
via distributional RL over the return distribution) and systematic model
misspecification~(handled via distributionally robust RL with Wasserstein
ambiguity sets over the transition dynamics).
The robust Bellman operator incorporating this ambiguity set is:
\begin{equation}
    (\mathcal{T}^{\rm rob}V)(s) = \max_a \min_{P \in \mathcal{U}_\varepsilon(s,a)}
    \!\left\{ r(s,a) + \gamma\,\mathbb{E}_{s' \sim P(\cdot|s,a)}[V(s')]\right\},
    \label{eq:robust_bellman}
\end{equation}
where $\mathcal{U}_\varepsilon(s,a) = \{P(\cdot|s,a) : W_c(P(\cdot|s,a),
P_0(\cdot|s,a)) \leq \varepsilon\}$ is a Wasserstein ball of radius $\varepsilon$
around the nominal simulator dynamics $P_0$, and $W_c$ is the optimal-transport
cost induced by a cost function $c$.
Such an operator would yield policies that remain stable under alternative
physics tunes or calibration envelopes, a desirable property for detector
designs intended to operate over multi-year run periods.

% -----------------------------------------------------------------------------
\subsection{Open Problems and Future Research Directions}
\label{sec:detectors:future}

\paragraph{Co-optimization of geometry and reconstruction.}
Current RL detector design work optimizes geometry with a fixed reconstruction
algorithm.
A joint optimization, where the detector geometry and the reconstruction
algorithm~(e.g., clustering parameters, track seeding strategy) are co-optimized
by a single RL policy, could discover configurations that are suboptimal for any
fixed reconstruction approach but excel when the algorithm is adapted to the
geometry.
Figalli et al.~\cite{Figalli2026DesigningAI} identify this as
one of the primary open challenges for AI-assisted detector design, noting that
multiple sub-detector co-optimization requires maintaining a consistent simulation
of the full detector so that cross-subsystem correlations relevant to global
performance are preserved.
A bi-level formulation, in which an outer policy selects the geometry and an
inner reconstruction policy adapts to it from a small number of simulated
events, could draw on meta-RL~(Section~\ref{sec:rl:paradigms}); the
meta-objective would be the reconstruction performance after adaptation rather
than for a single fixed algorithm.

\paragraph{Connection to differentiable programming.}
The differentiable detector framework of the MODE Collaboration~\cite{Dorigo2023MODEWhitePaper}
and RL-based design are complementary: differentiable methods are efficient for
continuous parameters, RL is natural for discrete choices.
A hybrid framework using differentiable optimization for continuous parameters
conditioned on discrete decisions made by an RL agent is an unexplored direction
that could combine the best of both approaches.

\paragraph{Multi-objective RL for detector design.}
Real detector design involves multiple competing objectives: physics performance,
cost, power consumption, radiation tolerance, and maintainability.
Multi-objective RL~(Section~\ref{sec:rl:paradigms}) could provide
a family of design options spanning the performance-cost tradeoff, giving
engineers principled choices rather than a single optimized point.
Distributional RL~(Section~\ref{sec:rl:valuebased}) is a natural complement: by optimizing
tail-sensitive criteria such as CVaR over the return distribution, the agent can
explicitly target designs that perform well in the worst-case physics or
calibration scenarios rather than only on average.

\paragraph{Multi-agent RL for joint sub-detector co-optimization.}
The constrained collaborative MARL framework developed for particle
tracking by Kortus et al.~\cite{Kortus2026MARL} suggests a natural extension to
design-time detector optimization.
A multi-agent formulation in which each agent is responsible for optimizing a
specific sub-detector~(tracker, calorimeter, muon system) under a shared
material budget and geometric acceptance constraint would allow the joint design
to be decomposed into manageable sub-problems while preserving the global
coupling between subsystems through the shared constraint and a centralized
critic.
Competitive dynamics between agents~(each sub-detector competing for a share of
the material budget) and cooperative dynamics~(the global physics score rewards
all agents jointly) reproduce the structure of real detector design review
processes, where subsystem groups negotiate resource allocation under experiment-wide
performance targets.

\paragraph{Detector design aided by language models.}
The study of Zoccheddu et al.~\cite{Zoccheddu2026LLMDetector} compares large
language models~(LLMs) with RL on the two benchmarks of Qasim et al.
Using only prompting, LLMs propose valid, resource-aware, and physically
meaningful designs, but RL still yields stronger final designs, although the
gap is modest for the best models.
Refining the proposals of the LLM with a trust-region optimizer over the
continuous layer positions narrows the gap without closing it, which the
authors attribute to the discrete design choices, such as sensor types, that
the refinement cannot change.
This finding suggests a hybrid approach, which the authors propose as future
work, in which an LLM proposes an initial design, possibly guided by human
knowledge, which RL then refines in both its continuous and discrete choices.
This hybrid paradigm, in which LLMs reduce the human effort required to structure
and supervise optimization while RL performs the reward-driven search, offers a
concrete path toward closed-loop, end-to-end detector design at the scale of
complexity required for next-generation facilities.

%% file: sections/9_analysis.tex
% =============================================================================
%  9. Analysis, Event Processing, and Theoretical Computation
% =============================================================================

This section covers a heterogeneous set of RL applications that do not fit
neatly into the preceding sections but share a common structural property: they
formulate combinatorial or sequential decision problems in HEP, spanning
real-time triggering, data management, data analysis, event reconstruction, and
formal QFT computation, where the feedback is complex, expensive, or delayed.
Because the sub-topics differ substantially in domain and methodology,
traditional methods and open problems are discussed locally for each
sub-section rather than in a unified background and a closing section.

% -----------------------------------------------------------------------------
\subsection{Overview and Unifying Thread}
\label{sec:analysis:background}

The applications in this section span four distinct levels of the HEP workflow.
At the real-time data-taking level~(Section~\ref{sec:analysis:trigger}), RL
controls the trigger system that decides which collision events to record.
At the data management level~(Section~\ref{sec:analysis:caching}), RL decides
which files the distributed computing infrastructure caches for analysis.
At the analysis level~(Sections~\ref{sec:analysis:bkg}--\ref{sec:analysis:reco}),
RL addresses problems in extracting physics signals from recorded data:
background and systematics modeling, and combinatorial event reconstruction.
At the theoretical computation level~(Section~\ref{sec:analysis:feynman}), RL
optimizes the algorithmic procedures of formal QFT computation, both perturbative
(integral reduction and polylogarithm simplification) and non-perturbative (the
conformal bootstrap).
The unifying thread is that all of these problems involve discrete or sequential
choices with a reward that depends on the full decision sequence, precisely the
setting for which RL is designed.
Combinatorial event reconstruction, integration-by-parts~(IBP) reduction, and polylogarithm
simplification follow the construction pattern of Figure~\ref{fig:construction},
in which the agent edits an assignment, a system of identities, or an
expression; trigger control and data caching instead follow the control loop of
Figure~\ref{fig:accelerator_rl}.
Together, they illustrate that the reach of RL in HEP extends well beyond the
operational and instrumentation domains covered in earlier sections.

% -----------------------------------------------------------------------------
\subsection{Real-Time Trigger Control}
\label{sec:analysis:trigger}

\paragraph{Problem and traditional methods.}
Particle collisions at the LHC occur at a rate of 40~MHz, producing a raw data
volume of approximately 40~TB/s that cannot be stored in full.
An event-selection system, the \textbf{trigger}, must therefore decide in real
time which collisions to record and which to discard, reducing the rate by about
four orders of magnitude before writing to permanent storage.
The trigger must be simultaneously sensitive to a wide range of physics processes
at varying rates of occurrence; it must reject nearly all background while
preserving events that may contain new physics signatures.

Modern LHC triggers operate in two stages.
The \textbf{Level-1}~(L1) trigger is implemented in custom FPGAs and
application-specific integrated circuits~(ASICs)
and must process each crossing within $\mathcal{O}(\mu\text{s})$, reducing the
rate from 40~MHz to roughly 100~kHz using simple object-based selections~(e.g.,
transverse energy $H_T > $ threshold, isolated muon or electron $p_T$ cuts).
The \textbf{High-Level Trigger}~(HLT) then runs a software-based reconstruction
on the accepted events, reducing the rate further to $\mathcal{O}$(kHz) for
permanent storage.
In practice, \textit{trigger menus}, i.e., the complete sets of physics selection criteria
and their associated rate thresholds, are largely static: thresholds are set
during commissioning and updated only infrequently, requiring expert
optimization against simulated event samples.
As pileup increases and background compositions shift over a fill, static menus
become miscalibrated, reducing signal efficiency or causing rate violations that
induce detector dead-time.

\paragraph{Why RL?}
We identify three structural reasons why RL is well suited to trigger control.
First, decisions must be fast, and compact policies with few parameters can be
trained to fit within the latency budget.
Second, the objective is indirect: the trigger should maximize the yield of
physics analyses, a goal that is difficult to encode as a differentiable loss;
policy-gradient methods estimate the gradient of the expected reward from
sampled actions and therefore do not need a differentiable objective.
Third, operating conditions shift across runs; policies trained over a range of
conditions can remain robust to moderate shifts, and model-based RL can plan for
anticipated changes.
The indirect objective also makes reward design the central challenge.
Because analysis yield is not observable online, existing controllers optimize
proxies: tracking a target background rate within a tolerance band and
maximizing the efficiency on a few benchmark signal processes, with weights
tuned by grid search~\cite{Ding2026LearningTrigger}.
A second proxy rewards events that are unlikely under a model of known
background processes, as anomaly-detection triggers already do with autoencoder
scores~\cite{Emami2026SelfDrivingTrigger}.
A third route learns from logged operations: the threshold and prescale changes
made by trigger experts during data taking are demonstrations for inverse
RL~(Section~\ref{sec:rl:irl}), and logged trigger decisions can pretrain a
policy with offline RL~(Section~\ref{sec:rl:offline}).
Logged data contain only events accepted by the deployed trigger, so both routes
must correct for this selection when evaluating thresholds looser than those
actually used.

\paragraph{RL approach.}
\textbf{Ding et al.}~\cite{Ding2026LearningTrigger} cast online trigger threshold tuning as
a streaming MDP and train a policy to adapt trigger thresholds in real time.
The state at step $t$ is:
\begin{equation}
    s_t = (h_t,\, u_t),
    \label{eq:trigger_state}
\end{equation}
where $h_t \in \mathbb{R}^d$ is a fixed-size sequential summary of the most
recent $K$ events~(trigger scores and pileup context $N_{pv}$), and
$u_t = (c_t, \hat{r}_t, \Delta c_{t-1})$ encodes the current threshold $c_t$,
the observed background rate, and the last threshold change
$\Delta c_{t-1} = c_t - c_{t-1}$\footnote{The description of the RL formulation
in the paper additionally lists the previous observed rate $\hat{r}_{t-1}$ in
the controller context.}.
The action $a_t \in \mathcal{A}$ is a discrete threshold increment, updating
the threshold as $c_{t+1} = \mathrm{clip}(c_t + a_t, c_{\min}, c_{\max})$.
The reward balances three competing objectives:
\begin{equation}
    R_t = \lambda_1
    \begin{cases}
        1 - \!\left(\dfrac{|r_{t+1} - r^\star_B|}{\delta_B}\right)^{\!2}
            & \text{if } |r_{t+1} - r^\star_B| \leq \delta_B, \\[4pt]
        -\!\left(\dfrac{|r_{t+1} - r^\star_B|}{\delta_B} - 1\right)
            & \text{otherwise,}
    \end{cases}
    + (1-\lambda_1)\!\left(\alpha\,\epsilon^{t\bar{t}}_{t+1}
    + (1-\alpha)\,\epsilon^{h\to 4b}_{t+1}\right)
    - \lambda_2\,\frac{|\Delta c_{t-1}|}{\Delta c_{\max}},
    \label{eq:trigger_reward}
\end{equation}
where $r^\star_B$ is the target background rate, $\delta_B$ is the tolerance,
$\epsilon^i$ are the signal efficiencies for the two representative processes
$t\bar{t}$ and $h\to 4b$, and $\lambda_1, \lambda_2 > 0$ trade off rate
stability against signal efficiency and action smoothness; following the
notation of the paper, the last term is a move penalty that charges each step
for the magnitude of its threshold change.
A safety shield clips threshold updates to $[-\Delta c_{\max}, \Delta c_{\max}]$
to satisfy operational constraints arising from detector readout bandwidth.
The work develops and evaluates three policy families: a baseline DQN \cite{Mnih2015DQN}, a
critic-free Group Relative Policy Optimization~(GRPO) \cite{Shao2024GRPO} adapted to the streaming
setting, and an adaptation of \textbf{Group Filtered Policy
Optimization}~(GFPO)~\cite{Shrivastava2026GFPO}, originally proposed to shorten
LLM reasoning, which here enforces background-rate feasibility \textit{before}
advantage normalization by filtering the group of candidate actions according
to whether their implied rate lies within the tolerance band $[r^-, r^+]$.
Two filtering rules are introduced: GFPO-F keeps the candidates with the smallest
absolute rate error, regardless of feasibility, while GFPO-FR keeps feasible
candidates first, ranked by signal efficiency, and pads the group with the
out-of-band candidates closest to feasibility when too few are feasible.
GRPO without filtering fails under distribution shift because the sampled group
of candidate actions frequently contains zero feasible actions; group-relative
normalization then reinforces out-of-band actions.
GFPO resolves this by filtering the normalization group toward rate-feasible
candidates, yielding consistent gains in the InBand fraction, the fraction of
time steps in which the background rate lies within the tolerance band, and, on
Monte Carlo
streams, gains of up to 2\% in signal efficiency during in-tolerance intervals.
The approach is evaluated on Monte Carlo streams and, crucially, on real CMS
collision data~(Run 283408), to which the policies trained on Monte Carlo are
transferred without any fine-tuning.
On real data, GFPO-F and GFPO-FR raise the InBand
fraction of the $H_T$ trigger from 0.43 for the PID baseline to 0.99 and 0.95,
respectively, and that of the anomaly-detection trigger from 0.41 to 0.69 and
0.61.
The gain in rate stability comes with a trade-off in signal efficiency: on the
$H_T$ trigger, the overall $t\bar{t}$ efficiency increases by 0.2 to 0.35
percentage points, whereas the $h\to 4b$ efficiency within the tolerance band
decreases by about 2 percentage points, and a similar decrease is observed for
GFPO-F on the anomaly-detection trigger.
The authors describe this, to their knowledge, as the first demonstration of
RL-based trigger control on real LHC collision data, and as such it represents a
significant milestone for RL in HEP beyond simulation.
A closely related precursor by Emami, Tosciri, Salvi
et al.~\cite{Emami2026SelfDrivingTrigger} introduced the self-driving trigger
concept on CMS Open Data, combining an $H_T$ selection with autoencoder-based
anomaly detection and adapting the thresholds in real time with a
proportional-derivative controller, whose gains are tuned by grid search; the
integral term of a full PID loop was found to have no significant effect.
For menus with several trigger paths, the authors define cost functions that
trade off background rate, signal efficiency, and bandwidth, and evaluate them
with an idealized controller that knows the upcoming data.
The work does not train an RL agent, but it defines the control problem, the
classical baselines against which RL controllers can be compared, and cost
functions that could serve as their training objective.

\paragraph{Open problems.}
Ding et al.\ regulate a single scalar threshold per trigger and train an
independent policy for each, while real LHC trigger menus contain hundreds of
items whose rates share a global bandwidth budget, so a change to one threshold
affects the total accept rate.
The authors identify multi-trigger coordination, with the feasibility filter
applied over a joint threshold vector under a shared budget, as the next step;
it requires a policy that acts on all thresholds simultaneously, and the
emulated-collider benchmark of~\cite{Ding2026LearningTrigger} provides a natural
testbed.
Transfer to other experiments such as ATLAS and LHCb has not yet been tested,
and deployment within the latency budget of hardware triggers faces the
constraints discussed for accelerator control in
Section~\ref{sec:accelerators:future}.

% -----------------------------------------------------------------------------
\subsection{Data Caching for Distributed Analysis}
\label{sec:analysis:caching}

\paragraph{Problem and traditional methods.}
LHC analysis data are distributed across a worldwide computing infrastructure,
and regional caches close to the users reduce access latency and network load.
A cache must decide which requested files to store and which stored files to
evict, under a fixed storage budget and a request pattern that drifts as
analyses evolve.
Classic replacement policies such as least recently used~(LRU), least frequently
used~(LFU), or size-based eviction are fixed heuristics: they write every
requested file and differ only in what they evict, so they cannot trade cache
throughput against the cost of writing and deleting data.

\paragraph{RL approach.}
\textbf{Tedeschi et al.}~\cite{Tedeschi2021QCACHE} introduced QCACHE, in which two
Q-learning agents manage a CMS cache: an admission agent decides whether to
store a requested file, and an eviction agent decides which cached files to
keep.
The reward is defined by cache hits and misses over a time window, and the
agents are trained on historical CMS request data for the Italian region.
A follow-up~\cite{Tedeschi2022DataLake} develops this into three policies for a
data lake model of distributed analysis.
SCDL uses a single admission agent with LRU eviction triggered by occupancy
watermarks, which the authors describe as closer to a contextual bandit than to
full RL.
SCDL2 adds a tabular eviction agent whose actions range from keeping all files
of a category to deleting all of them.
The DQN variant replaces both tables with deep Q-networks over a state of file
size, request frequency, time since the last request, data type, cache
occupancy, and hit rate.
The policies are evaluated on 2018 CMS analysis-job monitoring data for the
Italian region with simulated caches of 100 to 1000~TiB.
Because hit rate alone ignores the cost of writing data, the figure of merit is
a score defined as throughput, i.e., data read on hits normalized to an infinite
cache, divided by cost, i.e., data written plus deleted.
The RL policies rank first on this score at every cache size, for example 0.34
for the DQN variant against 0.19 for LRU at 100~TiB, while LRU retains the
highest raw throughput; the simulator represents the bandwidth limit only
through a daily threshold on requests, and access latency is not part of the
figure of merit.

\paragraph{Open problems.}
Both studies evaluate a single regional cache in a simplified simulation that
does not model access latency.
Open steps are latency-aware simulation, validation in a production cache, and
coordinated admission across the sites of a data lake, where the decisions of
one cache change the request patterns seen by others and the problem becomes
multi-agent.
Because request patterns drift as analyses evolve, the policies also face the
non-stationarity discussed for data quality monitoring in
Section~\ref{sec:dqm:future}.

% -----------------------------------------------------------------------------
\subsection{Background and Systematics Modeling in Analyses}
\label{sec:analysis:bkg}

\paragraph{Problem and traditional methods.}
In searches for rare decay modes or new physics, the dominant challenge is
accurate determination of the background: the contributions from known SM
processes that mimic the signal.
Traditional background estimation relies on data-driven
techniques~\cite{Navas2024PDG}: control regions~(signal-depleted phase-space
regions where the background model is normalized to data), sideband
fits~(interpolating the background shape from mass sidebands into the signal
window), and the ABCD method~(extrapolating background rates from three
control regions defined by inverting two uncorrelated selection criteria).
Before any of these techniques is applied, the analyst must decide which
background processes to model.
In beauty hadron decay analyses, for example at LHCb~\cite{LHCb2008Detector},
many decay channels can mimic a given signal through similar final states,
misidentified particles, or kinematic overlaps; because simulating each
candidate background is expensive, only those judged most relevant are
simulated, and the choice relies on the intuition and expertise of the
physicist rather than on a systematic method.
A missed background biases the measurement and complicates its interpretation.
Systematic uncertainties are traditionally handled through nuisance parameters
and validated on control channels, but an effect missing from the uncertainty
budget is, by construction, not covered by either.

\paragraph{RL approach.}
\textbf{Hijano Mendizabal et al.}~\cite{HijanoMendizabal2026BkgRL} introduced a
hybrid RL and genetic algorithm~(GA) framework that, given a signal decay,
determines the set of relevant background decays.
A transformer-based agent, trained with a single-player variant of
AlphaZero~\cite{Silver2018AlphaZero}, builds each candidate background as a sequence of tokens representing
the decay chain.
Because the reward is available only once a complete decay has been built and
the trajectory space is large, pure RL struggles to find any rewarded
trajectory; the GA therefore searches the trajectory space first, and the
successful trajectories it finds serve as demonstrations for the agent, which
combines a policy-gradient objective with an imitation term.
In a toy scenario with 40 particles and $B$-meson signal decays, the GA finds
most of the relevant backgrounds of the training signals, and the fine-tuned
agents find all 94 of them and essentially all 42 relevant backgrounds of
signals not seen during training, which a GA would have to search again from
scratch.

\textbf{Golutvin et al.}~\cite{Golutvin2023DLAdvocate} use RL to stress-test a
measurement against systematic effects absent from its uncertainty budget.
They invert the usual logic: assuming the SM, they ask what hidden mismodeling
would be needed to reproduce an observed deviation.
For a branching-fraction measurement they use gradient-based optimization; for
the angular observable $P_5'$ of $B^0 \to K^{*0}\mu^+\mu^-$, measured by LHCb as
$-0.439 \pm 0.117$ in the $q^2 \in [4, 6]~\mathrm{GeV}^2$ bin, they train a DQN
agent.
The hypothesized mismodeling is a per-event efficiency weight $w = w^+ w^-$, in
which each factor is linear in the transverse momentum of one muon, separated by
the muon charge relative to the kaon, and floored at 0.1 to remain physical.
The state is the parameter vector $(k^+, k^-)$ of this weight, each action
increases or decreases one parameter by a step that shrinks from 0.1 to 0.01 as
the fit improves, and episodes last at most 200 steps.
The reward is piecewise in $\chi^2/N_{\rm meas}$ against the measurements:
negative and proportional to it above 3, increasing as it falls between 0.1 and
3, and a large terminal bonus below 0.1.
The study uses a RapidSim~\cite{Cowan2017RapidSim} fast simulation of $10^6$ decays rather than a full
detector simulation.
Fitting the signal channel alone, the agent finds a mismodeling that shifts
$P_5'$ by about 0.24, comparable to the observed tension and to the 0.28 shift
found by direct $\chi^2$ minimization, but this mismodeling shifts the
$B^0 \to J/\psi K^{*0}$ control channel far beyond its uncertainty.
Adding the control channel to the reward caps the achievable shift at about
0.01, so an efficiency mismodeling of this form cannot explain the $P_5'$
anomaly.
The authors note that RL is more flexible than the optimization-based approach
for non-differentiable simulators but computationally more expensive, and that
a definitive statement requires a full detector simulation.

\paragraph{Open problems.}
The stress test of Golutvin et al.\ searches a two-parameter efficiency model,
linear in muon $p_T$, on a fast simulation.
Extending it to high-dimensional families of mismodeling, spanning several
detector effects at once, and to full detector simulation would make each
evaluation far more expensive, so sample efficiency becomes the limiting factor
and learned surrogates of the simulation~(Section~\ref{sec:rl:modelbased}) a
natural remedy.
One formulation is a two-player game, in which the analyst chooses
control channels and the advocate searches for the mismodeling that best
survives them; turning such a search into a routine robustness check for
measurements that deviate from the SM is an open goal.

% -----------------------------------------------------------------------------
\subsection{Combinatorial Event Reconstruction}
\label{sec:analysis:reco}

\paragraph{Problem and traditional methods.}
At hadron colliders, the assignment of final-state reconstructed objects~(jets,
leptons, photons) to the hypothesized decay products of a hard scatter is a
combinatorial problem.
For a top-quark pair event with fully hadronic decay, there are $\mathcal{O}(10^3)$
possible assignments of jets to quarks.
Traditional solutions rely on $\chi^2$ minimization over kinematic constraints
(e.g., $W$ boson and top quark mass windows) or on learned discriminants trained
on simulation with known truth assignments.

\paragraph{RL approach.}
\textbf{Dillon and Spannowsky}~\cite{Dillon2026TheoryRL} formulate final-state
particle assignment as an MDP solved by a DQN with a permutation-invariant
transformer backbone.
The state is a complete one-to-one mapping between final-state four-momenta and
the partons of the matrix element; each action either swaps two
particle-parton assignments or does nothing, swaps already used in the episode
are masked, and an episode lasts at most $N-3$ steps for $N$ partons, the
maximum number of swaps needed to reorder the final state.
The reward is the logarithmic change in the magnitude of the tree-level matrix
element produced by the action, evaluated without any per-event truth labels:
\begin{equation}
    r_t = \log \frac{|\mathcal{M}(s_{t+1})|}{|\mathcal{M}(s_t)|},
    \label{eq:reco_reward}
\end{equation}
where $|\mathcal{M}(s)|$ is the magnitude of the matrix element for the mapping
$s$, so that swaps that increase the matrix element are rewarded and the return
of an episode equals the total logarithmic gain.
On parton-level $t\bar{t}$, $t\bar{t}W$, and $t\bar{t}t\bar{t}$ events with fully
hadronic decays, where the number of possible mappings grows from
$\mathcal{O}(10^3)$ to $\mathcal{O}(10^8)$, the agent finds mappings whose matrix
elements closely match those of the true assignments; reconstruction
efficiencies decrease with multiplicity, from most $t\bar{t}$ events being fully
reconstructed to about a third of the $t\bar{t}t\bar{t}$ events.
Momentum smearing, used as a proxy for detector effects, degrades the
performance only moderately.
The paper reports no comparison with $\chi^2$ minimization or supervised
assignment methods, and a separate network is trained for each process.
This work demonstrates that physics knowledge can be encoded in the reward
function via matrix element evaluation, replacing the need for supervised truth
labels with a first-principles theoretical calculation.

\paragraph{Open problems.}
Full event reconstruction (assigning all final-state objects, inferring
invisible particles, and estimating the primary vertex) is a much harder
combinatorial problem than the assignments considered
in~\cite{Dillon2026TheoryRL}.
The continuous part of this problem is already addressed by generative models:
conditional normalizing flows~\cite{Papamakarios2021NF} reconstruct neutrino
kinematics in semileptonic $t\bar{t}$ events~\cite{Leigh2023NuFlows} and in
dileptonic final states with two neutrinos~\cite{Raine2024Nu2Flows}.
Combining RL for the discrete assignment decisions with such flows for the
continuous kinematic completion is a promising direction for holistic
event-level inference, and the $t\bar{t}$, $t\bar{t}W$, and $t\bar{t}t\bar{t}$
samples of~\cite{Dillon2026TheoryRL} offer a starting benchmark.

% -----------------------------------------------------------------------------
\subsection{Formal QFT Computation}
\label{sec:analysis:feynman}

\paragraph{Problem and traditional methods.}
Multi-loop perturbative calculations in QFT require the reduction of large numbers
of Feynman integrals to a smaller basis of \textit{master integrals} via
IBP identities~\cite{Chetyrkin1981IBP}.
The \textbf{Laporta algorithm}~\cite{Laporta2001IBP} performs this reduction by
constructing and solving a system of linear equations among the integrals.
The efficiency of the algorithm depends critically on the \textit{ordering} in
which integrals are generated and eliminated: a good ordering can reduce the
number of arithmetic operations by orders of magnitude compared to a naive one.
Finding good orderings has historically required expert heuristics and is a
bottleneck for state-of-the-art multi-loop calculations.

\paragraph{RL approach.}
\textbf{Zeng}~\cite{Zeng2025IBPRL} proposes two methods for optimizing IBP
reduction ordering, demonstrated on one-loop massive bubble integrals as a
tractable benchmark.
The problem is formulated as a combinatorial optimization over the ordering of
IBP equation generation~(from pairs of seed integrals and IBP operators) and of
integral elimination, with the objective of minimizing the total number of
arithmetic operations.
In the RL formulation, the state encodes the status of a fixed board of
candidate seed integrals, the actions alternate between selecting a
seed--operator pair and selecting an integral to eliminate, and the reward is
the negative cost of each step:
\begin{equation}
    r_t = -\,c(s_t, a_t),
    \label{eq:ibp_reward}
\end{equation}
where $c(s_t, a_t)$ counts the arithmetic operations performed at step $t$, so
that the return equals minus the total cost of the reduction.
The first method uses RL: a neural network policy interacts with the IBP reduction
environment and learns the ordering by trial and error.
The second uses metaheuristics: simulated annealing~\cite{Kirkpatrick1983SA} minimizes the computational
cost as a black-box function of priority values that control the ordering, with
a small neural network compressing the parameter space for large-scale problems.
The resulting schemes are competitive with the most efficient human-designed
algorithms: the RL agent reaches this level on the smallest benchmark, while
the larger cases are handled by the metaheuristic approach.
The discovered schemes are interpretable and resemble existing human-designed
seeding strategies.
Together with earlier work on symbolic simplification, discussed next, this
establishes formal QFT computation as an application domain orthogonal to the
experimental and phenomenological applications in other sections.

\paragraph{Symbolic simplification and the conformal bootstrap.}
Feynman integrals often evaluate to polylogarithms, whose analytic expressions
are simplified by applying functional identities in a well-chosen order.
\textbf{Dersy, Schwartz, and Zhang}~\cite{Dersy2022Polylog} cast the
simplification of dilogarithm expressions as a game in which each move applies
an inversion, reflection, or duplication identity, and train an RL agent to
reduce the expression; comparing it with a transformer-based approach, they find
both effective, but the latter more powerful and more promising for practical
symbolic manipulation.
Beyond perturbation theory, \textbf{K\'antor, Niarchos, and
Papageorgakis}~\cite{Kantor2022CFTAI, Kantor2022BootstrapRL} applied SAC to the
conformal bootstrap, searching for approximate solutions of the truncated
crossing equations of two-dimensional conformal field theories~(CFTs) and recovering
known theories such as the 2D Ising model and the compactified free boson.
With Richmond~\cite{Kantor2023Bootstrap6D}, they extended the method to the
six-dimensional $(2,0)$ superconformal bootstrap, using SAC as a stochastic
optimizer for four-point functions of scalar superconformal primaries in the
energy-momentum multiplet.
Starting from the supergravity limit and varying the central charge
adiabatically, the searches produce two curves built from 80 CFT data points
(70 for unprotected long multiplets and 10 for protected short multiplets),
which the authors conjecture to describe the A- and D-series $(2,0)$ theories.
The results are competitive with standard numerical bootstrap bounds and with
data from the operator product expansion~(OPE) inversion formula~\cite{CaronHuot2017Inversion}, and the implementation is released as the
BootSTOP code.

\paragraph{Open problems.}
Natural extensions of the IBP work include RL for sector decomposition ordering
in numerical integration and for the selection of integral basis elements in
the differential equations method.
For analytic simplification, the dilogarithm result above suggests that RL is
more likely to help as a search component combined with learned proposals than
as a standalone simplifier.
Learned heuristics for IBP reduction are also being pursued without RL:
concurrently with Zeng, von Hippel and Wilhelm~\cite{VonHippel2025IBP} and Song
et al.~\cite{Song2025ExplainableIBP} used FunSearch~\cite{RomeraParedes2024FunSearch},
a program search guided by large language models~\cite{Brown2020GPT3}, together
with genetic programming to discover
interpretable priority functions, and evaluated them on multi-loop integrals,
whereas RL has so far been demonstrated on the one-loop bubble.
A head-to-head comparison of RL and program search on shared multi-loop
benchmarks, and hybrids in which an RL policy proposes reductions that program
search then distills into explicit heuristics, remain open.

%% file: sections/10_conclusions.tex
% =============================================================================
%  10. Conclusions and Outlook
% =============================================================================

% -----------------------------------------------------------------------------
\subsection{Summary: What RL Has Achieved in HEP}
\label{sec:conclusions:summary}

This survey has reviewed the application of reinforcement learning across seven
domains of HEP, in the broad sense defined in Section~\ref{sec:intro:motivation}.
The picture that emerges is one of a field that has moved from isolated
proof-of-concept demonstrations to a growing body of methodologically diverse
and physically grounded work, tested on real machines and, in a few cases, on
real collision data.

In \textbf{accelerator control}~(Section~\ref{sec:accelerators}), RL agents have
been trained or tested on real machines at CERN, DESY, Elettra, KIT, Brookhaven,
Forschungszentrum J\"ulich, and the Institute of Modern Physics in Lanzhou, and
trained policies have been implemented on
FPGA hardware at Fermilab.
At KARA, an agent trained online on the machine controls a feedback loop with
microsecond inference on dedicated hardware.
The same paradigm tunes the operating point of cryogenic dark matter detectors
in the Gran Sasso underground laboratory.
This is the most mature application domain, with the largest number of studies.

In \textbf{data quality monitoring}~(Section~\ref{sec:dqm}), two complementary
approaches have been proposed: a human-in-the-loop multi-agent PPO system that
queries the shifter only when needed, developed on synthetic data with a
preliminary test on real LHCb data, and an adversarial semi-bandit that monitors
ALICE TRD anode currents on real data and flags deviations hours before a trip.
Systematic validation across runs and operating conditions remains the main
open step.

In \textbf{jet physics}~(Section~\ref{sec:jets}), RL has produced learned
algorithms for grooming, clustering, and input ordering that match or improve
on hand-crafted baselines, with the grooming policy adapting its decisions to
the kinematic configuration of each jet rather than applying a fixed rule; the
clustering results are obtained on a simplified parton shower model.

In \textbf{charged particle tracking}~(Section~\ref{sec:tracking}), RL has
established a label-free track-finding formulation that uses the Coulomb
scattering likelihood as reward, competitive with supervised and heuristic
baselines on a proton imaging detector and robust across imaging conditions.
Constrained multi-agent extensions enforce unique hit assignments, a
prerequisite for dense environments, although scaling to HL-LHC occupancies has
yet to be shown.

In \textbf{BSM model building}~(Section~\ref{sec:bsm}), RL has emerged as a
principled tool for navigating the discrete, combinatorially vast space of
consistent theories.
Applications span Froggatt--Nielsen flavor models, axion model building,
graph-based exploration of extended matter sectors with GNN policies,
grammar-constrained generation of Lagrangians fitted to experimental data, and
string compactifications, where RL was applied first.
The connection to RL for combinatorial optimization explains why RL suits this
setting, but direct comparisons remain scarce: the two available comparisons
with genetic algorithms, on flux vacua and on heterotic monad-bundle models,
found the methods complementary rather than one dominating the other.

In \textbf{detector and instrument design}~(Section~\ref{sec:detectors}), the
first systematic RL-based optimization studies have been published, improving
on a hand-designed reference segmentation for calorimeters and extending the
formulation to spectrometer tracker placement; head-to-head comparisons with Bayesian optimization and differentiable
approaches are still missing.
Distributionally robust RL with Wasserstein ambiguity sets has been proposed to
handle simulator misspecification, and hybrid pipelines in which LLMs propose
designs and RL refines them are a promising direction to further reduce human
engineering effort.

In \textbf{analysis, event processing, and theoretical computation}~(Section
\ref{sec:analysis}), RL has been applied to trigger threshold control, evaluated
on recorded CMS Open Data without fine-tuning; file caching for distributed CMS
analysis; background determination in beauty hadron decays; the stress-testing
of a measurement against systematic effects missing from its uncertainty budget;
combinatorial event reconstruction rewarded by matrix element evaluation; and
formal QFT computation, namely the ordering of IBP integral reduction, the
simplification of polylogarithm expressions, and the search for solutions of
the conformal bootstrap.

Across domains, the reviewed works follow three recurring formulations.
In the first, the agent controls a physical system or a data stream over time,
as in accelerator tuning and feedback, trigger control, and data
caching~(Figure~\ref{fig:accelerator_rl}).
In the second, the agent monitors a detector data stream and decides when to
involve a human expert, who either labels the queried data or acts on the
raised alarms~(Figure~\ref{fig:dqm_hitl}).
In the third, which emerges across Sections~\ref{sec:jets}
to~\ref{sec:analysis}, the agent constructs an object edit by edit under
feasibility constraints, enforced by a mask or by reward penalties, and an
evaluator scores the
result~(Figure~\ref{fig:construction}): the object can be a jet clustering
tree, a track candidate, a BSM model, a detector configuration, an assignment of
reconstructed objects, or an integral reduction ordering.
The formulation largely determines the bottleneck: machine time and safety in
control, expert attention in monitoring, and the cost of the evaluator in
construction.

Table~\ref{tab:summary} collects the RL works reviewed in
Sections~\ref{sec:accelerators} to~\ref{sec:analysis} that report empirical
results, together with the baseline each is compared against and the outcome of
that comparison.
Of its 50 entries, 24 compare RL quantitatively with a traditional method, such
as an optimizer, a controller, a heuristic, or a non-RL learned model: 16 find
RL better on the primary metric, in some cases with a trade-off on a secondary
one~\cite{Ding2026LearningTrigger, Tedeschi2022DataLake} or only in part of the
studied settings~\cite{Rajput2025DDRL, Velotti2023AWAKEElectron}, 6 comparable or
complementary, one worse, and one reports mixed results.
Works with several baselines are classified by the comparison against the
traditional ones, ignoring human, sampling, and RL baselines, and counted as
mixed when RL outperforms some traditional baselines but not
others~\cite{Brehmer2021MCTS}.
Eleven further works compare only against human operators or their
labels~(five), random or Markov chain sampling~(four), or other RL
agents~(two); two validate against a reference configuration or a known
solution; and 13 report no comparison.
Evaluation on a real machine or on recorded data appears in 15 works: 12 in
accelerator and detector operation, and the remaining three in data quality
monitoring and trigger control.
Head-to-head evidence against strong traditional baselines is therefore
concentrated in accelerator control, jet physics, and analysis, and is largely
absent from BSM model building, where most studies report no baseline or
compare only against random sampling.

% Cross-domain summary table
\begingroup
\footnotesize
\setlength{\tabcolsep}{3pt}
\renewcommand{\arraystretch}{1.15}
\begin{longtable}{@{}p{2.6cm} p{2.8cm} p{2.0cm} p{2.6cm} p{1.9cm} p{3.0cm}@{}}
\caption{Summary of the RL works reviewed in Sections~\ref{sec:accelerators}
to~\ref{sec:analysis} that report empirical results.
\textit{Baseline}: the method the RL approach is compared against in the
reviewed work or, where cited, in a closely related study.
\textit{Env.}: environment in which the reported results are obtained.
Sim: simulation; Surr: learned or empirical surrogate of the system; Real: real
machine or detector; Data: recorded experimental data; Formal: exact
theoretical evaluation; Data fit: likelihood against experimental measurements;
$\to$: training in the first, evaluation in the second; FT: fine-tuning.
\textit{Outcome} of RL relative to the baseline:
$+$ better; $=$ comparable or complementary; $-$ baseline better;
R: validated only against a reference configuration or a known solution;
n/a: no comparison reported.
Position papers and studies validated only on standard RL benchmarks are
omitted.}
\label{tab:summary}\\
\toprule
\textbf{Work} & \textbf{Task} & \textbf{RL method} & \textbf{Baseline} &
\textbf{Env.} & \textbf{Outcome} \\
\midrule
\endfirsthead
\multicolumn{6}{@{}l}{\textit{Table~\ref{tab:summary}, continued}}\\
\toprule
\textbf{Work} & \textbf{Task} & \textbf{RL method} & \textbf{Baseline} &
\textbf{Env.} & \textbf{Outcome} \\
\midrule
\endhead
\midrule
\multicolumn{6}{@{}r}{\textit{Continued on next page}}\\
\endfoot
\bottomrule
\endlastfoot

\multicolumn{6}{@{}l}{\textbf{Accelerator control}~(Section~\ref{sec:accelerators})}\\
\midrule
Kain et al.~\cite{Kain2020AWAKE}
  & AWAKE, LINAC4 trajectory correction & PER-NAF & none
  & Real; Sim$\to$Real & n/a \\
O'Shea et al.~\cite{OShea2020FERMIPG}
  & FERMI FEL intensity; THz source & Policy gradient & none
  & Real & n/a \\
Bruchon et al.~\cite{Bruchon2020FERMIBasicRL}
  & FERMI seed laser alignment & Q-learning; NPG REINFORCE & none
  & Real & n/a \\
Kaiser et al.~\cite{Kaiser2022ICML}
  & ARES transverse tuning & TD3 & classical optimizers incl.\ Nelder--Mead; human operators
  & Sim$\to$Real & $+$ (Nelder--Mead equal final error, more steps); $=$ vs.\ operators, faster \\
Kaiser et al.~\cite{Kaiser2024RLOvsBO}
  & ARES focusing and steering & TD3 (RLO) & Bayesian optimization
  & Sim$\to$Real & $+$ (lower final error in most real trials; much faster inference) \\
Xu et al.~\cite{Xu2023LatticeAgnostic}
  & Transverse tuning across lattice sections & SAC & none
  & Sim & n/a \\
Chen et al.~\cite{Chen2023TBSAC}
  & CAFe~II orbit correction; RFQ transmission & Trend-based SAC & human experts
  & Sim$\to$Real & $+$ (about $10\times$ faster) \\
Pang et al.~\cite{Pang2020LANSCE}
  & LANSCE linac section & Deep RL & human operators
  & Sim & $+$ \\
Awal et al.~\cite{Awal2024Injection}
  & COSY injection & SAC & human operators
  & Sim$\to$Real & $=$ quality, faster \\
Ibrahim et al.~\cite{Ibrahim2025RL}
  & VEPP-5 transmission & SAC & none
  & Sim & n/a \\
Ibrahim et al.~\cite{Ibrahim2026RLABC}
  & VEPP-5 beamline (RL-ABC) & DDPG & differential evolution; BO
  & Sim & $+$ (same best transmission, more consistent; far above BO) \\
Hirlaender and Bruchon~\cite{Hirlaender2020FERMIFEL}
  & FERMI FEL intensity & Model-based (BNN ensemble) & model-free RL
  & Real & $=$ (more sample-efficient, lower asymptotic performance) \\
Dong et al.~\cite{Dong2026NSLSII}
  & NSLS-II orbit feedback & Model-based (online model) & SVD orbit feedback; supervised learning
  & Sim; Real & $+$ (about 1 vs.\ 5~$\mu$m RMS on the real ring) \\
Velotti et al.~\cite{Velotti2023AWAKEElectron}
  & AWAKE electron line setup & TD3 & numerical optimizers
  & Real & $+$ (a handful of steps vs.\ up to hundreds); degrades under drift \\
Madysa et al.~\cite{Madysa2022LEIR}
  & LEIR intensity & Model-free RL & none
  & Surr & n/a \\
Meier et al.~\cite{Meier2022SRFGun}
  & SRF photoinjector beam properties & RL through neural network surrogate & common optimizers
  & Surr & $+$ (fewer evaluations) \\
Rajput et al.~\cite{Rajput2025DDRL}
  & CEBAF heat load and trip rate & DDRL & model-free RL; BO; GA
  & Surr & $+$ on the full linac; $=$ on smaller problems \\
Colen et al.~\cite{Colen2026ExplainableRL}
  & CEBAF heat load and trip rate & TD3 with learned constraint & TD3 with reward penalty; DDRL
  & Surr & $+$ vs.\ TD3; $-$ vs.\ DDRL \\
St.~John et al.~\cite{StJohn2021Booster, Kafkes2021BoosterTwin}
  & Booster magnet power supply regulation & Double DQN & PID
  & Surr & $+$ (about $2\times$ in reward) \\
Scomparin et al.~\cite{Scomparin2026KARA}
  & KARA betatron oscillation damping & PPO & FIR bunch-by-bunch feedback
  & Real & $+$ (higher damping rate) \\
Grech et al.~\cite{Grech2022LHCTune}
  & LHC tune feedback & NAF; PPO & PI controller
  & Sim & $+$ (fewer steps; degrades more gracefully with stuck actuators) \\
CRESST~\cite{CRESST2023RL}
  & TES operating point tuning & SAC & human experts
  & Sim; Real & $=$ \\
\midrule

\multicolumn{6}{@{}l}{\textbf{Data quality monitoring}~(Section~\ref{sec:dqm})}\\
\midrule
Jullian Parra et al.~\cite{jullian2025human}
  & Histogram DQM with shifter queries & Multi-agent PPO & noisy shifter labels (shifter only)
  & Sim; Data (LHCb) & $+$ (higher accuracy than the noisy labels) \\
Bower~\cite{Bower2020ATSB}
  & ALICE TRD anode current monitoring & Adversarial semi-bandit & none
  & Data & n/a \\
\midrule

\multicolumn{6}{@{}l}{\textbf{Jet physics}~(Section~\ref{sec:jets})}\\
\midrule
Carrazza and Dreyer~\cite{Carrazza2019GroomRL}
  & Jet grooming & DQN & Recursive Soft Drop
  & Sim & $+$ \\
Brehmer et al.~\cite{Brehmer2021MCTS}
  & Hierarchical jet clustering & MCTS with learned policy & greedy; beam search; $A^*$~\cite{Greenberg2021Astar}
  & Sim & $+$ vs.\ greedy, beam search; $-$ vs.\ $A^*$ \\
Erdmann et al.~\cite{Erdmann2020ReinforcedSorting}
  & Jet ordering for $t\bar{t}H$ classification & RL sorting network & $p_T$ ordering
  & Sim & $+$ \\
\midrule

\multicolumn{6}{@{}l}{\textbf{Charged particle tracking}~(Section~\ref{sec:tracking})}\\
\midrule
Kortus et al.~\cite{Kortus2023Tracking}
  & Track finding (pCT) & PPO & heuristic search; supervised model
  & Sim & $=$ \\
Kortus et al.~\cite{Kortus2026MARL}
  & Multi-track hit assignment & Constrained MARL & sequential track follower; single- and multi-agent RL
  & Sim & $+$ \\
\midrule

\multicolumn{6}{@{}l}{\textbf{BSM model building}~(Section~\ref{sec:bsm})}\\
\midrule
Harvey and Lukas~\cite{Harvey2021FNModels}
  & Froggatt--Nielsen charges & Policy gradient & none
  & Formal & n/a \\
Nishimura et al.~\cite{Nishimura2023Flavor}
  & Quark and lepton flavor models & DQN & none
  & Formal & n/a \\
Nishimura et al.~\cite{Nishimura2024Axion}
  & Flavorful axion models & RL & random search (estimated)
  & Formal & $+$ (6 vs.\ at least 55 days) \\
Wojcik et al.~\cite{Wojcik2024BSMGraph, Wojcik2025BSMGraphPRD}
  & Vector-like lepton models & H-PPO with GNN & none
  & Data fit & n/a \\
Alexander et al.~\cite{Alexander2026Albert}
  & Grammar-based theory generation & Transformer with GRPO & none
  & Data fit & n/a \\
Halverson et al.~\cite{Halverson2019BranesBrains}
  & Type IIA D6-brane vacua & A3C & random walk
  & Formal & $+$ (up to about $200\times$) \\
Larfors and Schneider~\cite{Larfors2020LineBundle}
  & $SU(5)$ line bundle models & A3C & random walk
  & Formal & $+$ (up to $1\,700\times$) \\
Yip et al.~\cite{Yip2026LBExplorer}
  & Line bundle standard models & Transformer-based RL & none
  & Formal & n/a \\
Constantin et al.~\cite{Constantin2022HeteroticRL, Abel2022GAvsRL}
  & $SO(10)$ monad bundle models & Actor-critic & random sampling; GA
  & Formal & $+$ vs.\ random; $=$ vs.\ GA \\
Cole et al.~\cite{Cole2021GAandRL}
  & Type IIB flux vacua & A3C & GA
  & Formal & $=$ (complementary) \\
Krippendorf et al.~\cite{Krippendorf2021FluxRL}
  & Type IIB flux vacua & RL & Metropolis; random sampling
  & Formal & $+$ \\
\midrule

\multicolumn{6}{@{}l}{\textbf{Detector design}~(Section~\ref{sec:detectors})}\\
\midrule
Qasim et al.~\cite{Qasim2025DetectorRL}
  & Calorimeter segmentation; tracker layout & PPO & hand-designed reference
  & Sim & R \\
\midrule

\multicolumn{6}{@{}l}{\textbf{Analysis and theoretical computation}~(Section~\ref{sec:analysis})}\\
\midrule
Ding et al.~\cite{Ding2026LearningTrigger}
  & Trigger threshold control & DQN; GRPO; GFPO & PID controller
  & Sim$\to$Data (CMS) & $+$ rate stability; $-$ $h\to 4b$ efficiency \\
Tedeschi et al.~\cite{Tedeschi2021QCACHE, Tedeschi2022DataLake}
  & CMS cache admission and eviction & Q-learning; DQN & LRU and other heuristics
  & Sim (logged requests) & $+$ throughput per cost; $-$ raw throughput \\
Hijano Mendizabal et al.~\cite{HijanoMendizabal2026BkgRL}
  & Background determination in $b$-hadron decays & RL with GA demonstrations & GA
  & Sim (toy) & $+$ (all training backgrounds; generalizes to unseen signals) \\
Golutvin et al.~\cite{Golutvin2023DLAdvocate}
  & Systematics stress test of $P_5'$ & DQN & direct $\chi^2$ minimization
  & Sim (fast) & $=$, at higher cost \\
Dillon and Spannowsky~\cite{Dillon2026TheoryRL}
  & Jet-parton assignment & Transformer DQN & none
  & Sim (parton level) & n/a \\
Zeng~\cite{Zeng2025IBPRL}
  & IBP reduction ordering & RL; simulated annealing & human-designed orderings
  & Formal & $=$ (RL on the smallest case only) \\
Dersy et al.~\cite{Dersy2022Polylog}
  & Dilogarithm simplification & RL & transformer
  & Formal & $-$ \\
K\'antor et al.~\cite{Kantor2022CFTAI, Kantor2022BootstrapRL}
  & 2D conformal bootstrap & SAC & known solutions
  & Formal & R \\
K\'antor et al.~\cite{Kantor2023Bootstrap6D}
  & 6D $(2,0)$ superconformal bootstrap & SAC & numerical bootstrap bounds; OPE inversion
  & Formal & $=$ \\
\end{longtable}
\endgroup

% -----------------------------------------------------------------------------
\subsection{Open Challenges}
\label{sec:conclusions:challenges}

Despite significant progress across seven domains, a set of fundamental
challenges recurs throughout the HEP RL literature; addressing them is necessary
for RL to be adopted widely in the field.

\paragraph{Reward design.}
The reward function is often the most critical and difficult design choice in
RL applications.
Rewards that capture both the underlying physics and the constraints of the
system or equipment can be difficult to design or prohibitively expensive to
evaluate, so surrogate objectives that are easier to define and compute are
commonly used instead.
Such rewards are computed from measured or simulated proxies of the physical
objective, and an optimizer will exploit any divergence between the proxy and
the objective.
The reviewed works offer concrete examples.
At ARES, the beam parameters measured on a diagnostic screen are false once the
beam leaves the screen: a Bayesian optimizer required an explicit penalty for
this case and still took many steps to recover the beam, and aggressive
optimizer settings overfit to falsely positive noisy samples of the
objective~\cite{Kaiser2024RLOvsBO}.
In the LHC tune feedback, the reward depends only on the tune error, which is
consistent with the observation that a trained agent terminates episodes with
non-zero corrections where the PI controller settles near zero~\cite{Grech2022LHCTune}.
In graph-based BSM exploration, the distribution of sampled models reflects the
reward design as well as the physics~\cite{Wojcik2025BSMGraphPRD}.
Inverse RL~(Section~\ref{sec:rl:irl}) can recover a reward from expert behavior,
and multi-objective RL~(Section~\ref{sec:rl:paradigms}) defers the choice of
weights between competing objectives to deployment time rather than fixing it
during training.

\paragraph{Sim-to-real transfer.}
In HEP, machine and deep learning models are commonly trained on simulated data,
which can be generated in large amounts and for precisely controlled scenarios,
whereas real data are scarce and expensive.
RL agents are likewise trained in simulation and then deployed on real systems,
and the resulting distribution shift can degrade their performance.
The reviewed works show that this gap can be bridged:
agents trained only in simulation have been deployed without retraining at
AWAKE~\cite{Kain2020AWAKE}, domain randomization over unobserved machine
parameters enables zero-shot transfer at ARES and COSY~\cite{Kaiser2022ICML,
Awal2024Injection}, trigger policies trained on Monte Carlo streams transfer to
recorded CMS data without fine-tuning~\cite{Ding2026LearningTrigger}, and
refitting a learned dynamics model to the current machine state adapts the
agent to the real system~\cite{Dong2026NSLSII}.
Pretrained physics foundation models~\cite{Birk2024OmniJet, Golling2024MPM,
Mokhtar2026MLPFFM} are natural candidates for policy encoders that could be
fine-tuned with RL, but none has yet been used as an RL backbone; the closest
evidence is the pretrain-then-RL recipe used in theory
generation~\cite{Alexander2026Albert} and early work on RL fine-tuning of
language models for accelerator tasks~\cite{Sulc2025LLMAccelerator}.

\paragraph{Sample efficiency.}
Every domain has a scarce resource that limits how many interactions an agent
can afford.
In accelerator control it is machine time: the ARES agents are trained for
millions of steps, which is feasible only with a fast simulator~\cite{Kaiser2022ICML}.
In DQM it is expert attention, since every query to the shifter has a
cost~\cite{jullian2025human}.
In detector design it is full simulation, which becomes affordable only by
reusing stored Geant4 output across designs~\cite{Qasim2025DetectorRL}.
In BSM model building it is the evaluation of each candidate model, with about
1.6 million models sampled in a single graph-based scan~\cite{Wojcik2025BSMGraphPRD}.
In tracking and trigger control, Monte Carlo is cheap, and policies are so far
trained in simulation only~\cite{Kortus2023Tracking, Ding2026LearningTrigger}:
for tracking, the label-free scattering reward could in principle be evaluated
on recorded hits, whereas recorded trigger data contain only the events
accepted by the deployed trigger.
Model-based RL, offline RL from archived logs, and pretrained representations
are the main remedies; none has yet been demonstrated at the scale of a full
LHC experiment.

\paragraph{Non-stationarity and adaptation.}
Physics equipment changes over time, and policies must follow it.
Accelerators drift: agents trained on the AWAKE electron line need more steps,
or fail, weeks or months after training~\cite{Velotti2023AWAKEElectron}, which
motivates online model learning at NSLS-II~\cite{Dong2026NSLSII}, meta-RL adaptation to drifting hidden parameters
at AWAKE~\cite{Hirlaender2024MetaRL}, and continual learning for deployed
accelerator systems~\cite{Rajput2025ContinualAccel}.
In DQM, the human-in-the-loop agent queries the shifter mainly after operating
conditions change~\cite{jullian2025human}, and the bandit formulation is
adversarial precisely because detector currents are not
stationary~\cite{Bower2020ATSB}.
In trigger control, pileup and background composition change within a fill, and
group-relative policy optimization fails under this shift unless candidate
actions are filtered for rate feasibility~\cite{Ding2026LearningTrigger}.
Detector designs must remain valid across alternative physics tunes and
calibrations over multi-year operation, which motivates distributionally robust
RL~\cite{Figalli2026DesigningAI}.
In model building, the reward is defined by current measurements, so updated
experimental inputs change the target and require retraining or adaptation of
the policy.

\paragraph{Safety.}
The RL applications reviewed here present safety constraints of different
nature. In particular, we distinguish among:
\begin{itemize}
    \item \textbf{Operational safety.} In accelerator control and detector
    design, violations are physical and potentially irreversible: unsafe actions
    on a machine may cause equipment downtime, radiation damage, or destroyed
    equipment, whereas material budget violations or mechanical interference
    discovered late in the design process may be irreversible given the
    multi-year lead times of custom detector components.

    \item \textbf{Informational safety.} In trigger control and DQM, violations
    are informational and cumulative: rate violations cause detector dead-time
    that permanently discards physics data, and over-certifying bad data
    corrupts downstream physics analyses in ways that may not be immediately
    detectable.

    \item \textbf{Theoretical safety.} In jet physics and BSM model building,
    violations are theoretical: an IRC-unsafe policy produces results that are
    formally meaningless in perturbation theory, regardless of their empirical
    performance on simulation, and a BSM model violating exclusion limits is
    physically inconsistent and must be discarded.
\end{itemize}
The reviewed works already contain the first dedicated mechanisms: a safety
layer that guarantees unique hit assignments in multi-agent
tracking~\cite{Kortus2026MARL}, rate-feasibility filtering and bounded threshold
updates in trigger control~\cite{Ding2026LearningTrigger}, and grammars that
exclude inconsistent theories from the action space of model-building
agents~\cite{Wojcik2025BSMGraphPRD, Alexander2026Albert}.
Except for those in trigger control, these mechanisms act as the feasibility
mask of the construction pattern~(Figure~\ref{fig:construction}), and they
enforce their constraints exactly because the constraints are known
analytically: grammar validity and hit uniqueness can be checked before an edit
is applied; no reviewed jet algorithm yet enforces IRC safety in this way.
In control problems, the constraints depend on the response of the machine,
which is known only approximately.
Action clipping and reward penalties alone are practical mitigations that
provide no formal safety guarantee; constrained MDPs~\cite{Altman1999CMDP} and
safe model-based RL~(Section~\ref{sec:rl:safety}) offer principled solutions to
this challenge.

\paragraph{Benchmarks and baselines.}
Several of the reviewed results lack a classical baseline: the graph-based and
grammar-based BSM scans are not compared against random or Markov chain Monte
Carlo sampling~\cite{Wojcik2025BSMGraphPRD, Alexander2026Albert}, and RL-based
detector design has not been compared head-to-head with Bayesian
optimization~\cite{Qasim2025DetectorRL}.
Comparisons at equal budget, measured in machine interactions or simulator
calls and repeated over several random seeds, are needed to establish where RL
helps.
For construction problems, the shared structure of
Figure~\ref{fig:construction} suggests a common environment interface, in which
the state, the edits, the feasibility mask, and the evaluator are specified per
domain, so that the same baselines, such as random sampling, Markov chain Monte
Carlo, beam search, and genetic algorithms, can be run on every task.
Open resources on which such comparisons can be built already exist: the Cheetah
beam dynamics code~\cite{Kaiser2024Cheetah}, the RL-ABC beamline
environments~\cite{Ibrahim2026RLABC}, the Ginkgo jet model~\cite{Cranmer2019Ginkgo}
and libGroomRL~\cite{Carrazza2019LibGroomRL}, the MEDIC fault-injection
framework for DQM~\cite{Bassa2026MEDIC}, CMS Open Data together with the public
trigger-control code of~\cite{Ding2026LearningTrigger}, and the ACTS Open Data
Detector with the ColliderML high-luminosity dataset~\cite{Acts2022,
Elitez2025ColliderML}.

\paragraph{Statistical validity of policy-driven decisions.}
When an RL policy makes decisions on which data are recorded or certified,
physics analyses inherit them.
An adaptive trigger changes the event selection over time~\cite{Ding2026LearningTrigger},
and automated DQM decides which data enter an analysis~\cite{jullian2025human}:
the resulting selection efficiencies must remain measurable, and their
systematic uncertainties controlled, for cross-section measurements to remain
valid.
This requires, for example, logging the policy state alongside the data and
monitoring the selection with independent reference samples.
Adversarial stress tests of measurements against hidden systematic
effects~\cite{Golutvin2023DLAdvocate} indicate the kind of validation that
policy-driven selections will need.

% -----------------------------------------------------------------------------
\subsection{Outlook}
\label{sec:conclusions:closing}

The applications reviewed in this survey span a gradient of maturity.
Accelerator and trigger control have produced results on real machines and on
real collision data; data quality monitoring, tracking, and detector design have
validated prototypes on simulated or recorded data; BSM model building and
formal computation remain exploratory, although RL-based string model building
has already produced new candidate models and structural results.
Across this gradient, the common first obstacle is formalizing a physics task as
a Markov decision process, which requires a shift in how HEP problems are
approached: 1)~the task must be framed as a decision process in which the agent
interacts with the problem, rather than as a static learning problem; 2)~instead
of collecting labeled data, as in supervised learning, a reward function must be
defined that guides the agent toward the desired solution; and 3)~since particle
physics problems are not standard RL tasks, the lack of off-the-shelf simulators
and learning environments requires researchers to build their own before
addressing the target problem.

\paragraph{Conditions under which RL is advantageous.}
The evidence reviewed in this survey does not support RL as a universal
replacement for established methods; rather, it identifies the conditions under
which RL is the more suitable choice, anticipated in
Section~\ref{sec:intro:motivation} and supported here by the reviewed results:
\begin{itemize}
    \item \textbf{Repeated tasks with a fast simulator.} When a task recurs on a
    stable system, the training cost is amortized over many deployments: at
    ARES, RL-trained optimizers reach a lower final beam error than Bayesian
    optimization in most real-machine trials, at an inference cost orders of
    magnitude lower~\cite{Kaiser2024RLOvsBO}.

    \item \textbf{Decisions inside fast control loops.} Where iterative
    optimizers cannot run within the loop period, a trained policy can, as in
    the microsecond-latency feedback at KARA~\cite{Scomparin2026KARA} and the
    FPGA implementation of a regulation policy for the Fermilab Booster, tested
    on a bench setup~\cite{StJohn2021Booster}.

    \item \textbf{Robustness to faults and corrupted measurements.} Policies
    trained over a distribution of conditions degrade gracefully when the
    system does: with stuck actuators, the LHC tune feedback agents stay closer
    to the target than the PI controller, which fails~\cite{Grech2022LHCTune},
    and RL-trained optimizers
    recover a beam lost from the diagnostic screen within a few
    steps~\cite{Kaiser2024RLOvsBO}.

    \item \textbf{Physics as the training signal in place of labels.} The reward
    can encode first-principles knowledge where ground truth is unavailable,
    from the Coulomb scattering likelihood in track
    finding~\cite{Kortus2023Tracking} to tree-level matrix elements in event
    reconstruction~\cite{Dillon2026TheoryRL}.

    \item \textbf{Sequential discrete construction with a reusable policy.} When
    the objective is defined only through a sequence of discrete
    choices~(Figure~\ref{fig:construction}), a
    trained policy can be reused across instances: in string model building,
    agents remain efficient on manifolds they were not trained
    on~\cite{Larfors2020LineBundle}.
\end{itemize}
Conversely, Bayesian optimization remains preferable for new setups, rarely
repeated tasks, and systems without a fast simulator~\cite{Kaiser2024RLOvsBO};
optimization through a differentiable surrogate, used directly or inside RL
training, is more efficient than model-free RL when the objective is smooth and
high-dimensional~\cite{Rajput2025DDRL}; genetic algorithms perform comparably
when a set of solutions rather than a policy is needed~\cite{Abel2022GAvsRL,
Cole2021GAandRL, Rajput2025DDRL}; and transformers proved more powerful than RL
for symbolic simplification~\cite{Dersy2022Polylog}.

\paragraph{Transfer within the construction pattern.}
Because the construction problems of Sections~\ref{sec:jets}
to~\ref{sec:analysis} share the structure of Figure~\ref{fig:construction},
techniques developed in one domain apply to the others, although the
corresponding communities rarely cite each other.
Tree search, through MCTS guided by a learned policy or $A^*$ with an analytic
heuristic, has so far been used only in jet
clustering~\cite{Brehmer2021MCTS, Greenberg2021Astar}, and could
equally guide BSM model building or detector design, where each evaluation is
expensive.
Detector design penalizes infeasible layer placements in the
reward~\cite{Qasim2025DetectorRL}, whereas BSM model building excludes
infeasible edits through grammar masks~\cite{Wojcik2025BSMGraphPRD,
Alexander2026Albert}, and multi-agent tracking through a safety
layer~\cite{Kortus2026MARL}, so that the agent does not need to learn the
constraint.
Pretraining the policy before RL fine-tuning, used in theory
generation~\cite{Alexander2026Albert} and in jet
ordering~\cite{Erdmann2020ReinforcedSorting}, has not yet been tried in tracking
or detector design.

\paragraph{Metrics for future facilities.}
As the field moves toward the HL-LHC and future colliders such as FCC-ee, RL
systems will be judged on metrics that reflect operational and physics
requirements rather than on the return of a benchmark environment.
In accelerator control, the relevant quantities are the time and number of
machine interactions to convergence, the final beam quality, the smoothness of
actuator trajectories~\cite{Kaiser2024RLOvsBO}, the inference latency relative
to the loop period~\cite{Scomparin2026KARA}, the performance retained under
drift without retraining~\cite{Dong2026NSLSII}, and the number of constraint
violations.
In trigger control, they are the fraction of time within the target rate band,
the signal efficiency on benchmark processes, the smoothness of threshold
updates~\cite{Ding2026LearningTrigger}, and the stability and measurability of
the trigger efficiency over time.
In DQM, they are the number of expert queries per run, the rates of missed bad
data and of good data wrongly flagged, and the detection lead
time~\cite{Bower2020ATSB}.
In tracking, they are efficiency, fake rate, and time per event at high pileup,
compared with the combinatorial Kalman filter.
In detector design, they are the physics figure of merit against cost and
material budget, the number of full-simulation calls, and the robustness of the
design across physics tunes and calibrations~\cite{Figalli2026DesigningAI}.
In model building and formal computation, they are the number of viable models
per expensive evaluation, the coverage of the model space relative to random or
Markov chain Monte Carlo sampling, the rediscovery of known models, and the
computational cost relative to standard heuristics~\cite{Zeng2025IBPRL}.
Across domains, the compute spent on training must be weighed against the
operational savings it enables.

\paragraph{Convergence with foundation models.}
A methodological convergence with large-scale machine learning is also emerging.
Group-relative policy optimization, developed for language
models~\cite{Shao2024GRPO}, now appears in trigger
control~\cite{Ding2026LearningTrigger} and in theory generation, where it
fine-tunes a pretrained transformer~\cite{Alexander2026Albert}, mirroring the
pretrain-then-RL recipe of language model alignment~\cite{Ouyang2022RLHF}.
Combined with the physics foundation models discussed above, this suggests a
path toward policies that start from broad pretrained knowledge and are
specialized to a facility or task with RL.

Throughout the survey, RL has often found decision rules that match or improve
on those designed by domain experts.
This does not mean that traditional approaches, which are deeply grounded in
physics knowledge, should be abandoned; rather, they should be complemented by
RL where the conditions above apply.
Conversely, traditional methods and domain knowledge are a source of
inspiration, and their combination with RL, through physics-informed rewards,
constrained action spaces, and differentiable or learned models of the system,
may mitigate or even resolve many of the theoretical and practical challenges
discussed here and in the previous sections.
Beyond reviewing which high-energy and particle physics problems can be
addressed with RL, this survey aims to enable physics researchers to become
familiar with the RL formalism quickly, so as to foster future research on both
existing and new particle physics problems; to this end, we reviewed not only
the basics of deep reinforcement learning but also its sub-fields, such as
offline, inverse, and model-based RL, that are most relevant to the surveyed
applications.